\documentclass[lettersize,onecolumn]{IEEEtran}
\normalsize
\usepackage{graphicx}
\usepackage{float}
\usepackage{subfigure}
\usepackage{lineno}
\usepackage{verbatim}
\usepackage{cite}
\usepackage{latexsym}
\usepackage{amsmath}
\usepackage{amssymb}
\usepackage{mathrsfs}
\usepackage{balance}
\usepackage{setspace}
\usepackage{epstopdf}
\usepackage{graphics}

\usepackage{caption}
\usepackage{graphicx}
\usepackage{float} 
\usepackage{subcaption}

\usepackage{hyperref}

\usepackage[usenames]{color}
\usepackage{mdwmath}
\usepackage{mdwtab}
\usepackage{array}
\usepackage{picins}
\usepackage{eqparbox}
\usepackage{lettrine}
\usepackage{bbm}
\usepackage{dsfont}
\usepackage[noend]{algpseudocode}
\usepackage{algorithmicx,algorithm}

\usepackage{tikz}
\usetikzlibrary{matrix}
\usetikzlibrary{decorations.pathreplacing}

\newtheorem{remark}{Remark}
\newtheorem{theorem}{Theorem}
\newtheorem{lemma}{Lemma}
\newtheorem{definition}{Definition}
\newtheorem{prop}{Proposition}

\newtheorem{example}{Example}
\newtheorem{corollary}{Corollary}

\allowdisplaybreaks[4]
\newcommand*{\QEDA}{\hfill\ensuremath{\blacksquare}}  
\begin{document}

	\title{On the Stochastic Analysis of Random Linear Streaming Codes in Multi-Hop Relay Networks}

    \author{Kai Huang, Xinyu Xie, Chunpeng Chen, Wenjie Guan, Xiaoran Wang
    , Jinbei Zhang
    \\

School of Electronic and Communication Engineering, Sun Yat-sen University, Shenzhen, China\\

Email: \{huangk56, xiexy55, chenchp6, guanwj9, wangxr77\}@mail2.sysu.edu.cn
, zhjinbei@mail.sysu.edu.cn
}
    {}
	\maketitle

	\begin{abstract}
		\boldmath
        In this paper, we aim to explore the stochastic performance limit of large-field-size Random Linear Streaming Codes (RLSCs) in multi-hop relay networks. 
        In our model, a source transmits a sequence of streaming messages to a destination through multiple relays subject to a delay constraint. 
        Most previous research focused on \textit{deterministic adversarial channel} which introduces only restricted types of erasure patterns, and aimed to design the optimal capacity-achieving codes.        
        In this paper, we focus on \textit{stochastic channel} where each hop is subject to i.i.d. packet erasures, and carry out stochastic analysis on the error probability of multi-hop RLSCs.
        Our contributions are three-folds.
        Firstly, the error event of large-field-size RLSCs is characterized in two-hop relay network with a novel framework, which features quantification of information flowing through each node in the network.
        Due to the erasures in different hops, some source symbols can be ``detained" at the source or relay while others have arrived at the destination.
        By iteratively computing the number of detained symbols at each node, this framework extends the concept ``information debt" from point-to-point network [Pinwen Su et al. 2022] into two-hop relay networks. 
        Secondly, based on the error event, the expression of average error probability in two-hop network is derived by carefully analyzing the expectation terms. 
        To handle the expectation over all possible erasure patterns along two hops of the network, the transition matrices of the detained symbols are novelly constructed in a ``band fashion" with \textit{nested structure}.
        Thirdly, the derived results in two-hop network are further generalized into relay networks with arbitrary number of hops. 
        The generalization is majorly due to the extendibility of the proposed framework as well as the nested structure of matrix construction.
        Furthermore, simulations are conducted to verify the accuracy of our stochastic analysis on the error probabilities, and compare with some existing streaming codes for the adversarial channels.        
	\end{abstract}

\begin{IEEEkeywords}
Random linear streaming codes, multihop relay networks, stochastic analysis, large finite field.
\end{IEEEkeywords}

    \IEEEpeerreviewmaketitle

	\section{Introduction} \label{section:introduction}

The demand for low-latency and reliable communication has surged in recent years, driven by applications such as video conferencing, live streaming, autonomous vehicles, virtual reality and telemedicine. In these scenarios, data packets must be transmitted and decoded within strict deadlines, making traditional retransmission-based mechanisms like Automatic Repeat Request (ARQ) unsuitable due to their inherent round-trip delays. To address this challenge, Forward Error Correction (FEC) techniques, particularly streaming codes, have emerged as a promising solution. These codes are designed to recover packet erasures in real-time while adhering to stringent decoding delay constraints, ensuring seamless communication even in unreliable networks.

\subsection{Point-to-point Streaming Codes for Adversarial Channel}

Streaming codes were initially investigated in point-to-point networks, where the focus was on correcting burst erasures and random packet losses under delay constraints. The seminal work by Martinian and Sundberg \cite{bursty only}, laid the foundation for burst erasure correction with low decoding delay. 
Particularly, \cite{bursty only} first investigated a bursty $B$-erasure channel with decoding delay $T$. 
Subsequent research extended the results from only bursty erasures \cite{b1,b2,b3} to both bursty and isolated erasures \cite{bn1,bn2,bn3,adversarial channel,SLFTCOM,SLFTIT,variable size}.
Specifically, \cite{bn1,bn2,bn3,adversarial channel,SLFTCOM,SLFTIT,variable size} generalized the original bursty erasure model into the $(W,B,M)$-\textit{sliding window packet erasure channels (SWPEC)}, which introduces either one burst erasure with length no longer than $B$ or multiple arbitrary erasures with total count no larger than $M$ within any window of length $W$. 
Since then, SWPEC has become a main stream while investigating the low-latency streaming codes. 
The main focuses of the above research are to design error-free construction of streaming codes  (mostly employing the diagonal interleaving technique) in order to match the capacity of SWPEC, and possibly reduce the scale of the operation finite field.

\subsection{Multi-hop Streaming Codes for Adversarial Channel}

Recently, the scope of streaming codes has expanded to multi-hop networks, especially the three-node (or equivalently, two-hop) relay network, which consists of a source, a relay, and a destination. This topology is prevalent in content delivery networks and other practical communication systems. The relay network introduces additional challenges, as erasures can occur in both the source-to-relay and relay-to-destination links, each with distinct erasure patterns. Fong et al. \cite{SLF 3hop first paper} pioneered the study of streaming codes in the three-node relay network setting.
Particularly, \cite{SLF 3hop first paper} considered SWPEC that introduces no more than $N_1$ and $N_2$ erasures in the windows of the first-hop channel and second-hop channel, respectively, and proposed a Symbol-Wise Decode-Forward (SWDF) strategy where the source symbols within the same message are decoded by the relay with different delays.
\cite{MNK adaptive relaying,GKF adaptive relaying TIT} improved upon the construction of \cite{SLF 3hop first paper} by adapting the relaying strategy based on the erasure patterns appeared in the first hop, which allows the relay to forward information about symbols before it can decode, and variable-rate encoding, which decreases the rate used to encode a packet as more erasures affect that packet.
Moreover, the bursty erasure in decoding window was further considered in \cite{VR 3hop burst,SS 3hop burst and arbitrary,JJH 3hop burst and arbitrary}.
Furthermore, \cite{ED multihop,ED multihop TIT} extended the three-node relay network into a more general model, i.e., the multi-hop relay network, where each link is subjected to a certain maximum number of packet erasures.
\cite{ED multihop,ED multihop TIT} also employed adaptive relaying strategy (referred to as the ``state-dependent scheme" therein), which exploits the ability of each relay node to adapt the encoded transmission depending on the erasures on its previous link. 
Although an additional header should be appended to the encoded relaying transmission, \cite{ED multihop,ED multihop TIT} showed that this overhead vanishes as the size of the finite field increases.
Beyond that, the linear relaying model was also investigated under difference topologies, e.g., multi-link multi-hop network\cite{ED multilink multihop}, multicast relayed networks\cite{GKF 3hop multicast}, and multi-access relayed networks\cite{GKF 3hop multiaccess}.
Focusing on the delay-vs-throughput perspective, \cite{DOO multihop DAF} defined a metric called the Delay Amplification Factor, i.e.,  DAF$(R)$ as a function of the throughput $R$, which characterizes the growth rate of the asymptotic delay with respect to the number of hops. 
\cite{DOO multihop DAF} implied that the existing Decode-and-Forward designs can lead to a linearly growing delay and it could be circumvented with a new delay-centric solution.

The aforementioned investigations on streaming codes mainly focused on the worst-case performance over a predefined deterministic subset of erasure patterns, and thus is referred to as the \textit{adversarial channel}.
Adversarial channel is broadly regarded as tractable approximation of the \textit{stochastic channel}. 
For example, a SWPEC introducing at most $N$ arbitrary packet erasures can model the i.i.d. packet erasure channel, while a SWPEC introducing at most $N$ arbitrary or $B$ packet erasures can be viewed as a tractable approximation to the commonly-accepted Gilbert-Elliott channel model.
However, the works on adversarial channel majorly concentrate on the type of the erasure patterns (e.g., the number and ordering of erasures in a window), while ignoring the statistics (probability distribution of the erasure patterns) of the channel. 
In above works, the verification of expected performance in practical stochastic scenarios majorly relies on numerical simulation. 


\subsection{Point-to-point Streaming Codes for Stochastic Channel}

\cite{RLSCs ISIT,RLSCs,asymptotics1,asymptotics2,asymptotics3,Analysis in Stochastic Channel} carried out stochastic analysis on the average error probability of random linear streaming codes. 
In \cite{RLSCs ISIT,RLSCs,asymptotics1,asymptotics2,asymptotics3}, RLSCs under sufficiently large finite size regime were intensively studied in a point-to-point i.i.d. Symbol Erasure Channel (SEC), where the occurrence of erasures are probabilistic. 
Particularly, \cite{RLSCs ISIT,RLSCs} generalized the concept of \textit{information debt}, which was first proposed in \cite{phd} and used to describe how many linear equations the destination still needs for successful decoding.
Under the Generalized Maximum Distance Separable (GMDS) condition, \cite{RLSCs ISIT} and \cite{RLSCs} characterized the error event of large-finite-field RLSCs for any finite memory length $\alpha < \infty$ and any finite decoding deadline $\Delta < \infty$. Then the closed-form expression of the exact error probability was derived with a novel random-walk-based analysis framework. 
In \cite{asymptotics1,asymptotics2,asymptotics3}, asymptotic results (with some parameters being asymptotically large) were developed.
More recently, \cite{Analysis in Stochastic Channel} extended the results of \cite{RLSCs} from i.i.d. SEC into the more practical Gilbert-Elliott SEC \cite{Gilbert,Elliott}, which can capture both burst and arbitrary errors. The performance of systematic and non-systematic RLSCs are both derived analytically.


\subsection{Challenges and Our Contributions}

In this paper, we investigate the performance limit of large-field-size RLSCs in Multi-hop Relay Networks (MRN) with i.i.d. packet erasure channel.
When stochastic channel with multi-hops is considered, the performance analysis of RLSCs will become more challenging, majorly due to the following two aspects.
(1) Without delicately designed generator matrix, the error events (including successful decoding event and the error event of both decoding failure and exceeding the latency constraint) of RLSCs are challenging to characterize, especially in multi-hop networks. In multi-hop scenario, due to the erasures in different hops, the delivery of a portion of source symbols could be delayed.
More importantly, due to the distinct erasure patterns in different hops, it is possible that some source symbols have been able to be decoded at the destination, while other source symbols are still detained at the relay or source nodes. 
The information detained at each relay can be distinct in each timeslot.
Thus, how to characterize the information detained at each node in each timeslot, and furthermore, characterize the error event accordingly is so far unsolved. 
(2) When the channel erasure is stochastic, the expected performance should be averaged over all possible erasure patterns.
The derivation will be even more analytically challenging in multi-hop networks, since the dimensions of erasure patterns increase exponentially along with the number of hops. 

Overcoming the above challenges, our main contributions can be summarized as follows.
\begin{itemize}
    \item First, we characterize the error event of large-field-size RLSCs in the Two-hop Relay Networks (TRN), with a novel framework featuring quantification of information flowing through each node in the network.
    Due to the erasures in different hops, the packets received at relay or destination not necessarily contain information of all latest source symbols. 
    In other words, the information of some source symbols can be detained at source and relay (in the form of linear combination mixed with other source symbols). 
    To quantify the information flowing through each node, we introduce a novel framework to iteratively compute the number of source symbols detained at each node. 
    The framework is a generalization of the concept ``information debt" from point-to-point network \cite{RLSCs} to the multi-hop relay network.

    \item Second, we derive the expression of error probability of RLSCs in TRN based on the characterization of error event.
    To handle the expectation over all possible erasure patterns along multiple hops of the network, the transition matrices of the detained symbols are novelly constructed in a ``band fashion" with \textit{nested structure}, which can be generated by a series of matrix-embedding operations.
    Then the average error probability is derived by carefully analyzing the expectation terms. 
        
    \item  Third, we show that the results in TRN can be extended to arbitrary $L$-node relay networks.
    The extendibility is majorly based on the proposed nested structure of the transition matrices.
    The correctness of our theoretical results is also numerically verified via Monte-Carlo simulations.  
    
\end{itemize}

The rest of the paper is organized as follows. In Section \ref{section:model statement}, we describe the considered model of multi-hop relay networks and random linear streaming codes. In Section \ref{section:Main Results1}, we present the characterization of error event of RLSCs in TRN. 
In Section \ref{section:Main Results2}, we present the characterization of error probability of RLSCs in TRN. 
In Section \ref{section:Main Results3}, we extend the results of TRN to MRN with $L$ hops. 
The numerical comparisons are presented in Section \ref{section:numerical results}. We conclude in Section \ref{section:conclusion}.

\textit{Notations:} In this paper, for some integers $a$ and $b$, $\{a,a+1,\dots,b\}$ is denoted as  $[a,b]$ and $\{1,2,\dots,a\}$ is denoted as $[a]$. $\Phi$ represents the empty set. 
$\mathbb{N}$ represents natural numbers.
For simplicity, the complement of $x\in [0,1]$ is represented as $\bar{x}\triangleq 1-x$.
The probability is denoted by $\text{Pr}(\cdot)$, and the expectation is denoted by $\mathbb{E}\{\cdot\}$. We use $(\cdot)^\top$ to represent the transpose of a matrix or a vector. We use $\mathbf{s}(a:b) \triangleq [\mathbf{s}^\top(a),\mathbf{s}^\top(a+1),\dots, \mathbf{s}^\top(b)]^\top$ to represent the \textit{cumulative} column vector. 
Submatrix of a matrix $A$ generated by slicing of the rows $a$ to $b$ and columns $c$ to $d$ is denoted as $A(a:b,c:d)$. Slicing of all the rows/columns is denoted by ``:'', e.g., $A(1:3,:)$ denotes the first three rows of matrix $A$.
$\mathds{1}\{\cdot\}$ is the indicator function. In the presented partitioned matrices, the omitted entries are all zeros.
$\vec{\mathbf{1}}$ and $\vec{\mathbf{0}}$ are used to represent column vectors of all 1s or 0s, respectively.
$\vec{\delta}_{k}$ is a column vector where the $k$-th entry is one and all other entries are zeros.
Identity matrix of size $n$ is denoted by $\mathbf{I}_n$. 
Let $diag[x(1),\cdots,x(n)]$ be the $n$-by-$n$ diagonal matrix with $x(1),\cdots,x(n)$ listed sequentially in its main diagonal.


\section{System Model and Definitions} \label{section:model statement}

In this section, we describe the model of multi-hop relay networks and the random linear streaming codes.

\begin{figure*}[!hbtp]
    \centering
    \includegraphics[width=1\linewidth]{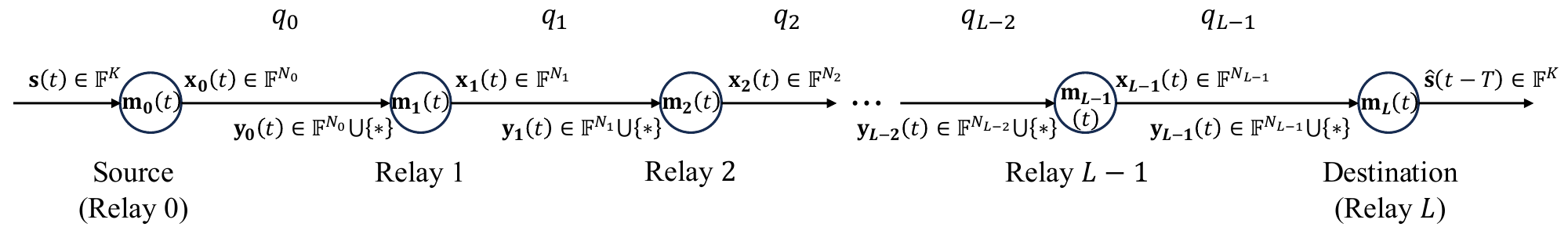}
    \caption{Illustration of a $L$-hop relay network (with $L+1$ nodes).}
    \label{figure:Illustration_of_multihop}
\end{figure*}

Consider a $L$-hop relay network as shown in Fig. \ref{figure:Illustration_of_multihop}, which consists of a source node, $L-1$ relays and a destination node. Denote the $l$-th relay as $r_l$ for shorthand, $l\in[1,L-1]$. 
For consistency, the source node and the destination node are labeled as relay $r_0$ and relay $r_L$, respectively.
The link connected between any two adjacent relay is regarded as an i.i.d. packet erasure channel. Denote the probability of successful delivery on the link $(r_l,r_{l+1})$ as $q_l$, $\forall l\in [0,L-1]$.
Denote the indicator function of erasure on link $(r_l,r_{l+1})$ at timeslot $t$ as $e_l(t)$. Let $e_l(t) = 1$ if the erasure occurs on $(r_l,r_{l+1})$ at timeslot $t$ and $e_l(t) = 0$ if the packet is delivered perfectly.
In addition to the basic transmission and reception functions, each node also has encoding and storage functions. 
We assume that the storage size of each node is large enough to cache all the received packets until all source symbols are successfully decoded or determined as error\footnote{This assumption is line with \cite{asymptotics2}. However, in practical application the memory size does not need to be asymptotically large. This argument will be discussed in Remark \ref{remark:infinite memory}.}.
At timeslot $t$, the encoded symbols transmitted by $r_l$ and the symbols received by $r_{l+1}$ are denoted as $\mathbf{x}_l(t)$, $\mathbf{y}_l(t)$, respectively, and the storage of $r_l$ is denoted as $\mathbf{m}_l(t)$.
The length of the packet encoded at $r_l$ equals to $N_l$, i.e., $\mathbf{x}_l(t)\in \mathbb{F}^{N_l}$.
In the following, we detail the transmission process of the network.

\textbf{Source Encoder:} For any timeslot $t\ge 1$, $K$ source symbols $\mathbf{s}(t)=[s_1(t),\dots, s_K(t)]^\top$ will arrive at $r_0$. Each symbol $s_k(t), k\in [K]$ is an i.i.d. sample from a large finite field $\mathbb{F}$. 
The arrived source symbols will be cached into the memory of $r_0$ in order, i.e., 
\begin{equation}
    \mathbf{m}_0(t) = [\mathbf{s}^\top(1),\mathbf{s}^\top(2),\dots, \mathbf{s}^\top(t)]^\top \triangleq \mathbf{s}(1:t).\label{equation:m=s}
\end{equation}
In each timeslot $t$, the source encoder uses all the received symbols until time $t$ as input and outputs one encoded packet with $N_0$ symbols $\mathbf{x}_0(t) = [x_1(t),\dots,x_{N_0}(t)]^\top \in \mathbb{F}^{N_0}$ for transmission.
Let $\mathbf{G}_0(t)$ be the \textit{generator matrix} at $r_0$ for timeslot $t$, thus 
\begin{equation}\label{equation:source encoding}
    \mathbf{x}_0(t) = \mathbf{G}_0(t) \cdot \mathbf{m}_0(t).
\end{equation}

\textbf{I.i.d. Packet Erasure Channel and Relay Encoder:}
In every timeslot $t$, the $N_0$ encoded symbols of $\mathbf{x}_0(t)$ will be transmitted into the channel $(r_0,r_1)$ by the transmitter. 
$\mathbf{x}_0(t)$ can be erased in the channel or perfectly received by $r_1$ according to probability.
Denote the signal received by $r_1$ as $\mathbf{y}_0(t) \in \mathbb{F}^{N_0}\cup \{*\}$, where $\{*\}$ represents the erased symbol.
With probability $q_0$, $\mathbf{y}_0(t) = \mathbf{x}_0(t)$ and with probability $1 - q_0$, $\mathbf{y}_0(t) = \{*\}$, i.e., 
\begin{equation}
    \mathbf{y}_0(t) = \left\{
    \begin{aligned}
        &\mathbf{x}_0(t) \qquad\quad\quad\ \text{if} \  e_0(t) = 0\\
        &\{*\} \qquad\qquad\quad  \text{if} \ e_0(t) = 1.
    \end{aligned}
    \right
    .
\end{equation}  
Then we have 
\begin{equation}\label{equation:y0}
    \mathbf{y}_0(t) = \mathbf{H}_0(t)\cdot \mathbf{m}_0(t),
\end{equation}
where $\mathbf{H}_0(t)$ is the \textit{receiver matrix} generated by erasing the rows of $\mathbf{G}_0(t)$ which correspond to the erasure timeslots.
Note that $\mathbf{H}_0(t)$ contains both the information of generator matrix $\mathbf{G}_0(t)$ and the information of channel erasures in the first hop $e_0(t)$.

Once $r_1$ receives the symbols perfectly, it will cache them into its storage in order. 
The storage of $r_1$ is composed of all perfectly received packets from $r_0$.
For ease of presentation, denote 
$\mathbf{m}_1(t) = [\mathbf{y}_0^\top(1),\mathbf{y}_0^\top(2),\dots, \mathbf{y}_0^\top(t)]^\top \triangleq \mathbf{y}_0(1:t)$. We note that for any erased packet $\mathbf{y}_i^\top(t) = \{*\}$, the symbol $\{*\}$ will be completely evicted from $\mathbf{m}_1(t)$.
Similar to \cite{DOO multihop DAF}, we assume that each relay node only directly encodes the packets it has received and stored, and then forwards without any decoding process. 
Let $\mathbf{G}_1(t)$ be the generator matrix at $r_1$ for timeslot $t$, thus 
\begin{equation}
    \mathbf{x}_1(t) = \mathbf{G}_1(t) \cdot \mathbf{m}_1(t).
\end{equation}
Then $r_2$ will receive $\mathbf{y}_1(t) \in \mathbb{F}^{N_1}\cup \{*\}$, cache it into its memory, encode $\mathbf{x}_2(t) \in \mathbb{F}^{N_2}$, and further transmit it. Then $r_3$ will receive $\mathbf{y}_2(t) \in \mathbb{F}^{N_2}\cup \{*\}$ and so forth.
The same process continues in the same timeslot $t$ until the destination $r_L$ receives $\mathbf{y}_{L-1}(t) \in \mathbb{F}^{N_{L-1}}\cup \{*\}$.
Therefore, for any $l\in [1,L-1]$, we have
\begin{equation}\label{equation:my}
    \mathbf{m}_l(t) = \mathbf{y}_{l-1}(1:t),
\end{equation}
\begin{equation}
    \mathbf{x}_l(t) = \mathbf{G}_l(t) \cdot \mathbf{m}_l(t),
\end{equation}
\begin{equation}
    \mathbf{y}_l(t) = \mathbf{H}_l(t) \cdot \mathbf{m}_l(t).
\end{equation}

After properly shifting and stacking the $\mathbf{G}_l(t)$ and $\mathbf{H}_l(t)$ along with the timeslots respectively, we can obtain the \textit{cumulative generator matrices} $\mathbb{G}_l(t)$ and \textit{cumulative receiver matrices} $\mathbb{H}_l(t)$ satisfying that 
\begin{align}
    \mathbf{x}_l(1:t) &= \mathbb{G}_l(t) \cdot \mathbf{m}_l(t),\label{equation:encoding1}\\
    \mathbf{y}_l(1:t) &= \mathbb{H}_l(t) \cdot \mathbf{m}_l(t).\label{equation:encoding2}
\end{align}
Illustrations of equations (\ref{equation:encoding1}) and (\ref{equation:encoding2}) can be found in Example \ref{Example:3-node}. 
All entries of $\mathbb{G}_l(t)$ and $\mathbb{H}_l(t)$ in the white space are zeros, while other entries are non-zero.
By alternately plugging equations (\ref{equation:my}) and (\ref{equation:encoding2}) into each other, together with (\ref{equation:m=s}) and (\ref{equation:y0}), the observations at the destination $r_L$ can be given by
\begin{align}
    \mathbf{y}_{L-1}(1:t) &= \prod_{l=0}^{L-1} \mathbb{H}_l(t) \cdot \mathbf{s}(1:t) \label{equation:muliplication of receiver matrix}\\ 
    &\triangleq \mathcal{H}(t) \cdot \mathbf{s}(1:t), \label{equation:overall receiver matrix}
\end{align}
where $\mathcal{H}(t)$ is the \textit{overall cumulative receiver matrix} which is the result of all encoding operations and channel erasures until timeslot $t$ through the network.

\textbf{Decodability at the Destination:} 
At timeslot $t$, the destination $r_L$  can observe $\mathbf{y}_{L-1}(1:t)$. The decoder should try to decode $\mathbf{s}(t)$ at timeslot $t+\Delta$ with observations $\mathbf{y}_{L-1}(1:t+\Delta)$, where $\Delta$ is the decoding delay.
The decodability at the destination is defined as follows.
\begin{definition}\label{Definition decodability 1}
The symbol $s_k(t)$ is $\Delta$-decodable if the vector $\vec{\delta}_{(t-1)K+k}^\top$ is in the row space of $\mathcal{H}(t+\Delta)$, where $\vec{\delta}_{(t-1)K+k}$ is a column vector such that its $((t-1)K+k)$-th element is one and all the other elements are zeros. 
\end{definition}
\begin{definition}\label{Definition decodability 2}
The vector $\mathbf{s}(t)$ is $\Delta$-decodable if all symbols $\{s_k(t):k\in[K]\}$ are $\Delta$-decodable. 
\end{definition}

For the simplest multi-hop relay network, i.e., two-hop relay network with $L=2$, an example is given as follows.

\begin{example}\label{Example:3-node}
Assume that eight timeslots are considered. At the source $r_0$, the overall encoding process is given by equation (\ref{equation:encoding1}), i.e., $\mathbf{x}_0(1:8) = \mathbb{G}_0(8) \cdot \mathbf{m}_0(8)$, as shown in Fig. \ref{fig:Example_1_1}, wherein the multiplication framed in the blue boxes stands for equation (\ref{equation:source encoding}) that $\mathbf{x}_0(4) = \mathbf{G}_0(4) \cdot \mathbf{m}_0(4)$. 
Assume that in the first hop, encoded packets $\mathbf{x}_0(3),\mathbf{x}_0(4),\mathbf{x}_0(6),\mathbf{x}_0(7)$ are erased.
Thus the receiver matrix $\mathbb{H}_0(8)$ can be derived from  $\mathbb{G}_0(8)$ by eliminating the corresponding rows to the erasure timeslots $3,4,6,7$.
The erasure process is given by equation (\ref{equation:encoding2}) as shown in Fig. \ref{fig:Example_1_2}, wherein the multiplication framed in the blue boxes stands for equation (\ref{equation:y0}) that $\mathbf{y}_0(5) = \mathbf{H}_0(5) \cdot \mathbf{m}_0(5)$.
Due to the erasures in the first hop, the storage of relay $r_1$ are filled sequentially with the received packets $\mathbf{y}_0(1),\mathbf{y}_0(2),\mathbf{y}_0(5),\mathbf{y}_0(8)$ at timeslots 1,2,5 and 8, respectively.
Therefore, $\mathbf{m}_1(3)$ and $\mathbf{m}_1(4)$ remain equal to $[\mathbf{y}_0^\top(1),\mathbf{y}_0^\top(2)]^\top$, 
since no new packets are received and added into the storage at timeslots 3,4 for subsequent encoding.
As a result, at timeslots $3,4$, $r_1$ can only keep encoding the packets with the information of $[\mathbf{y}_0^\top(1),\mathbf{y}_0^\top(2)]^\top$ (with different generator matrices).
The corresponding encoding process at relay $r_1$ is presented in Fig. \ref{fig:Example_1_3}, wherein the white blanks are due to the channel erasures in the first hop $(r_0,r_1)$.
After removing the entries of erasures, Fig. \ref{fig:Example_1_4} can be obtained, wherein the multiplication framed in the red boxes stands for the equation $\mathbf{x}_1(6) = \mathbf{G}_1(6) \cdot \mathbf{m}_1(6)$.
Assume that in the second hop, encoded packets $\mathbf{x}_1(2),\mathbf{x}_1(5),\mathbf{x}_1(7)$ are erased.
After removing the corresponding rows in $\mathbb{G}_1(8)$, $\mathbb{H}_1(8)$ can be derived as shown in Fig. \ref{fig:Example_1_6}.
Fig. \ref{fig:Example_1_7} stands for equation (\ref{equation:muliplication of receiver matrix}) which shows the received symbols at the destination can be presented as a multiplication of receiver matrices, and Fig. \ref{fig:Example_1_8} stands for the overall cumulative receiver matrix $\mathcal{H}(8)$ in (\ref{equation:overall receiver matrix}).
Then $\mathcal{H}(8)$ can be used to determine the decodability at the destination $r_2$ with \textit{Definitions \ref{Definition decodability 1} and \ref{Definition decodability 2}}.

    \begin{figure}[!hbtp]
        \centering
        \includegraphics[width=0.7\linewidth]{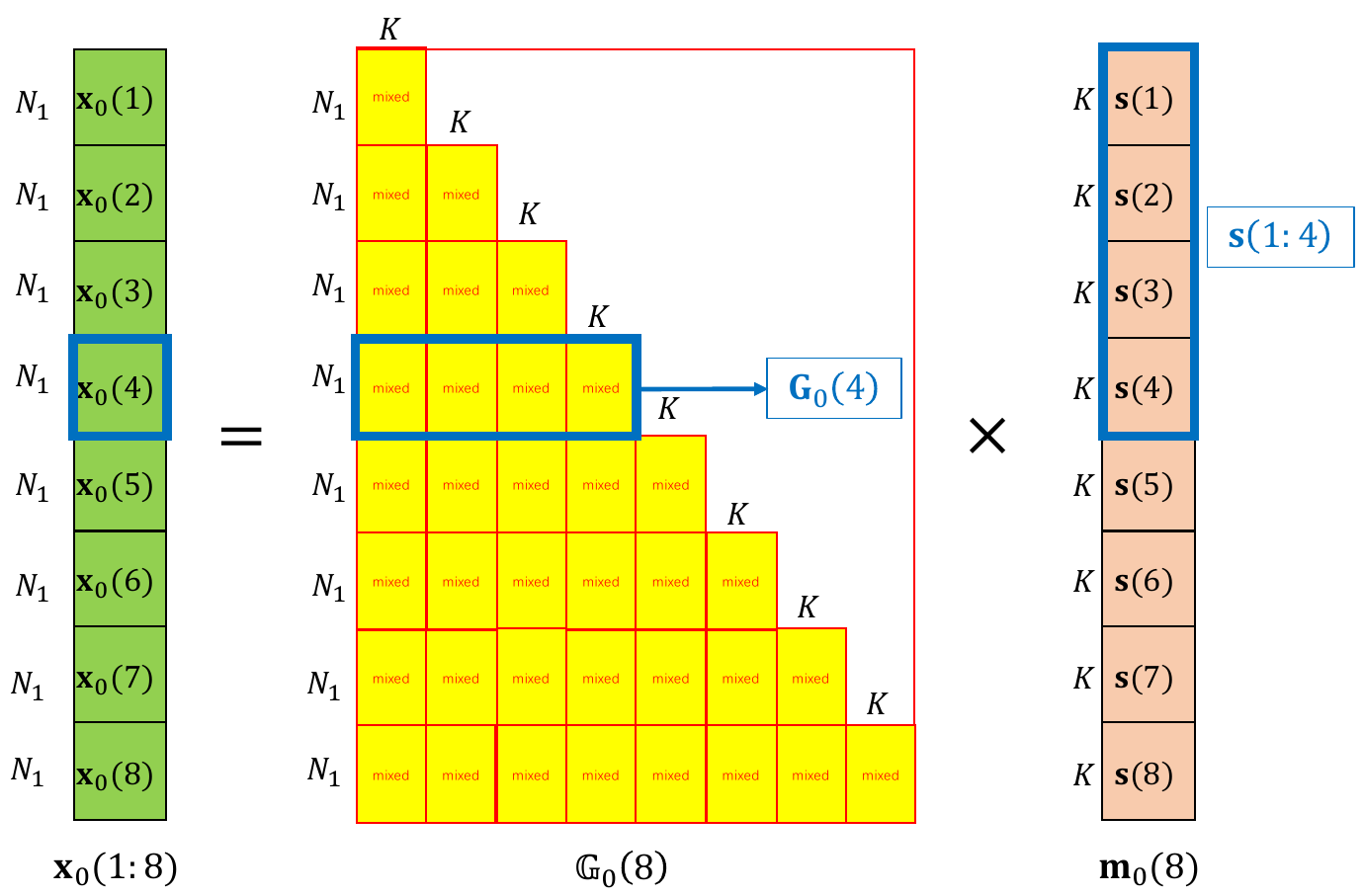}
        \caption{The encoding process at the source $r_0$.}
        \label{fig:Example_1_1}
    \end{figure}
    \begin{figure}
        \centering
        \includegraphics[width=0.7\linewidth]{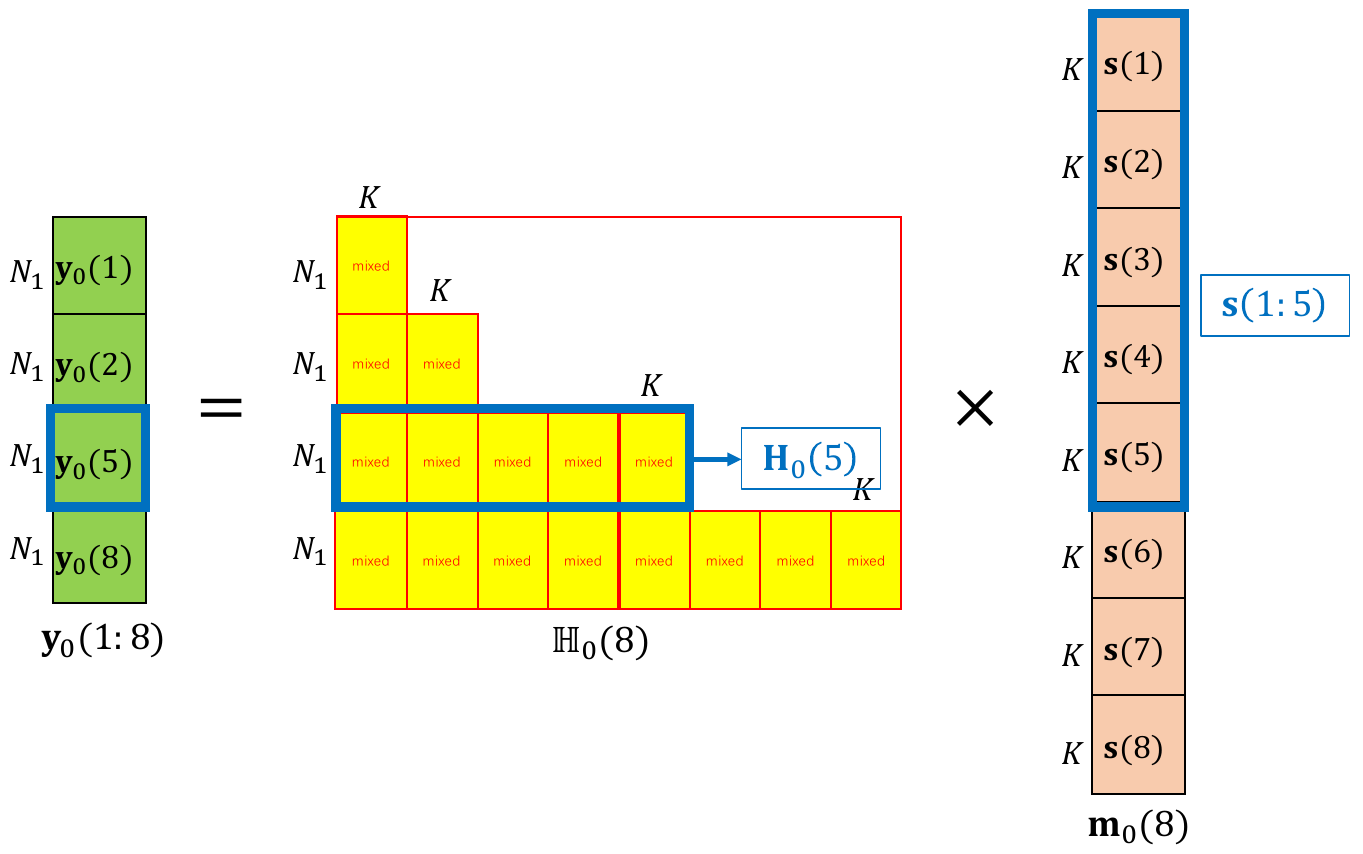}
        \caption{$\mathbf{x}_0(3),\mathbf{x}_0(4),\mathbf{x}_0(6),\mathbf{x}_0(7)$ are erased in $(r_0,r_1)$.}
        \label{fig:Example_1_2}
    \end{figure}
    \begin{figure}
        \centering
        \includegraphics[width=0.62\linewidth]{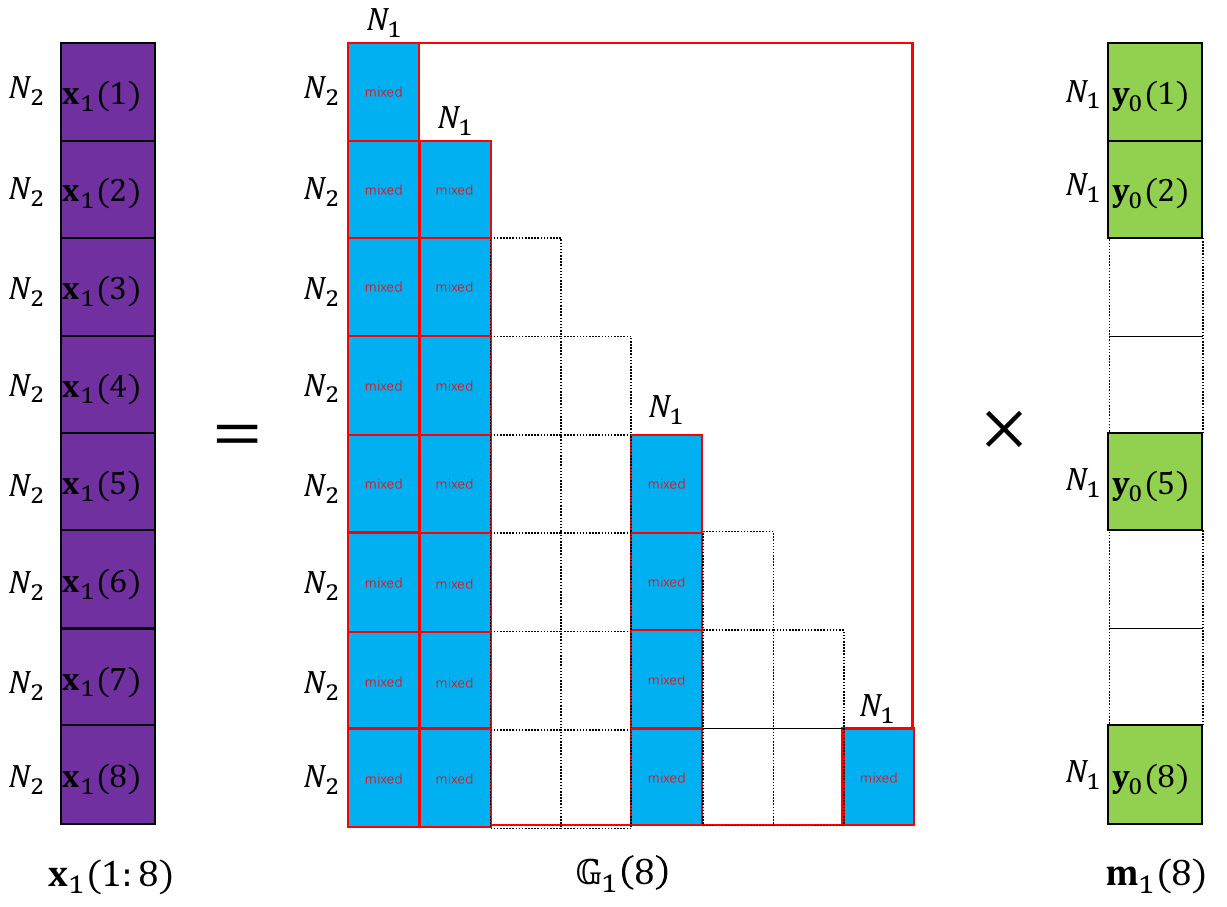}
        \caption{The encoding process at the relay $r_1$. The white blanks are due to the channel erasures in the first hop $(r_0,r_1)$.}
        \label{fig:Example_1_3}
    \end{figure}
    \begin{figure}
        \centering
        \includegraphics[width=0.6\linewidth]{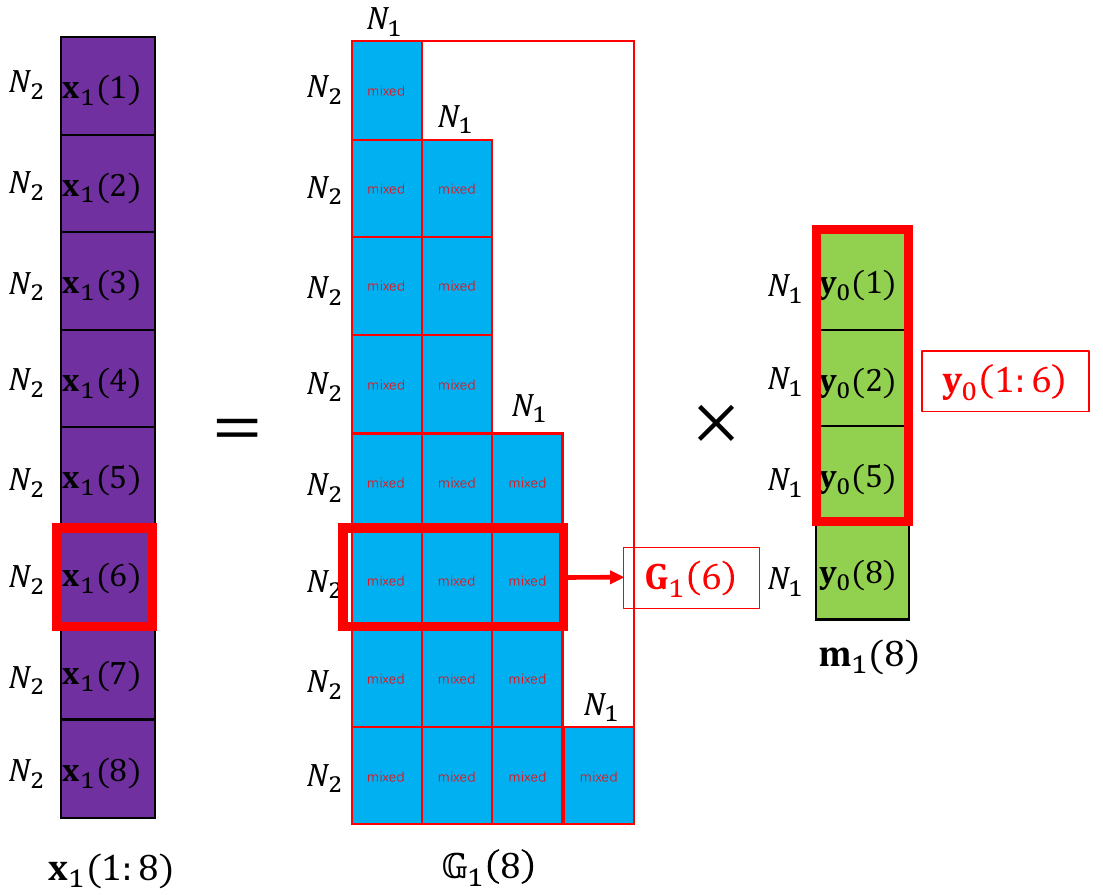}
        \caption{The encoding process at the relay $r_1$ after eliminating the erased entries.}
        \label{fig:Example_1_4}
    \end{figure}
    \begin{figure}
        \centering
        \includegraphics[width=0.55\linewidth]{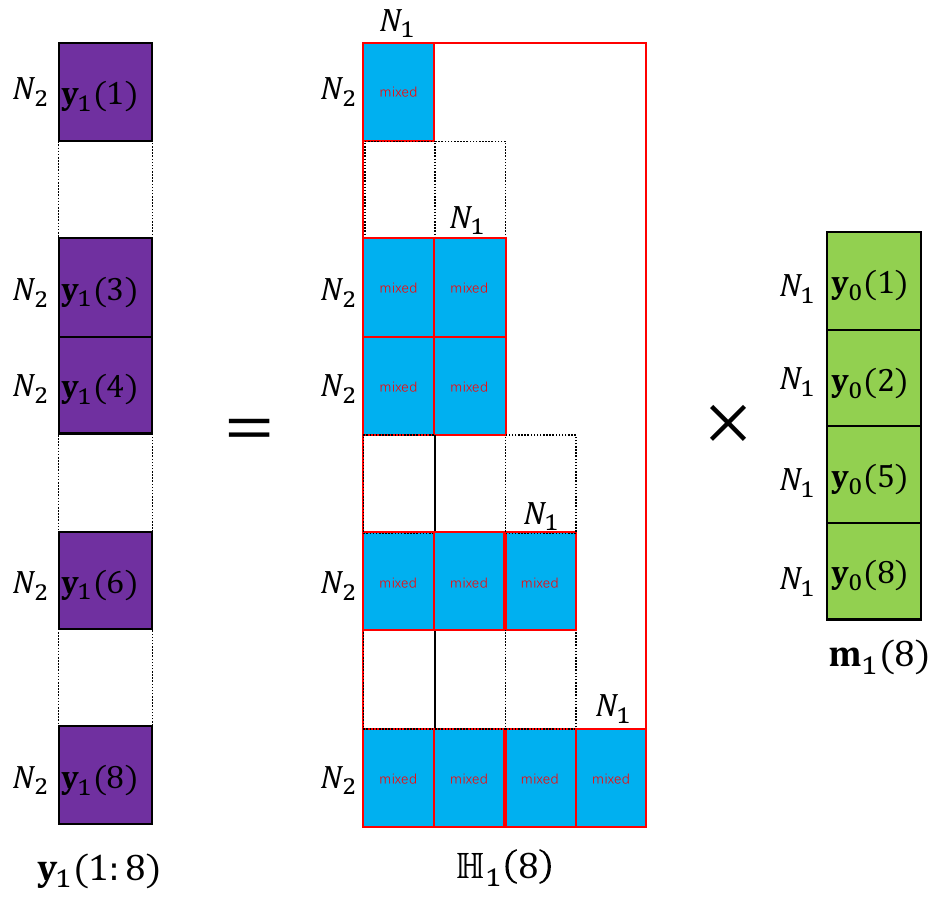}
        \caption{$\mathbf{x}_1(2),\mathbf{x}_1(5),\mathbf{x}_1(7)$ are erased in $(r_1,r_2)$.}
        \label{fig:Example_1_5}
    \end{figure}
    \begin{figure}
        \centering
        \includegraphics[width=0.62\linewidth]{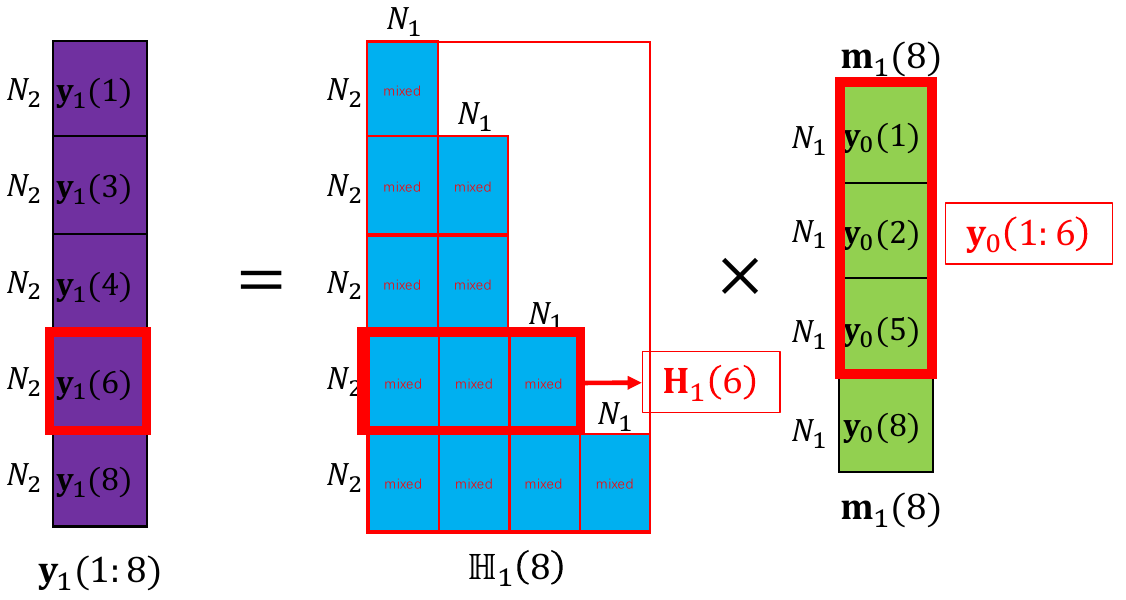}
        \caption{$\mathbf{x}_1(2),\mathbf{x}_1(5),\mathbf{x}_1(7)$ are erased in $(r_1,r_2)$ after eliminating the erased entries.}
        \label{fig:Example_1_6}
    \end{figure}
    \begin{figure}
        \centering
        \includegraphics[width=0.7\linewidth]{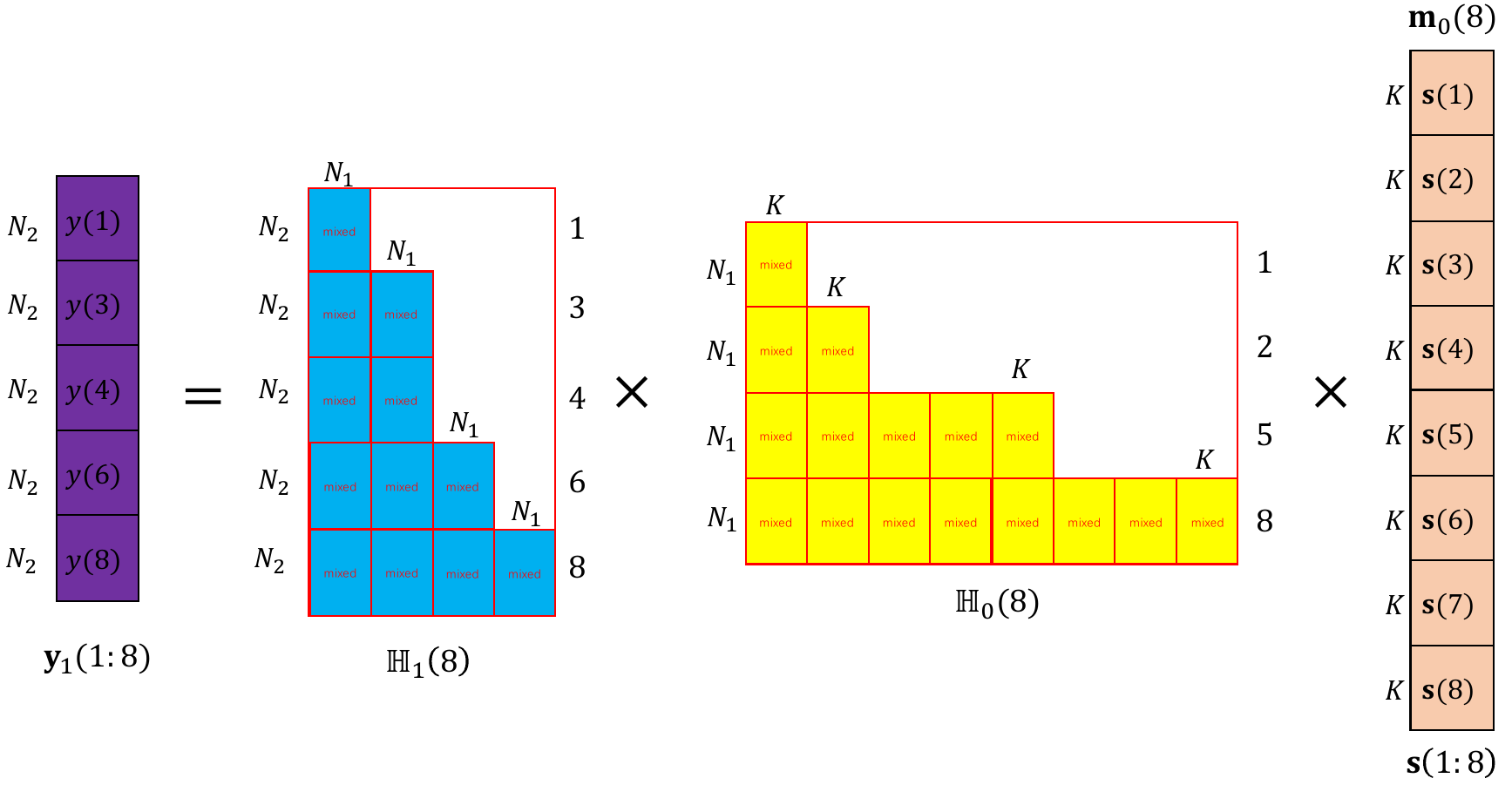}
        \caption{The received symbols at the destination.}
        \label{fig:Example_1_7}
    \end{figure}
    \begin{figure}
        \centering
        \includegraphics[width=0.65\linewidth]{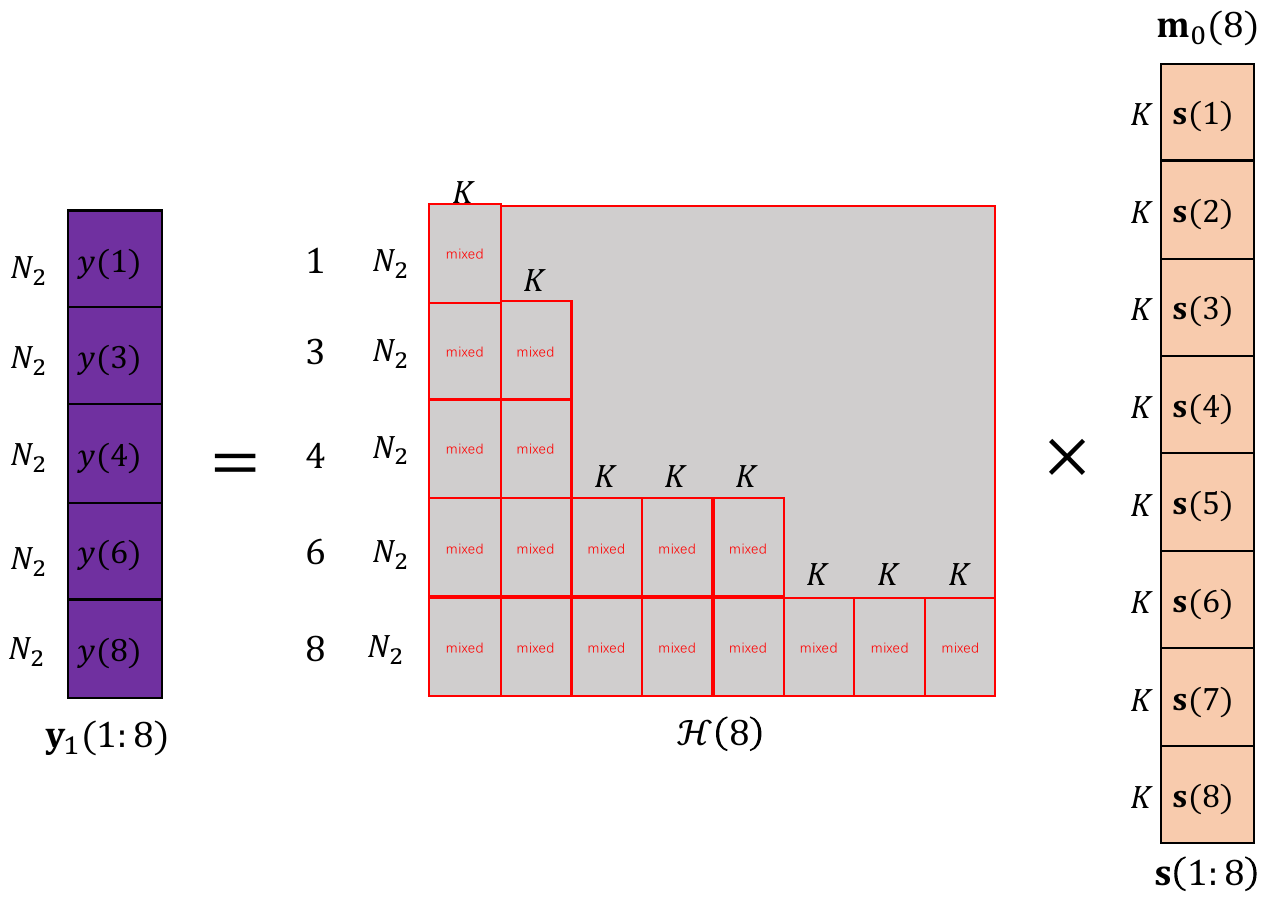}
        \caption{The overall cumulative receiver matrix $\mathcal{H}(8)$.}
        \label{fig:Example_1_8}
    \end{figure}
    
\end{example}

In this paper, we aim at characterizing the exact value of the slot error probability of RLSCs in the multi-hop relay network, defined as
\begin{equation}
    p_{e,[1,T]}^{\text{RLSC}(q)} \triangleq \frac{1}{T}\!\sum_{t=1}^T \text{Pr}(\mathbf{s}(t) \text{ is not } \Delta\text{-decodable}).
\end{equation}
We are exclusively interested in the \textit{long term} slot error probability under the \textit{sufficiently large finite field size regime}, which is defined by
\begin{equation}
    p_e \triangleq \lim_{T\rightarrow \infty}\lim_{q\rightarrow \infty}  p_{e,[1,T]}^{\text{RLSC}(q)}.
\end{equation}

To simplify the discussion, we impose two assumptions on randomness of the generator matrix.

\noindent \textbf{(I) Non-systematic Random linear streaming codes:}
All entries in the per-slot generator matrix $\mathbf{G}_l(t),\forall l\in[0,L-1]$, represented by the yellow and blue blocks in the figures of Example \ref{Example:3-node}, are chosen uniformly and randomly from $\mathbb{F}$, excluding 0. 
This assumption indicates the Non-systematic RLSCs, where the $N$ symbols sent in each timeslot is a mixture (linear combination) of symbols from the present and all previous timeslots. 

\noindent \textbf{(II) The Generalized MDS Condition (GMDS):} Let $B$ be any positive integer. For any $t$ and any finite sequence of pairs $\{(i_b,j_b):b\in[B]\}$ satisfying the following two conditions: \textbf{(a)} $i_{b_1}\neq i_{b_2}$ and $j_{b_1}\neq j_{b_2}$ for any $b_1\neq b_2$ and \textbf{(b)} the $(i_b, j_b)$-th entry of $\mathbb{G}_l(t)$ is non-zero for all $b \in [B]$, define the corresponding row and column index sets $S_R \triangleq \{i_b:b \in [B]\}$ and $S_C \triangleq \{j_b:b \in [B]\}$.
The GMDS condition requires that the submatrix of any cumulative generator matrix $\mathbb{G}_l(t),l\in [1,L-1]$ induced by $S_R$ and $S_C$ is always invertible.

These two assumptions jointly ensure all successfully delivered symbols can carry as much information as possible for decoding, and thus avoid the discussion on some corner cases. 
In this way, all the randomness is a result of random channel realization, not the random code construction.

\section{Characterization of Error Event of RLSCs in TRN}\label{section:Main Results1}

In this section, we consider the simplest multi-hop relay network, i.e., TRN.
In the following, we will first characterize the transmissions in TRN model by a framework of stochastic process in Definition \ref{definition:informtion debt}.
Then the error event of RLSCs in TRN is derived by careful analysis of the framework.

In TRN, due to the erasures in different hops, the delivery of a portion of source symbols could be delayed.
More importantly, due to the distinct erasure patterns in different hops, it is possible that some source symbols have been able to be decoded at the destination, while other source symbols are still detained at the relay or source nodes. 
The information detained at each relay can be distinct in each timeslot.
In example \ref{Example:3-node}, for timeslots 3,4, since that $\mathbf{y}_0(3),\mathbf{y}_0(4)$ are erased in the first hop while $\mathbf{y}_1(3),\mathbf{y}_1(4)$ are perfectly delivered in the second hop, the destination receives $\mathbf{y}_1(3),\mathbf{y}_1(4)$ containing only information of $\mathbf{s}(1),\mathbf{s}(2)$ while the information of $\mathbf{s}(3),\mathbf{s}(4)$ are still detained at the source $r_0$. 
Similarly, for timeslots 5, since that $\mathbf{y}_0(5)$ is perfectly delivered in the first hop while $\mathbf{y}_1(5)$ is erased in the second hop, the information of $\mathbf{s}(3),\mathbf{s}(4),\mathbf{s}(5)$ are jointly transmitted to and then detained at the relay $r_1$. 
To characterize the information flowing through each node of the network, we first generalize the concept of \textit{information debt} $I_d(t)$, which was originally introduced by E. Martinian \cite{phd}, into TRN. 
In \cite{RLSCs,Analysis in Stochastic Channel}, the information debt had been generalized and used to describe the error event in i.i.d. SEC and G-ESEC, respectively. 
The concept $I_d(t)$ is used to describe how many linear equations the destination still needs for successful decoding.

For TRN, let $\mathcal{A}$ be an arbitrary linear combination of some source symbols. Define the \textit{source information function} $S(\mathcal{A})$ as the number of terms (i.e., the number of source symbols contained) in the linear combination $\mathcal{A}$. 
Take Example \ref{Example:3-node} and focus on the first hop, since there are $K$ symbols arrive at the source encoder in each timeslot, $S(\mathbf{m}_0(t)) = K\cdot t$. 
At timeslot 2, since that $\mathbf{x}_0(2)$, a linear combination of $\mathbf{s}(1)$ and $\mathbf{s}(2)$, is delivered and appended to the storage of $r_1$, we have $S(\mathbf{m}_1(2)) = 2K$. 
However, at timeslots 3,4, since that $\mathbf{x}_0(3)$ and $\mathbf{x}_0(4)$ are both erased, no information of source symbols $\mathbf{s}_0(3)$ and $\mathbf{s}_0(4)$ is delivered to $r_1$, such that $S(\mathbf{m}_1(3)) = S(\mathbf{m}_1(4)) = S(\mathbf{m}_1(2)) = 2K$. At timeslot 5, $\mathbf{x}_0(5)$, a linear combination of the first five source symbols $\mathbf{s}(1:5)$, is delivered and cached into the storage of $r_1$, therefore $S(\mathbf{m}_1(5)) = 5K$. With this definition, $S(\mathbf{m}_l(t))$ represents the number of source symbols that are contained in the storage of $r_l$ at timeslot $t$, or in other words, the number of terms in the linear combination that is latest received at $r_l$.  

Then further denote $D_l(t) \triangleq S(\mathbf{m}_l(t)) - S(\mathbf{m}_{l+1}(t)), l\in \{0,1\}$ as the number of source symbols that are \textit{detained} at the $r_l$ in timeslot $t$.
Specifically, $D_0(t) \triangleq S(\mathbf{m}_0(t)) - S(\mathbf{m}_{1}(t))$ denotes the number of the source symbols that have arrived at $r_0$ from the information source, but have not been received\footnote{Here, ``received" doesn't mean $r_1$ can decode or recover those source symbols. It only means at least one encoded packet $\mathbf{x}_0(t)$ containing information of these source symbols have been successfully delivered to relay $r_1$.} by relay $r_1$ due to the erasures in the first hop.
Similarly, $D_1(t) \triangleq S(\mathbf{m}_1(t)) - S(\mathbf{m}_{2}(t))$ denotes the number of the source symbols that have been contained in $r_1$'s storage, but have not been received by the destination $r_2$  due to the erasures in the second hop.
Particularly, denote $D_2(t) \triangleq S(\mathbf{m}_{2}(t))$ as the number of source symbols contained in the received encoded packets $\mathbf{y}_1(1:t)$ at the destination. Since $r_2$ is the last node in the information flow, it doesn't have the concept of ``detain''. 
Due to the causality, information of the detained unknowns at $r_l$ has not been transmitted to any downstream nodes $\{r_{l'}|l'>l\}$. 
Note that the number of unknowns (uncoded source symbols) in the network equals to the summation of $D_l(t)$, i.e., $S(\mathbf{m}_0(t)) = \sum_{l=0}^2 D_l(t)$.

Denote $W(t) \triangleq |\mathbf{y}_1(1:t)|$ as the number of encoded symbols (corresponding to the $D_2(t)$ source symbols) that have been received at the destination $r_2$. 
Take Example \ref{Example:3-node}, since that $e_1(1)=0,e_1(2)=1,e_1(3)=0$, we have $W(1) = W(2) = N_2$ and $W(3) = 2N_2$.
Then we generalize the concept of information debt as follows.

\begin{definition}\label{definition:informtion debt}
    Let $D_0(0)=D_1(0)=D_2(0)=W(t)=I_d(0)=0$. For any $t\ge 1$, the information debt $I_d(t)$ of RLSCs in TRN is calculated iteratively by 
     \begin{align}
        D_0(t) &= D_0(t-1)\cdot e_0(t-1) + K, \label{equation:D0}\\
        D_1(t) &= D_1(t-1)\cdot e_1(t-1) + D_0(t)\cdot [1-e_0(t)], \label{equation:D1}\\
        D_2(t) &= D_2(t-1)\cdot \mathds{1}\{I_d(t-1) \neq 0\} + D_1(t)\cdot [1-e_1(t)], \label{equation:D2}\\
        W(t) &= W(t-1)\cdot \mathds{1}\{I_d(t-1) \neq 0\} + N_2 \cdot [1-e_1(t)], \label{equation:W}\\
        I_d(t) &\triangleq \left[D_2(t) - W(t)\right]^+. \label{equation:Id}
    \end{align}   
\end{definition}
The derivation of Definition \ref{definition:informtion debt} is presented in Appendix \ref{Appendix:characterization of the error event}.
The physical meaning of \textit{Definition \ref{definition:informtion debt}} is explained as follows. 
\begin{itemize}
    \item (\ref{equation:D0}) represents the arrival of $K$ source symbols at the beginning of each timeslot.
    Term $+K$ means the $K$ source symbols arrived at $r_0$ will add to the detained number $D_0(t)$, since any linear combination containing them have not been transmitted to the downstream nodes. Term $\cdot e_0(t-1)$ means if in the previous timeslot, $\mathbf{x}_0(t-1)$, which is a linear combination of all source symbols $\mathbf{s}(1:t-1)$, is successfully delivered, then $D_0(t-1)$ will be clear to zero since no source symbol will be detained at $r_0$ at the beginning of timeslot $t$. Otherwise, $D_0(t-1)$ will remains and be added into $D_0(t)$.

    \item (\ref{equation:D1}) represents the transmission in the first hop. 
    Note that $\mathbf{x}_0(t)$ is a linear combination of all source symbols $\mathbf{s}_0(1:t)$.
    Term $D_0(t)\cdot [1-e_0(t)]$ means if $\mathbf{x}_0(t)$ is successfully delivered, then the no source symbols will be detained at $r_0$, and thus the original detained number at $r_0$ will be added to $r_1$. Term $\cdot e_1(t-1)$ means if in the previous timeslot, $\mathbf{x}_1(t-1)$ is delivered, then $D_1(t-1)$ will be clear to zero at the beginning of timeslot $t$.
    
    \item (\ref{equation:D2}) and (\ref{equation:W}) jointly represent the transmission in the second hop. Similar to the first hop, if $\mathbf{x}_1(t)$ is successfully delivered, $r_2$ will receive $N_2$ encoded packets, which contain information of the $D_1(t)$ detained source symbols.
    Thus, the detained number at $r_1$ will be added to the number at $r_2$, and $N_2$ will be added to $W(t)$, respectively.
    The difference is that when the $I_d(t-1)$ in (\ref{equation:Id}) equals to zero in the previous timeslot, the decoding process at the destination will start and $D_2(t-1)$ source symbols can be decoded from the $W(t-1)$ equations (this will be discussed in the following Proposition \ref{Proposition:characterization of the error event}). 
    When the $D_2(t-1)$ source symbols are decoded, $D_2(t-1)$ and $W(t-1)$ will be  clear to zero at the beginning of timeslot $t$.

    \item (\ref{equation:Id}) characterizes the information debt at the destination, which equals to the number of unknowns (undecoded source symbols) minus the number of equations (received packets with respect to the unknowns) at the destination. $I_d(t)$ is a non-negative integer.
\end{itemize}


The temporal variation of $I_d(t)$ forms a stochastic process.
Define the \textit{zero-hitting time sequence} of $I_d(t)$, i.e., $\{t_i:i\in[0,\infty]\}$ as follows.
\begin{definition}\label{definition:hitting time old}
    Initialize that $t_0\triangleq0$ and define iteratively 
    \begin{align}
        t_i &\triangleq \inf \{t':t'>t_{i-1}, I_d(t')=0\}
    \end{align}
    as the $i$-th time that $I_d(t)$ hits 0.
\end{definition}

In TRN, the error event of RLSCs is characterized in the following proposition.

\begin{prop}\label{Proposition:characterization of the error event}
     Assume \textbf{GMDS} holds. 
For RLSCs in TRN, for any fixed index $i_0 \ge 0$, \textbf{(a)} $\mathbf{s}(t)$ is $\Delta$-decodable for all timeslots $t$ that 
\begin{equation}\label{equation:error event}
    \forall t \in \left[t_{i_0} - \frac{D_0(t_{i_0})+D_1(t_{i_0})}{K}+1,t_{i_0} - \frac{D_0(t_{i_0})+D_1(t_{i_0})-D_2(t_{i_0+1})}{K}\right] \bigcap \Big[t_{i_0+1}-\Delta,t_{i_0+1}\Big].
    \end{equation}
\textbf{(b)} $\mathbf{s}(t)$ is not $\Delta$-decodable for the rest of $t$.
\end{prop}

The proof of \textit{Proposition \ref{Proposition:characterization of the error event}} is presented in Appendix \ref{Appendix:characterization of the error event}.
The insight of \textit{Proposition \ref{Proposition:characterization of the error event}} is discussed as follows.
The decoding event of RLSCs is directly determined by the value of $I_d(t)$. 
Roughly speaking, when $t \neq t_{i_0}, \forall i_0 \ge 0$ or equivalently, $I_d(t)>0$, the source symbols waiting to be decoded are mixed up with each other in the encoded packets, since the number of equations received at the destination are not sufficient for decoding.
When $t = t_{i_0}, \forall i_0 \ge 0$ or equivalently, $I_d(t)=0$, the decoding process can start and $D_2(t_{i_0})$ symbols can be decoded at the destination.
Due to the feature of multi-hop erasure, when the destination have reached the decoding condition $I_d(t_{i_0})=0$ for any index $i_0$, there can always be some source symbols detained at the source or intermediate relays, i.e., $D_0(t_{i_0}) > 0$ or $D_1(t_{i_0}) > 0$. 
Obviously, the detained symbols can not be decoded at the destination, since $r_2$ hasn't received any packets containing information of these detained symbols.
Therefore, at each decoding time $t_{i_0}$, there will be $D_0(t_{i_0})+D_1(t_{i_0})$ symbols detained and can not be decoded at $r_2$.
However, these $D_0(t_{i_0})+D_1(t_{i_0})$ symbols could still remain potential $\Delta$-decodability, and should wait until the next zero-hitting time of information debt, i.e., $t_{i_0+1}$ to determine.

\begin{figure*}[!hbtp]
    \centering
    \includegraphics[width=0.9\linewidth]{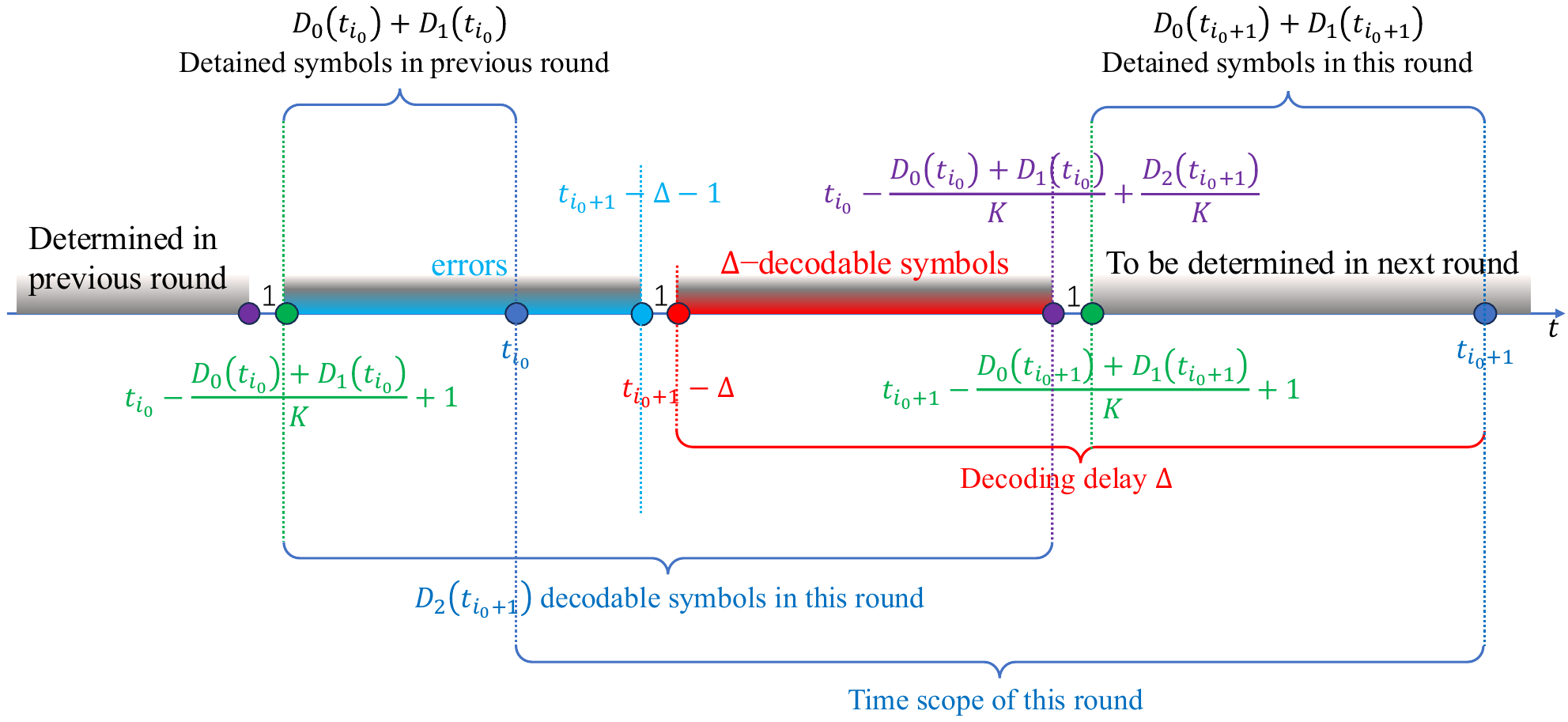}
    \caption{Illustration of error event in \textit{Proposition \ref{Proposition:characterization of the error event}}.}
    \label{fig:Characterization_of_error_event}
\end{figure*}

Let us focus on an interval of the two adjacent hitting times $[t_{i_0},t_{i_0+1})$, which will be referred to as a ``round" thereafter.
At timeslot $t_{i_0}$, the source symbols $\mathbf{s}\Big(t_{i_0}-\frac{D_0(t_{i_0})+D_1(t_{i_0})}{K}+1:t_{i_0}\Big)$ are still detained at the nodes previous to $r_2$ and waiting to be decoded in this round. 
At timeslot $t_{i_0+1}$, the source symbols $\mathbf{s}\Big(t_{i_0}-\frac{D_0(t_{i_0})+D_1(t_{i_0})}{K}+1 : t_{i_0}-\frac{D_0(t_{i_0})+D_1(t_{i_0})}{K}+\frac{D_2(t_{i_0+1})}{K}\Big)$, totally $D_2(t_{i_0+1})$ symbols, become able to be decoded at the destination (could possibly exceed the decoding delay $\Delta$). And the remaining source symbols $\mathbf{s}\Big(t_{i_0+1}-\frac{D_0(t_{i_0+1})+D_1(t_{i_0+1})}{K}+1:t_{i_0+1}\Big)$ are still detained at $r_0$ and $r_1$, waiting to be decoded in the next round.
Consider the decoding latency, $\Delta$ timeslots previous to $t_{i_0+1}$ are within the delay constraint.
Therefore, the $\Delta$-decodable symbols in this round are intersection of the two sets, as shown in (\ref{equation:error event}).
The first set represents the $D_2(t_{i_0+1})$ source symbols that are able to be decoded at $r_2$ at timeslot $t_{i_0+1}$ in this round. 
The second set stands for the delay constraint in this round.
A detailed illustration of \textit{Proposition \ref{Proposition:characterization of the error event}} can be found in Fig. \ref{fig:Characterization_of_error_event}.

It is also worthy noting that $D_0(t),D_1(t),D_2(t)$ are all multiples of $K$. This is due to the fact that $K$ source symbols will arrive at $r_0$ in each timeslot and thus $S(\mathbf{m}_0(t)),S(\mathbf{m}_1(t)),S(\mathbf{m}_2(t))$ can all be divided by $K$.
Since we are mainly interested in the number of error timeslots, we will normalize $D_0(t),D_1(t),D_2(t)$ by $K$ and denote them as $\widehat{D}_l(t)=\frac{D_l(t)}{K}, l\in[0,2]$ for simplicity of presentation thereafter. 

With \textit{Proposition \ref{Proposition:characterization of the error event}}, the following lemma holds directly, with which we can analyze the exact error probability.


\begin{lemma}\label{lemma:1} 
    Assume the transmission rate is within the capacity. In TRN, the error probability of RLSCs can be given by 
    \begin{equation}\label{equation:P_e}
        p_e = \frac{\mathbb{E}\left\{
        \min\bigg[
        \Big(t_{i_0+1} - t_{i_0} + \frac{D_0(t_{i_0}) + D_1(t_{i_0})}{K} - \Delta - 1\Big)^+ , \frac{D_2(t_{i_0+1})}{K}
        \bigg]
        \right\}}
        {\mathbb{E}\Big\{\frac{D_2(t_{i_0+1})}{K}\Big\}},
    \end{equation}
    where $i_0 \ge 0$ is any arbitrary but fixed index.
\end{lemma}

\textit{Proof:} Lemma \ref{lemma:1} holds from Proposition \ref{Proposition:characterization of the error event} by calculating the ratio of expected error timeslots to the expected interval of the zero-hitting times.
Notice that $t_i$ defined in (\ref{equation:Id}) is a Markov renewal process.
By \cite[Theorem 3.3]{Renewal process}, Lemma \ref{lemma:1} holds directly. 

\QEDA

\section{Characterization of Error Probability of RLSCs in TRN}\label{section:Main Results2}
In this section, we derive the error probability $p_e$ in TRN by characterizing the explicit expression of the denominator and the numerator of (\ref{equation:P_e}) separately.

One can notice that the transition of the number of detained symbols and the information debt can be modeled as a \textit{infinite Hidden Markov Model} (iHMM), which is a HMM with infinite number of hidden states and emission observations. 
The hidden state of iHMM can be expressed as a two-tuple $\big(\widehat{D}_0(t),\widehat{D}_1(t)\big)$, where $\widehat{D}_0(t),\widehat{D}_1(t) \in \mathbb{N}$.
And the emission observation $I_d(t)$ also transitions in natural numbers domain that $I_d(t)\in \mathbb{N}$.
Note that $\widehat{D}_0(t),\widehat{D}_1(t),I_d(t)$ can all be sufficiently large. 
However, the probability of the events that $\widehat{D}_0(\tau)\rightarrow \infty, \widehat{D}_1(\tau)\rightarrow \infty$ and $I_d(\tau)\rightarrow \infty$ for some $\tau > 0$ will asymptotically approach zero, respectively. 
Therefore, to handle the intractable infinite states/observations of iHMM, we preset the maximum values of $\widehat{D}_0(t),\widehat{D}_1(t),I_d(t)$ such that $\widehat{D}_0(t)\in[0,m_0-1],\widehat{D}_1(t)\in[0,m_1-1],I_d(t)\in[0,m_2-1]$, respectively.
During the numerical evaluation, $m_0,m_1,m_2$ can be adjusted to sufficiently large values to have an accurate simulation on $p_e$.
First denote the \textit{probability distribution of the hidden states} at timeslot $t$ as a $m_0 \cdot m_1$ vector (the subscripts are sorted in lexicographical order)
\begin{align}\label{equation:order of pi}
        \pi^t =  \Big[\overbrace{\pi_{0,0}^t,\cdots,\pi_{0,m_1-1}^t}^{m_1},\cdots,\overbrace{\pi_{m_0,0}^t,\cdots,\pi_{m_0,m_1-1}^t}^{m_1}\Big],
\end{align}
where term $\pi_{i,j}^t \triangleq \text{Pr}\left(\widehat{D}_0(t) = i, \widehat{D}_1(t) = j\right), \forall i\in [0,m_0-1], \forall j\in [0,m_1-1]$.
Further denote the \textit{stationary initial distribution of the hidden states when every time the information debt initials from zero} as $\pi^\infty$. Specifically, $\pi^\infty \triangleq \lim_{i_0\rightarrow \infty} \pi^{t_{i_0}}$. 

As shown in (\ref{equation:D0}) to (\ref{equation:Id}), the transition of $\widehat{D}_0(t),\widehat{D}_1(t)$ and $I_d(t)$ are highly coupled with each other, such that one can not use separate transition matrices to describe their transition behavior. 
Therefore, we use joint transition matrices to characterize the overall transitions of $\widehat{D}_0(t),\widehat{D}_1(t)$ and $I_d(t)$.
\begin{definition}\label{definition:joint transition matrix}
    Let set $\phi = [1,m_2-1]$ denotes the domain of $I_d(t)$ excluding zero.
    Denote $\mathbf{T}_{0,0}$ as a $(m_0 \cdot m_1)  \times (m_0 \cdot m_1) $-joint transition matrix of $\widehat{D}_0(t),\widehat{D}_1(t)$ while $I_d(t)$ initials from zero and then hits back to zero in one timeslot.
    Denote $\mathbf{T}_{0,\phi}$ as a $(m_0 \cdot m_1) \times (m_0 \cdot m_1 \cdot (m_2-1))$-joint transition matrix of $\widehat{D}_0(t),\widehat{D}_1(t),I_d(t)$ while $I_d(t)$ initials from zero and  transitions to a value not equal to zero in one timeslot.
    Denote $\mathbf{T}_{\phi,\phi}$ as a $(m_0 \cdot m_1 \cdot (m_2-1)) \times (m_0 \cdot m_1 \cdot (m_2-1))$-joint transition matrix of $\widehat{D}_0(t),\widehat{D}_1(t),I_d(t)$ while $I_d(t)$  transitions between the values within $\phi$ in one timeslot.
    Denote $\mathbf{T}_{\phi,0}$ as a $(m_0 \cdot m_1 \cdot (m_2-1)) \times (m_0 \cdot m_1)$-joint transition matrix of $\widehat{D}_0(t),\widehat{D}_1(t),I_d(t)$ while $I_d(t)$ initials from a value not equal to zero and transitions to zero in one timeslot.
    More precisely, the joint transition matrices $\mathbf{T}_{0,0},\mathbf{T}_{0,\phi},\mathbf{T}_{\phi,\phi},\mathbf{T}_{\phi,0}$ are defined according to their entries as follows. 
    Denote the entries of $\mathbf{T}_{0,0}$ as $\big[\mathbf{t}_{0,0}(i,j;v,w)\big]$, where $i\in[0,m_0-1]$ is the row index for $\widehat{D}_0(t)$, $j\in[0,m_1-1]$ is the row index for $\widehat{D}_1(t)$, $k\in[0,m_0-1]$ is the column index for $\widehat{D}_0(t)$, $l\in[0,m_1-1]$ is the column index for $\widehat{D}_1(t)$.
    It is defined that  
    \begin{equation}
        \mathbf{t}_{0,0}(i,j;v,w) \triangleq \text{Pr}\Big(\widehat{D}_0(t+1)=v,\widehat{D}_1(t+1)=w,I_d(t+1)=0 \Big| \widehat{D}_0(t)=i,\widehat{D}_1(t)=j,I_d(t)=0\Big).
    \end{equation}
    Similarly, for $\mathbf{T}_{0,\phi}=\Big[\mathbf{t}_{0,\phi}(i,j;v,w,d)\Big]$,$\mathbf{T}_{\phi,\phi}=\Big[\mathbf{t}_{\phi,\phi}(i,j,g;v,w,d)\Big]$, and $\mathbf{T}_{\phi,0}=\Big[\mathbf{t}_{\phi,0}(i,j,g;v,w)\Big]$, where $d\in[0,m_2-1]$ is the row index for $\widehat{D}_2(t)$ and $g\in[0,m_2-1]$ is the column index for $\widehat{D}_2(t)$, we have
    \begin{align}
        \mathbf{t}_{0,\phi}(i,j;v,w,d) &\triangleq \text{Pr}\Big(\widehat{D}_0(t+1)=v,\widehat{D}_1(t+1)=w,I_d(t+1)=d \Big| \widehat{D}_0(t)=i,\widehat{D}_1(t)=j,I_d(t)=0\Big), \\
        \mathbf{t}_{\phi,\phi}(i,j,g;v,w,d) &\triangleq \text{Pr}\Big(\widehat{D}_0(t+1)=v,\widehat{D}_1(t+1)=w,I_d(t+1)=d \Big| \widehat{D}_0(t)=i,\widehat{D}_1(t)=j,I_d(t)=g\Big), \\
        \mathbf{t}_{\phi,0}(i,j,g;v,w) &\triangleq \text{Pr}\Big(\widehat{D}_0(t+1)=v,\widehat{D}_1(t+1)=w,I_d(t+1)=0 \Big| \widehat{D}_0(t)=i,\widehat{D}_1(t)=j,I_d(t)=g\Big).
    \end{align}
    Note that there are at most three row indices $i,j,g$ and at most three column indices $v,w,d$ for the joint transition matrices.
    The row/column entries in joint transition matrices are listed in lexicographical order according to the row/column indices similar to (\ref{equation:order of pi}).
\end{definition}

In the following, we first define a transition matrix $\mathbf{D}_{i,q,m}$. Then we show that the transition of the hidden states for any timeslot can be derived by nesting of matrix $\mathbf{D}_{i,q,m}$.
Finally, derivation of the joint transition matrices $\mathbf{T}_{0,0},\mathbf{T}_{0,\phi},\mathbf{T}_{\phi,\phi},\mathbf{T}_{\phi,0}$ can be given jointly by Proposition \ref{proposition:transition matrix of hidden states} and Proposition \ref{proposition:transition matrix of information debt}.
Proposition \ref{proposition:transition matrix of hidden states} forms a structure of the joint transition matrix of $\widehat{D}_0(t),\widehat{D}_1(t)$, while Proposition \ref{proposition:transition matrix of information debt} complements the transition of $I_d(t)$ upon Proposition \ref{proposition:transition matrix of hidden states}.

\begin{definition}
Define a stochastic matrix with three parameters $i,q,m$ for any $0\le i \le m,q\in [0,1]$ as follows.
    \begin{equation}\label{equation:D_iqm}
\mathbf{D}_{i,q,m} = 
    \begin{tikzpicture}[baseline = (M.west)]
    \tikzset{brace/.style = {decorate, decoration = {brace, amplitude = 5pt}, thick}}
    \matrix(M)
    [
        matrix of math nodes,
        left delimiter = (,
        right delimiter = )
    ]
    {
        \textcolor{green}{q} & 0 & \cdots & 0 & \textcolor{red}{1-q} &  0  &  \cdots & 0 \\
        \textcolor{green}{q} & 0 & \cdots & 0 &  0  & \textcolor{red}{1-q} &  \cdots & 0 \\
        \textcolor{green}{\vdots} & \vdots & \vdots & \vdots & \vdots & \vdots & \textcolor{red}{\ddots} & \vdots \\
        \textcolor{green}{q} & 0 & \cdots & 0 &  0  &  0  &  \cdots & \textcolor{red}{1-q} \\
        \textcolor{green}{\vdots} & \vdots & \vdots & \vdots & \vdots & \vdots & \vdots & \textcolor{red}{\vdots} \\
        \textcolor{green}{q} & 0 & \cdots & 0 &  0  &  0  &  \cdots & \textcolor{red}{1-q} \\
    };
    \draw[brace]
        (M-1-1.north west)
        -- node[above = 5pt]{$i+1$}
        (M-1-5.north east);
\end{tikzpicture}_{m \times m}.
\end{equation}
\end{definition}

Note that $\mathbf{D}_{i,q,m}$ is a $m\times m$ matrix with its first column being $q$ (marked in green) and its $+i$-th diagonal\footnote{For any $i\ge0$, the $+i$-th diagonal represents the $i$-th diagonal on the upper right of the main diagonal, while the $-i$-th diagonal represents the $i$-th diagonal on the lower left of the main diagonal. Particularly, $0$-th diagonal represents the main diagonal.} being $1-q$ (marked in red). 
We refer to the first column as \textit{``Deliver Band (DB)''}, and refer to the $+i$-th diagonal as \textit{``Erasure Band (EB)''} thereafter.
Matrix $\mathbf{D}_{i,q,m}$ accounts for the transition of a hidden state in a timeslot. 
Then we show in Proposition \ref{proposition:transition matrix of hidden states} that the joint transition of $\widehat{D}_0(t),\widehat{D}_1(t)$ in a timeslot can be represented by nesting of matrix $\mathbf{D}_{i,q,m}$.

\begin{prop}\label{proposition:transition matrix of hidden states}
    Denote $\mathbf{D}_{i,q,m}$ as in equation (\ref{equation:D_iqm}).
    In TRN, the joint transition matrix of $\widehat{D}_0(t),\widehat{D}_1(t)$ can be derived by the following two steps of construction.

    \begin{itemize}
        \item[1.] Construct a matrix $\mathbf{M}^{(0)} = \mathbf{D}_{1,q_0,m_0}$. 

        \item[2.] Recall that $\mathbf{M}^{(0)}$ has two bands, i.e., DB and EB. Denote the entry in DB of $\mathbf{M}^{(0)}$ with row index $d_0$ as $\mathbf{M}^{(0)}_{D}(d_0)$. Denote the entry in EB of $\mathbf{M}^{(0)}$ with row index $d_0$ as $\mathbf{M}^{(0)}_{E}(d_0)$.
        For each entry of $\mathbf{M}^{(0)}$, embed a transition matrix as follows. 

            \begin{itemize}
                \item[(a)] For $\mathbf{M}^{(0)}_{D}(d_0), d_0\in [0,m_0-1]$, embed a matrix $\mathbf{D}_{d_0+1,q_1,m_1}$, such that the original scalar term $q_0$ is expanded to a matrix $q_0 \cdot \mathbf{D}_{i+1,q_1,m_1}$.

                \item[(b)] For $\mathbf{M}^{(0)}_{E}(d_0), d_0\in [0,m_0-1]$, embed a matrix $\mathbf{D}_{0,q_1,m_1}$, such that the original term $1-q_0$ is expanded to a matrix $(1-q_0) \cdot \mathbf{D}_{0,q_1,m_1}$.

                \item[(c)] For each of the rest zero-entries, expand it into a $m_1 \times m_1$ zero matrix $\mathbf{0}_{m_1 \times m_1}$.
            \end{itemize}
    \end{itemize}
\end{prop}

\textit{Proof:} Step 1 and step 2 represent the transition of $\widehat{D}_0(t)$ and $\widehat{D}_1(t)$, respectively.
Consider the first hop. With probability $q_0$, the transmission will succeed, thus $\widehat{D}_0(t)$ will be reduced to zero since the source symbols detained at $r_0$ are all forwarded. With probability $1 - q_0$, the transmission will fail, thus $\widehat{D}_0(t)$ will be incremented by 1, since $K$ symbols will arrive from the source. 
Therefore, matrix $\mathbf{M}^{(0)} = \mathbf{D}_{1,q_0,m_0}$ can represent the transition of $\widehat{D}_0(t)$ in the first hop.
Then consider the second hop. With probability $q_1$, the transmission will succeed, thus $\widehat{D}_1(t)$ will be reduced to zero since the source symbols detained at $r_1$ are all forwarded. With probability $1 - q_1$, the transmission will fail.
However, the increment of $\widehat{D}_1(t)$ will depend on the number of source symbols newly forwarded from $r_0$ in this timeslot. If the first hop succeeded, $\widehat{D}_1(t)$ will increase by $\widehat{D}_0(t) + 1$, where the $+1$ is due to the $K$ source symbols newly come at this timeslot. 
In this case, the transition of $\widehat{D}_1(t)$  can be represented by matrix $\mathbf{D}_{\widehat{D}_0(t)+1,q_1,m_1}$.
If the first hop failed, $\widehat{D}_1(t)$ will remain the same.
In this case, the transition of $\widehat{D}_1(t)$  can be represented by matrix $\mathbf{D}_{0,q_1,m_1}$.
For different values of $\widehat{D}_0(t)$, the transition matrices of $\widehat{D}_1(t)$ can be embedded into the corresponding entries of $\mathbf{M}^{(0)}$ to account for the transition in the second hop.
Therefore, proposition \ref{proposition:transition matrix of hidden states} is proved.
\QEDA

\begin{remark}
    Since the space of $\widehat{D}_0(t),\widehat{D}_1(t)$ are infinite, $\mathbf{D}_{i,q,m}$ was supposed to be an infinite matrix. Thus the $1-q$ in the last column accounts for the boundary effect caused by the approximation for the infinite domain.
\end{remark}

\begin{figure}
    \centering
    \includegraphics[width=0.99\linewidth]{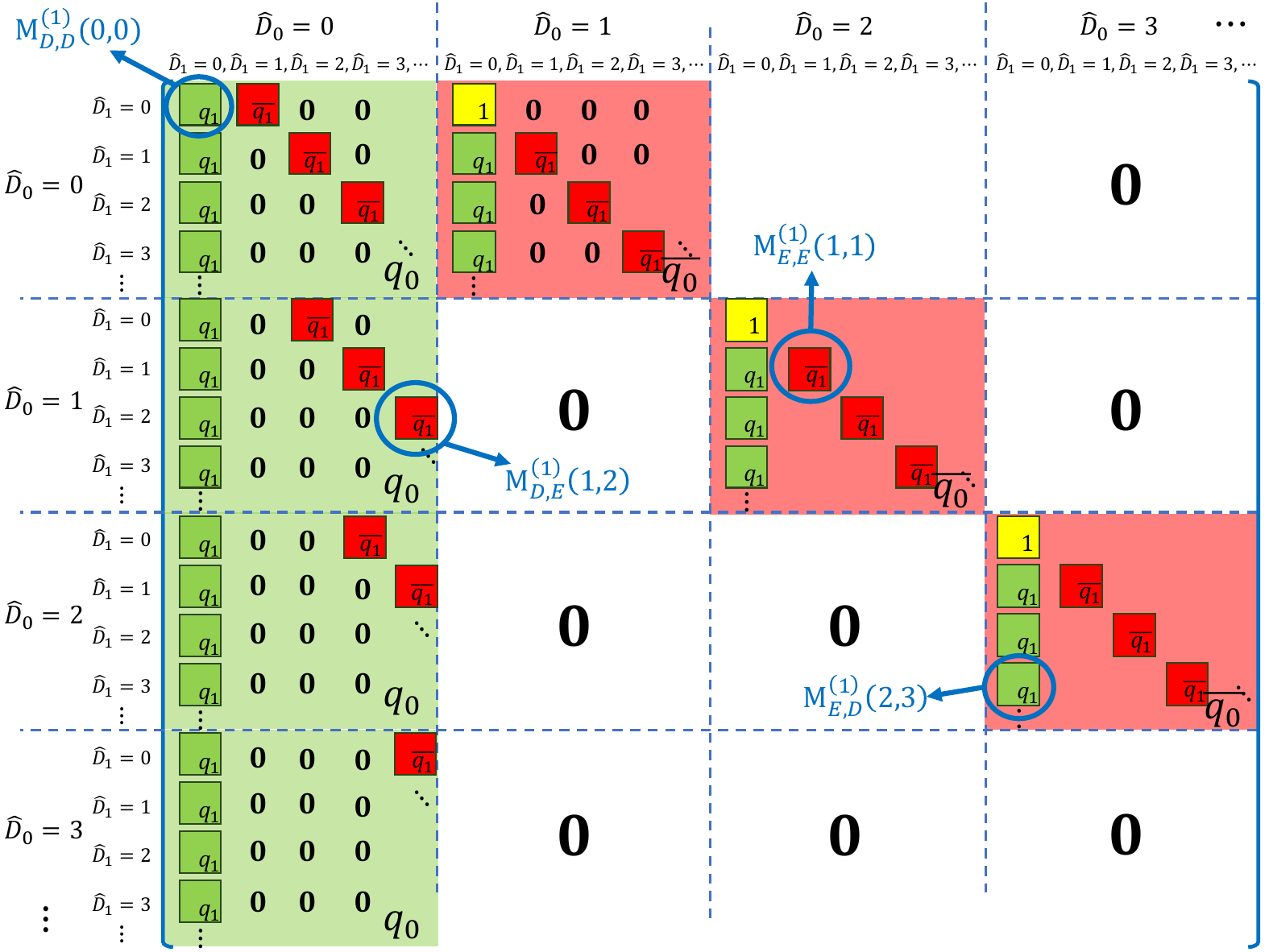}
    \caption{Transition structure of the hidden states $\widehat{D}_0(t)$ and $\widehat{D}_1(t)$. In the matrix, each block is attached with a probability. The probability of the big block will be effective for each small block within it. The notation $\overline{q_1} = 1-q_1$.
    Also note that in the matrices $\mathbf{D}_{0,q_1,m_1}$, two bands converge at the upper-left element, which is marked in yellow in the figure. Further embedding process in Proposition \ref{proposition:transition matrix of information debt} will still regard this yellow block as a green block and a red block overlapped with each other. The two blocks will be embedded separately according to the same row index $d_0=0$ and their final result will be summed up. }
    \label{fig:transition matrix of hidden states}
\end{figure}

Denote the $(m_0 m_1) \times (m_0 m_1)$ matrix after embedding as $\mathbf{M}^{(1)}$.
Since we are mainly interested in the non-zero entries in the ``bands", we label the entries of $\mathbf{M}^{(1)}$ according to the ``bands" as well as row indices $d_0$ and $d_1$ for $\widehat{D}_0(t)$ and $\widehat{D}_1(t)$, respectively as follows. 
Further denote the $m_1 \times m_1$-non-zero submatrices expanded from entries $\mathbf{M}^{(0)}_{D}(d_0)$ and $\mathbf{M}^{(0)}_{E}(d_0)$ as $\mathbf{M}^{(1)}_{D}(d_0)$ and $\mathbf{M}^{(1)}_{E}(d_0)$, respectively.
Denote the entries in DB and EB of submatrices $\mathbf{M}^{(1)}_{D}(d_0)$ with row index $d_1$ as $\mathbf{M}^{(1)}_{D,D}(d_0,d_1)$ and $\mathbf{M}^{(1)}_{D,E}(d_0,d_1), d_0\in [0,m_0-1], d_1\in [0,m_1-1]$, respectively.
And denote the entries in DB and EB of submatrices $\mathbf{M}^{(1)}_{E}(d_0)$ with row index $d_1$ as $\mathbf{M}^{(1)}_{E,D}(d_0,d_1)$ and $\mathbf{M}^{(1)}_{E,E}(d_0,d_1), d_0\in [0,m_0-1], d_1\in [0,m_1-1]$, respectively.
The resulting structure of Proposition \ref{proposition:transition matrix of hidden states} is presented in Fig. \ref{fig:transition matrix of hidden states}.
The green bands stand for DB while the red bands stand for EB.
One can notice that the transition structure of $\widehat{D}_0(t)$ and $\widehat{D}_1(t)$ exists a \textit{nested structure}, where one can recognize similar color pattern of \textit{``bands in bands"}.
With this nested structure, we will show in the next section that our results can be further extended readily to multi-hop networks with hidden states $\widehat{D}_0(t)$ to $\widehat{D}_L(t)$.
In Proposition \ref{proposition:transition matrix of information debt}, further embedding processes will transform $\mathbf{M}^{(1)}$ into different joint transition matrices $\mathbf{T}_{0,0},\mathbf{T}_{0,\phi},\mathbf{T}_{\phi,\phi},\mathbf{T}_{\phi,0}$, respectively.

\begin{prop}\label{proposition:transition matrix of information debt}
    Denote $\vec{\delta}_{k}^\top$ as a $1\times (m_2-1)$-row vector where the $k$-th entry is one and all other entries are zeros, $k\in [1,m_2-1]$.
    Particularly, for any $k\notin [1,m_2-1]$, define $\vec{\delta}_{k}^\top = \vec{\mathbf{0}}^\top$, which is a all zero row vector.
    Denote $\vec{\mathbf{1}}_{N_2}$ as a $(m_2-1) \times 1$-column vector where the first $N_2$ entries are all ones and all other entries are zeros.
    Denote $\mathbf{I}(k)$ as a $(m_2-1)\times (m_2-1)$-matrix with its $+k$-th diagonal being all ones. 
    For any $k\notin [1-m_2,m_2-1]$, $\mathbf{I}(k) = \mathbf{0}_{(m_2-1)\times (m_2-1)}$.
    In TRN, the joint transition matrices $\mathbf{T}_{0,0},\mathbf{T}_{0,\phi},\mathbf{T}_{\phi,\phi},\mathbf{T}_{\phi,0}$ can be derived from  $\mathbf{M}^{(1)}$ respectively as follows.

    \begin{itemize}
        \item[1.] Derivation of $\mathbf{T}_{0,0}$. 
        \begin{itemize}
            \item[a)] For $\mathbf{M}^{(1)}_{D,D}(d_0,d_1)$, embed a scalar $\vec{\delta}_{d_0+d_1+1}^\top \cdot \vec{\mathbf{1}}_{N_2}$, such that the entry becomes $q_0 q_1 \cdot \vec{\delta}_{d_0+d_1+1}^\top \cdot \vec{\mathbf{1}}_{N_2}$. 
            For $\mathbf{M}^{(1)}_{D,E}(d_0,d_1)$, embed a scalar $1$, such that the entry becomes $q_0(1-q_1)\cdot 1$.

            \item[b)] For $\mathbf{M}^{(1)}_{E,D}(d_0,d_1)$, embed a scalar $\vec{\delta}_{d_1}^\top \cdot \vec{\mathbf{1}}_{N_2}$, such that the entry becomes $(1-q_0) q_1 \cdot \vec{\delta}_{d_1}^\top \cdot \vec{\mathbf{1}}_{N_2}$. For $\mathbf{M}^{(1)}_{E,E}(d_0,d_1)$, embed a scalar $1$, such that that the entry becomes $(1-q_0) (1-q_1)\cdot 1$.
        \end{itemize}

        \item[2.] Derivation of $\mathbf{T}_{0,\phi}$. 
        \begin{itemize}
            \item[a)] For $\mathbf{M}^{(1)}_{D,D}(d_0,d_1)$, embed a row vector $\vec{\delta}_{d_0+d_1+1}^\top\cdot \mathbf{I}(-N_2)$, such that the entry becomes $q_0 q_1 \cdot \vec{\delta}_{d_0+d_1+1}^\top\cdot \mathbf{I}(-N_2)$. 
            For $\mathbf{M}^{(1)}_{D,E}(d_0,d_1)$, embed a $1\times (m_2-1)$-all zero row vector $\vec{\mathbf{0}}^\top$.

            \item[b)] For $\mathbf{M}^{(1)}_{E,D}(d_0,d_1)$, embed a row vector $\vec{\delta}_{d_1}^\top \cdot \mathbf{I}(-N_2)$, such that the entry becomes $(1-q_0) q_1 \cdot \vec{\delta}_{d_1}^\top \cdot \mathbf{I}(-N_2)$. For $\mathbf{M}^{(1)}_{E,E}(d_0,d_1)$, embed a $1\times (m_2-1)$-all zero row vector $\vec{\mathbf{0}}^\top$.
        \end{itemize}

        \item[3.] Derivation of $\mathbf{T}_{\phi,\phi}$. 
        \begin{itemize}
            \item[a)] For $\mathbf{M}^{(1)}_{D,D}(d_0,d_1)$, embed a matrix $\mathbf{I}\big(d_0+d_1+1\big)\cdot \mathbf{I}(-N_2)$, such that the entry becomes $q_0 q_1 \cdot \mathbf{I}\big(d_0+d_1+1\big)\cdot \mathbf{I}(-N_2)$. 
            For $\mathbf{M}^{(1)}_{D,E}(d_0,d_1)$, embed a matrix $\mathbf{I}(0)$, such that that the entry becomes $q_0(1-q_1)\cdot \mathbf{I}(0)$.

            \item[b)] For $\mathbf{M}^{(1)}_{E,D}(d_0,d_1)$, embed a matrix $\mathbf{I}(d_1)\cdot \mathbf{I}(-N_2) $, such that the entry becomes $(1-q_0) q_1 \cdot \mathbf{I}\big(d_1\big) \cdot \mathbf{I}(-N_2)$. For $\mathbf{M}^{(1)}_{E,E}(d_0,d_1)$, embed a matrix $\mathbf{I}(0)$, such that that the entry becomes $(1-q_0) (1-q_1)\cdot \mathbf{I}(0)$.
        \end{itemize}

        \item[4.] Derivation of $\mathbf{T}_{\phi,0}$.
        \begin{itemize}
            \item[a)] For $\mathbf{M}^{(1)}_{D,D}(d_0,d_1)$, embed a column vector $\mathbf{I}\big(d_0+d_1+1\big) \cdot \vec{\mathbf{1}}_{N_2}$, such that the entry becomes $q_0 q_1 \cdot \mathbf{I}(d_0+d_1+1) \cdot \vec{\mathbf{1}}_{N_2}$. 
            For $\mathbf{M}^{(1)}_{D,E}(d_0,d_1)$, embed a $(m_2-1)\times 1$-all zero column vector $\vec{\mathbf{0}}$.

            \item[b)] For $\mathbf{M}^{(1)}_{E,D}(d_0,d_1)$, embed a column vector $\mathbf{I}\big(d_1\big) \cdot \vec{\mathbf{1}}_{N_2}$, such that the entry becomes $(1-q_0) q_1 \cdot \mathbf{I}\big(d_1\big) \cdot \vec{\mathbf{1}}_{N_2}$. For $\mathbf{M}^{(1)}_{E,E}(d_0,d_1)$, embed a $(m_2-1)\times 1$-all zero column vector $\vec{\mathbf{0}}$.
        \end{itemize}
    \end{itemize}
\end{prop}

\textit{Proof:} 
The description of Proposition \ref{proposition:transition matrix of information debt} seems complex, while its insight is actually straight-forward. 
Upon $\mathbf{M}^{(1)}$, the embedding process in Proposition \ref{proposition:transition matrix of information debt} only accounts for transition of $I_d(t)$.
When the second hop fails, no equations and unknowns will be received at $r_2$, thus $I_d(t) = I_d(t-1)$ (corresponding to the scalar 1s in $\mathbf{T}_{0,0}$ and the identity matrix $\mathbf{I}(0)$ in $\mathbf{T}_{\phi,\phi}$). When the second hop succeeds, the transition depends on the number of source symbols transmitted from the first hop. If the first hop also succeeds, $(d_0+1)+d_1$ unknowns will be received at $r_2$ (corresponding to the $\vec{\delta}_{d_0+d_1+1}^\top$ and $\mathbf{I}(d_0+d_1+1)$ in the matrices). If the first hop fails, $(d_0+1)$ symbols are still detained at $r_0$ and thus only $d_1$ unknowns will be received at $r_2$ (corresponding to the $\vec{\delta}_{d_1}^\top$ and $\mathbf{I}(d_1)$ in the matrices).
Besides, when the second hop succeeds, $N_2$ equations will be received at $r_2$ (corresponding to the $\mathbf{I}(-N_2)$ and $\vec{\mathbf{1}}_{N_2}$ in the matrices).
Note that when the length of a round $k\ge 2$, the second hop of transmission in the first and the last timeslots of a round must succeed (corresponding to the $\vec{\mathbf{0}}^\top$ in $\mathbf{T}_{0,\phi}$ and $\vec{\mathbf{0}}$ in $\mathbf{T}_{\phi,0}$). 
We explain this statement separately as follows.
If the second hop of transmission fails at timeslot $t$ with the condition that $I_d(t-1)=0$, we immediately have $I_d(t)=0$ because no equations and unknowns are received at $r_2$ in this timeslot, which indicates the length of this round equals to $k=1$.
On the other hand, if the second hop of transmission fails at timeslot $t$ with the condition that $I_d(t-1)>0$, similarly we have $I_d(t)>0$, which indicates this is not the last timeslot of a round.
With the statements above, Proposition \ref{proposition:transition matrix of information debt} is proved by embedding the corresponding terms into the entries located by different band sequences (i.e. the subscripts $(D,D),(D,E),(E,D),(E,E)$) and row indices $d_0,d_1$.
\QEDA

With the joint transition matrices $\mathbf{T}_{0,0},\mathbf{T}_{0,\phi},\mathbf{T}_{\phi,\phi},\mathbf{T}_{\phi,0}$, the probability $\text{Pr}(t_{i_0+1}-t_{i_0} = k)$ and stationary initial distribution $\pi^\infty$ can be given in Proposition \ref{proposition:Pr(ti0+1-ti0)} and \ref{proposition:initial distribution}. 

\begin{prop}\label{proposition:Pr(ti0+1-ti0)}
    Assume that the stationary probability distribution of the hidden states starting from $I_d(t)=0$, i.e.,  $\pi^\infty$ is given. 
    The probability of event that time interval between any two adjacent decoding time equals to $k$, i.e., $\text{Pr}(t_{i_0+1}-t_{i_0} = k)$ can derived as follows.
    For $k=1$, we have 
    \begin{equation}\label{equation:Pr(T=1)}
        \text{Pr}(t_{i_0+1}-t_{i_0} = 1) = \pi^\infty \times \mathbf{T}_{0,0} \times \vec{\mathbf{1}}.
    \end{equation}
    For any $k\ge 2$, we have
    \begin{equation}\label{equation:Pr(T=k)}
        \text{Pr}(t_{i_0+1}-t_{i_0} = k) = \pi^\infty \times \mathbf{T}_{0,\phi} \times \mathbf{T}_{\phi,\phi}^{k-2} \times \mathbf{T}_{\phi,0}\times \vec{\mathbf{1}}.
    \end{equation}
\end{prop}
\textit{Proof:} Since the transition of detained symbols and information debt is a Markov process, Proposition \ref{proposition:Pr(ti0+1-ti0)} follows directly from the physical meaning of the joint transition matrices. 
Note that the $\vec{\mathbf{1}}$ in (\ref{equation:Pr(T=1)}) and (\ref{equation:Pr(T=k)}) are $(m_0 m_1)\times(m_0 m_1)$ column vectors of all 1s, which is to sum up the probabilities conditioning on each of the hidden states. \QEDA

\begin{prop}\label{proposition:initial distribution}
    Let $A = \left[\mathbf{I}_{m_0 \cdot m_1 \cdot (m_2-1)}-\mathbf{T}_{\phi,\phi}\right]^{-1}$.
    The stationary distribution of the hidden states when information debt initials from zero, i.e., $\pi^\infty$, is the solution of the following equations:
\begin{equation}\label{equation:solution of initial}
    \begin{bmatrix}
    (T_{0\rightarrow 0} - \mathbf{I}_{m_0 m_1})^\top \\
     \begin{array}{ccc}
        1 & \cdots & 1 
     \end{array}
    \end{bmatrix}_{(m_0 m_1+1)\times(m_0 m_1)} \cdot 
\begin{bmatrix}
    {\pi^\infty}^\top
\end{bmatrix}_{(m_0 m_1)\times1}
=\begin{bmatrix}
    0 \\
    \vdots \\
    0 \\
    1
\end{bmatrix}_{(m_0 m_1+1)\times1},
\end{equation}
where 
\begin{align}
    T_{0\rightarrow 0} = \mathbf{T}_{0,0} + \mathbf{T}_{0,\phi}\times A \times \mathbf{T}_{\phi,0}
\end{align}
is a $(m_0 m_1)\times (m_0 m_1)$ joint transition matrix of initial distribution of the hidden states $\widehat{D}_0(t)$ and $\widehat{D}_1(t)$ between any two adjacent times that $I_d(t)$ hits zero.
Specifically, the entries of $T_{0\rightarrow 0}$ are denoted as $\big[t(i,j;v,w)\big],i,v\in[0,m_0-1],j,w\in[0,m_1-1]$, where $t(i,j;v,w) \triangleq \text{Pr}\big(\widehat{D}_0(t_{i_0+1})=v,\widehat{D}_1(t_{i_0+1})=w \big| \widehat{D}_0(t_{i_0})=i,\widehat{D}_1(t_{i_0})=j\big)$. 
\end{prop}

The proof of Proposition \ref{proposition:initial distribution} is presented in Appendix \ref{Appendix:initial distribution}. 
Proposition \ref{proposition:initial distribution} shows that the initial probability distribution at any two adjacent decoding times $t_{i_0}$ and $t_{i_0+1}$ satisfies $\pi^{t_{i_0 + 1}} = \pi^{t_{i_0}} \cdot T_{0\rightarrow 0}$.
Therefore, the stationary initial distribution $\pi^\infty$ can be derived by solving equations (\ref{equation:solution of initial}). It is worthy noting that although there are $m_0\times m_1 + 1$ equations in (\ref{equation:solution of initial}), only $m_0\times m_1$ out of them are actually effective for the solution. This
is because the first $m_0\times m_1$ equations are linearly dependent, due to the feature of stochastic matrix $T_{0\rightarrow 0}$. 

With Proposition \ref{proposition:Pr(ti0+1-ti0)} and \ref{proposition:initial distribution}, the denominator and numerator of (\ref{equation:P_e}) can be derived in the following Lemmas, respectively.

\begin{lemma}\label{lemma:denominator}
    The denominator of (\ref{equation:P_e}) can be given as follows.
    \begin{align}
        \mathbb{E}\bigg\{\frac{D_2(t_{i_0+1})}{K}\bigg\} =\pi^\infty  \Big[\mathbf{T}_{0,0} + \mathbf{T}_{0,\phi}  A \big(\mathbf{I}_{m_0 m_1 (m_2-1)}+A\big) \mathbf{T}_{\phi,0}\Big]\vec{\mathbf{1}}.
    \end{align}
\end{lemma}

The proof of Lemma \ref{lemma:denominator} is given in Appendix \ref{Appendix:denominator}.

The characterization of the numerator is much more technical and involved. 
\begin{lemma}\label{lemma:numerator}
    Denote a $(m_0+m_1-1)\times 1$ column vector $\vec{\gamma} = [\overbrace{1,\cdots,1}^{\Delta + 2},0,-1,-2,\cdots]^\top$, such that its first $\Delta + 2$ elements are all ones, and each of the rest element equals to its previous element minus one.
    Denote $B = diag\big(0,1,2,3,\cdots\big)$ as diagonal matrix with proper size.
    Recall $A = \left[\mathbf{I}_{m_0 \cdot m_1 \cdot (m_2-1)}-\mathbf{T}_{\phi,\phi}\right]^{-1}$ and $\vec{\mathbf{1}}$ is a column vector of all 1s.
    The numerator of (\ref{equation:P_e}) can be given as follows.
    \begin{align}
        &\quad\mathbb{E}\left\{
        \min\bigg[
        \Big(t_{i_0+1} - t_{i_0} + \frac{D_0(t_{i_0}) + D_1(t_{i_0})}{K} - \Delta - 1\Big)^+ , \frac{D_2(t_{i_0+1})}{K}
        \bigg]
        \right\} \nonumber\\
        &= \pi^\infty \mathbf{Q} \mathbf{P} (\mathbf{T}_{0,0} + \mathbf{T}_{0,\phi} A \mathbf{T}_{\phi,0}) \mathbf{Q} \vec{\gamma} 
        + \pi^\infty \mathbf{Q} \mathbf{P} \mathbf{T}_{0,\phi} \mathbf{T}_{\phi,\phi}^{\Delta + 1} A^2 \mathbf{T}_{\phi,0} \mathbf{Q} \vec{\mathbf{1}}
        + \pi^\infty \mathbf{Q} B \mathbf{P} \mathbf{T}_{0,\phi}\mathbf{T}_{\phi,\phi}^{\Delta + 1} A\mathbf{T}_{\phi,0} \mathbf{Q} \vec{\mathbf{1}}
        +  \nonumber\\
        & \qquad\qquad \pi^\infty \mathbf{Q} \sum_{k=1}^{\Delta+2}\begin{bmatrix}
        \mathbf{0}_{\Delta+2-k} & \\
         & B \end{bmatrix} \mathbf{P} \mathbf{T}_{0,\phi} \mathbf{T}_{\phi,\phi}^{k-2} \mathbf{T}_{\phi,0} \mathbf{Q} \vec{\mathbf{1}} - \sum_{k=1}^{\Delta+1} \pi^{\infty}(0:\Delta + 1 - k) \mathbf{Q} \mathbf{P}  T_{0\rightarrow 0}^{(k)}(0:\Delta+1-k,:) \mathbf{Q}\vec{\gamma}, \label{equation:numerator}
    \end{align}
where $T_{0\rightarrow 0}^{(1)} = \mathbf{T}_{0,0}$ and $T_{0\rightarrow 0}^{(k)} = \mathbf{T}_{0,\phi} \times \mathbf{T}_{\phi,\phi}^{k-2} \times \mathbf{T}_{\phi,0},k\ge 2$ defined in the proof of Proposition \ref{proposition:initial distribution} are the joint transition matrices of initial distribution of the hidden states $\widehat{D}_0(t)$ and $\widehat{D}_1(t)$ between any two decoding times $t_{i_0}$ and $t_{i_0+1}$ while the event $t_{i_0+1} - t_{i_0} = k$ occurs.
And $T_{0\rightarrow 0}^{(k)}(0:\Delta+1-k,:)$ is a submatrix of $T_{0\rightarrow0}^{(k)}$ consisting of  its first $\Delta+2-k$ rows (starting from index 0).
$\mathbf{P},\mathbf{Q}$ are the left and right-summation matrices that convert transition matrix of hidden states to transition matrix of the sum of hidden states. More precisely, denote the transition matrix of the sum of hidden states as $\widetilde{T}_{0\rightarrow 0}=\left[\widetilde{t}(j,l)\right]$, where $\widetilde{t}(j,l) \triangleq \text{Pr}\big(\widehat{D}_0(t_{i_0+1}) + \widehat{D}_1(t_{i_0+1})=l \big| \widehat{D}_0(t_{i_0}) + \widehat{D}_1(t_{i_0})=j\big)$.
$\widetilde{T}_{0\rightarrow 0}$ can be generated from $T_{0\rightarrow 0}$ by multiplication $\widetilde{T}_{0\rightarrow 0} = \mathbf{P} \cdot T_{0\rightarrow 0} \cdot \mathbf{Q}$.
\end{lemma}

The proof of Lemma \ref{lemma:numerator} is given in Appendix \ref{Appendix:numerator}. 
The numerator is derived as summation of a series of matrix multiplication.
Note that (\ref{equation:numerator}) contains five terms. The first three terms are multiplications of the joint transition matrices, which are corresponding to summation over the length of each round $t_{i_0+1} - t_{i_0} = k$ from $\Delta + 1$ to infinity.
The last two terms are the remainder terms with respect to the case that $k$ is within the decoding delay $\Delta$.

\begin{theorem}\label{theorem:TRN}
    In TRN with delivery probability $q_0$ and $q_1$ in the first and the second hop, for any decoding delay $\Delta$, the asymptotic slot error probability $p_e$ of large-field-size NRLSCs can be computed by assembling Lemma \ref{lemma:1}, Lemma \ref{lemma:denominator} and \ref{lemma:numerator}.
\end{theorem}

\textit{Proof:} Theorem \ref{theorem:TRN} follows directly from Lemma \ref{lemma:1}, Lemma \ref{lemma:denominator} and Lemma \ref{lemma:numerator}.
\QEDA

\section{Generalization to Multi-hop Relay Networks}\label{section:Main Results3}
In this section, we generalize the analysis of RLSCs on TRN to Multi-hop Relay Networks (MRN), where there are $L+1$ nodes $r_0$ to $r_L$ in the linear system. 
Since the behavior of intermediate relay nodes $r_1$ to $r_{L-1}$ are highly homogeneous, previous results can be readily extended by considering more intermediate relay nodes.
Let $D_l(t)$ denote the number of source symbols detained at $r_l$ at timeslot $t$ and $\widehat{D}_l(t)$ denote the value normalized by $K$, $\forall l\in [0,L]$.
The maximum value of $\widehat{D}_l(t)$ is assumed to be $m_l$.
And $W(t)$ is the number of encoded symbols (corresponding to the $D_L(t)$ source symbols) that
have been received at the destination $r_L$.
In MRN, Definition \ref{definition:informtion debt} can be generalized as follows.  

\begin{definition}\label{definition:informtion debt for n hops}
    Let $D_l(0)=0,l\in [0,L]$. For any $t\ge 1$, the information debt $I_d(t)$ of RLSCs in MRN is calculated iteratively by 
     \begin{align}
        D_0(t) &= D_0(t-1)\cdot e_0(t-1) + K, \label{equation:D0_M}\\
        D_1(t) &= D_1(t-1)\cdot e_1(t-1) + D_0(t)\cdot [1-e_0(t)], \label{equation:D1_M}\\
        &\ \vdots \nonumber\\
        D_{L-1}(t) &= D_{L-1}(t-1)\cdot e_{L-1}(t-1) + D_{L-2}(t)\cdot [1-e_{L-2}(t)], \label{equation:DL-1_M}\\        
        D_L(t) &= D_L(t-1)\cdot \mathds{1}\{I_d(t-1) \neq 0\} + D_{L-1}(t)\cdot [1-e_{L-1}(t)], \label{equation:DL_M}\\
        W(t) &= W(t-1)\cdot \mathds{1}\{I_d(t-1) \neq 0\} + N_{L-1} \cdot [1-e_{L-1}(t)], \label{equation:W_M}\\
        I_d(t) &\triangleq \left[D_L(t) - W(t)\right]^+. \label{equation:Id_M}
    \end{align}   
\end{definition}

Definition \ref{definition:informtion debt for n hops} is derived from Definition \ref{definition:informtion debt} by iteratively adding the intermediate relay nodes with number of detained symbols that satisfies $D_{l}(t) = D_{l}(t-1)\cdot e_{l}(t-1) + D_{l-1}(t)\cdot [1-e_{l-1}(t)], l\in [1,L-1]$.
Equations (\ref{equation:D0_M}) to (\ref{equation:Id_M}) jointly quantify the information flowing through each node in MRN.
With Definition \ref{definition:informtion debt for n hops}, the error event in MRN can be generalized from Proposition \ref{Proposition:characterization of the error event}, and the expression of $p_e$ in Lemma \ref{lemma:1} can be generalized as follows.
\begin{corollary}\label{Proposition:characterization of the error event_M}
     Assume \textbf{GMDS} holds. 
For RLSCs in $L$-hops relay networks, for any fixed index $i_0 \ge 0$, \textbf{(a)} $\mathbf{s}(t)$ is $\Delta$-decodable for all timeslots $t$ satisfying that
\begin{equation}\label{equation:error event_M}
    \forall t \in \left[t_{i_0} - \frac{\sum_{l=0}^{L-1} D_{l}(t_{i_0})}{K}+1,t_{i_0} - \frac{\sum_{l=0}^{L-1} D_{l}(t_{i_0})-D_L(t_{i_0+1})}{K}\right] \bigcap \Big[t_{i_0+1}-\Delta,t_{i_0+1}\Big]
    \end{equation}
\textbf{(b)} $\mathbf{s}(t)$ is not $\Delta$-decodable for the rest of $t$.
\end{corollary}

\begin{corollary}\label{lemma:1_M} 
    Assume the transmission rate is within the capacity. The error probability of RLSCs can be given by 
    \begin{equation}\label{equation:P_e_M}
        p_e = \frac{\mathbb{E}\left\{
        \min\bigg(
        \Big(t_{i_0+1} - t_{i_0} + \frac{\sum_{l=0}^{L-1} D_{l}(t_{i_0})}{K} - \Delta - 1\Big)^+ , \frac{D_L(t_{i_0+1})}{K}
        \bigg)
        \right\}}
        {\mathbb{E}\Big\{\frac{D_L(t_{i_0+1})}{K}\Big\}}.
    \end{equation}
\end{corollary}

In MRN, at any decoding time $t_{i_0}$, $r_L$ is able to decode $D_L(t_{i_0})$ source symbols from $W(t_{i_0})$ equations in this round since its information debt equals to zero.
At $t_{i_0}$, there are also totally $\sum_{l=0}^{L-1} D_{l}(t_{i_0})$ source symbols detained at $r_0$ to $r_{L-1}$, which are undecodable at $r_L$ in this round and should wait for the next round to be determined.
By modifying these two terms in Proposition \ref{Proposition:characterization of the error event} and Lemma \ref{lemma:1}, Corollary \ref{Proposition:characterization of the error event_M} and Corollary \ref{lemma:1_M} can be obtained  directly.

Let $\mathbf{T}_{0,0},\mathbf{T}_{0,\phi},\mathbf{T}_{\phi,\phi},\mathbf{T}_{\phi,0}$ denote the joint transition matrices of $\widehat{D}_0(t),\cdots, \widehat{D}_{L-1}(t),I_d(t)$.
With the nested structure, in MRN the derivation of $\mathbf{T}_{0,0},\mathbf{T}_{0,\phi},\mathbf{T}_{\phi,\phi},\mathbf{T}_{\phi,0}$, can be readily extended from Proposition \ref{proposition:transition matrix of hidden states} and \ref{proposition:transition matrix of information debt}.
Specifically, in Proposition \ref{proposition:transition matrix of hidden states}, one can repeatedly add embedding steps similar to step 2 to construct the joint transition matrix of the hidden states $\widehat{D}_0(t),\cdots \widehat{D}_{L-1}(t)$. 
In general, the joint transition matrices of $\widehat{D}_0(t),\cdots \widehat{D}_{L-1}(t),I_d(t)$ in MRN can be given by the construction in the following corollaries. 

\begin{corollary}\label{proposition:transition matrix of hidden states_M}
    In MRN, the joint transition matrix of $\widehat{D}_0(t),\cdots \widehat{D}_{L-1}(t)$ can be derived by the following steps of construction.
    \begin{itemize}
        \item[1.] For the first hop, construct a matrix $\mathbf{M}^{(0)} = \mathbf{D}_{1,q_0,m_0}$. 

        \item[2.] Denote the matrix after embedding of the $l-1$-th hop as $\mathbf{M}^{(l-2)}$. For example, $\mathbf{M}^{(0)}$ denotes the $m_0 \times m_0$ matrix after embedding of the first hop.
        For the $l$-th hop, $l\ge 2$, each non-zero entry of $\mathbf{M}^{(l-2)}$ can be uniquely located by a sequence of bands as well as the row indexes of the hidden states $\widehat{D}_0(t),\cdots, \widehat{D}_{l-2}(t)$.
        Specifically, denote each non-zero entry of $\mathbf{M}^{(l-2)}$ as
        $\mathbf{M}^{(l-2)}_{i}(d_0,\cdots,d_{l-2})$, where $i\in [0,2^{l-1}-1]$ represents a band sequence with length $l-1$ and $d_0 \in [0,m_0-1],\cdots,d_{l-2}\in [0,m_{l-2}-1]$ are the row indices.
        For example, a band sequence $D,D,D,E,E$ with length $5$ can be represented by a binary string $11100$, where 1 denotes $D$ and 0 denotes $B$.
        Since $(11100)_2=(28)_{10}$, $i=28$ can represent the band sequence $D,D,D,E,E$.
        Denote the corresponding binary string of band sequence $i$ as $b(i)$. 
        Define a binary operation function $Z(*)$ which returns the position of the last zero\footnote{The last position of zero is counted from the left. For example, for string 11010111, the third and the fifth digits are zeros. 
        Thus the position of the last zero-digit is 5, i.e., $Z(11010111)=5$.
        For string 11111111, no zero is included, thus $Z(11111111)=0$. 
        For string 00000000, $Z(00000000)=8$.} in a binary string. 
        Thus, position of the last zero in $b(i)$ can be given by $Z(b(i))$.
        For each entry of $\mathbf{M}^{(l-2)}$, embed a transition matrix as follows. 
            \begin{itemize}
                \item[(a)] 
                For any non-zero entry $\mathbf{M}^{(l-2)}_{i}(d_0,\cdots,d_{l-2}), i\in [0,2^{l-1}-1]$, (1) if $Z(b(i)) = 0$, i.e., $i=2^{l-1}-1$ such that the corresponding binary string is all ones, embed $\mathbf{D}_{1 + \sum_{w = 0}^{l-1} d_w,q_{l-1},m_{l-1}}$; (2) otherwise, embed $\mathbf{D}_{\sum_{w = Z(b(i))}^{l-1} d_w,q_{l-1},m_{l-1}}$. 

                \item[(b)] For each of the rest zero-entries in $\mathbf{M}^{(l-2)}$, expand it into a $m_{l-1} \times m_{l-1}$ zero matrix $\mathbf{0}_{m_{l-1}}$.
            \end{itemize}  
            Repeatedly execute step 2 until $l=L$, such that $\mathbf{M}^{(L-1)}$, the $\left(\prod_{l=0}^{L-1} m_l\right) \times \left(\prod_{l=0}^{L-1} m_l\right)$ joint transition matrix of the hidden states $\widehat{D}_0(t),\cdots \widehat{D}_{L-1}(t)$ is established.
    \end{itemize}
\end{corollary}

In Corollary \ref{proposition:transition matrix of information debt_M}, $\mathbf{M}^{(L-1)}$ will be further converted to the joint transition matrices $\mathbf{T}_{0,0},\mathbf{T}_{0,\phi},\mathbf{T}_{\phi,\phi},\mathbf{T}_{\phi,0}$.

\begin{corollary}\label{proposition:transition matrix of information debt_M}
    The joint transition matrices of $\widehat{D}_0(t),\cdots \widehat{D}_{L-1}(t),I_d(t)$ in MRN can be constructed by embedding the non-zero entries of $\mathbf{M}^{(L-1)}$, i.e., $\mathbf{M}^{(L-1)}_i(d_0,\cdots,d_{L-1})$, $i\in [0,2^{L}-1]$ and $d_0 \in [0,m_0-1],\cdots,d_{L-1}\in [0,m_{L-1}-1]$ as follows.
    \begin{itemize}
        \item[1.] To derive $\mathbf{T}_{0,0}$, for each non-zero entry $\mathbf{M}^{(L-1)}_i(d_0,\cdots,d_{L-1})$,
        \begin{itemize}
            \item[a)] if $Z(b(i)) = L$, i.e., the last band in the band sequence $i$ is $E$, embed a scalar $1$,

            \item[b)] if $Z(b(i)) = 0$, i.e., $i=2^{L}-1$, such that every band in the band sequence $i$ is $D$, embed a scalar $\vec{\delta}_{1 + \sum_{w = 0}^{L-1} d_w}^\top \cdot \vec{\mathbf{1}}_{N_2}$,
            
            \item[c)] otherwise, embed a scalar $\vec{\delta}_{\sum_{w = Z(b(i))}^{L-1} d_w}^\top \cdot \vec{\mathbf{1}}_{N_2}$.
        \end{itemize}

        \item[2.] To derive $\mathbf{T}_{0,\phi}$, for each non-zero entry $\mathbf{M}^{(L-1)}_i(d_0,\cdots,d_{L-1})$, 
        \begin{itemize}
            \item[a)] if $Z(b(i)) = L$, embed a $1\times (m_{L}-1)$-all zero row vector $\vec{\mathbf{0}}^\top$,

            \item[b)] if $Z(b(i)) = 0$, embed a row vector $\vec{\delta}_{1 + \sum_{w = 0}^{L-1} d_w}^\top\cdot \mathbf{I}(-N_2)$
            
            \item[c)] otherwise, embed a row vector $\vec{\delta}_{\sum_{w = Z(b(i))}^{L-1} d_w}^\top \cdot \mathbf{I}(-N_2)$.
        \end{itemize}

        \item[3.] To derive $\mathbf{T}_{\phi,\phi}$, for each non-zero entry $\mathbf{M}^{(L-1)}_i(d_0,\cdots,d_{L-1})$, 
        \begin{itemize}
            \item[a)] if $Z(b(i)) = L$, embed a $(m_{L}-1) \times (m_{L}-1)$-matrix $\mathbf{I}(0)$,

            \item[b)] if $Z(b(i)) = 0$, embed a matrix $\mathbf{I}\big(1 + \sum_{w = 0}^{L-1} d_w\big)\cdot \mathbf{I}(-N_2)$
            
            \item[c)] otherwise, embed a matrix $\mathbf{I}\big(\sum_{w = Z(b(i))}^{L-1} d_w\big)\cdot \mathbf{I}(-N_2)$.
        \end{itemize}

        \item[4.] To derive $\mathbf{T}_{\phi,0}$, for each non-zero entry $\mathbf{M}^{(L-1)}_i(d_0,\cdots,d_{L-1})$, 
        \begin{itemize}
            \item[a)] if $Z(b(i)) = L$, embed a $(m_{L}-1)\times 1$-all zero column vector $\vec{\mathbf{0}}$,

            \item[b)] if $Z(b(i)) = 0$, embed a column vector $\mathbf{I}\big(1 + \sum_{w = 0}^{L-1} d_w\big) \cdot \vec{\mathbf{1}}_{N_2}$,
            
            \item[c)] otherwise, embed a column vector $\mathbf{I}\big(\sum_{w = Z(b(i))}^{L-1} d_w\big) \cdot \vec{\mathbf{1}}_{N_2}$.
        \end{itemize}
    \end{itemize}
\end{corollary}

\begin{remark}
    (\textit{Discussion on complexity of the constructions})
    Note that when $L$ increases by one, the embedding operation of the joint transition matrices should add one more dimension of hidden state $\widehat{D}_{L+1}(t)$, i.e., $m_{L+1}^2$. 
    With the increase of $L$, the size of joint transition matrices will increase exponentially. 
    For large scale linear networks with a great many of hops, two techniques can be exploited to reduce the complexity.
    (1) Choose appropriately small values for $\{m_i|i\in [L]\}$. Since when each time the transmission on link $(r_i,r_{i+1})$ succeeds $\widehat{D}_i(t)$ will be reset to zero, $\widehat{D}_i(t)$ can accumulate to a large value only when the transmission on link $(r_{i-1},r_{i})$ succeeds for a few timeslots and at the same time the transmission on link $(r_i,r_{i+1})$ consecutively fails for a few timeslots, which is a low-probability event. 
    For example, for $q_{i-1} = 0.6$ and $q_{i} = 0.8$, the probability that $\widehat{D}_i(t)$ increases by 5 in 5 timeslots equals to $(0.6)^5\times (1-0.8)^5  \approx  2.5\times 10^{-5}$.
    Therefore, for simulation with a large $L$, it can be set that $m_i = 5,\forall i\in [0,L-1]$. In Section \ref{section:numerical results} we will show that this setting is sufficient to obtain an accurate simulation result. 
    (2) Note that due to the structure of embedding operations, the resulting joint transition matrices are highly sparse. For example, when $L=5$ and $m_i=5$ for $i\in[0,L-1]$, the corresponding $\mathbf{T}_{0,0}$ is with size $3125 \times 3125$.
    However, there are only $2^5 = 32$ elements in each row of $\mathbf{T}_{0,0}$. Thus its density approximates $0.01\%$, which indicates significant potential gain that can be achieved by using the methods for sparse matrix \cite{sparse matrix 1,sparse matrix 2}.
    Since the matrices are constructed in a banded fashion, each row contains the same number of elements.
    This feature makes the matrices efficient in row access.
    Therefore, one can first convert the Coordinate (COO) format to the Compressed Sparse Row (CSR) format and exploit optimized multiplications to significantly reduce the complexity.
\end{remark}

With the joint transition matrices $\mathbf{T}_{0,0},\mathbf{T}_{0,\phi},\mathbf{T}_{\phi,\phi},\mathbf{T}_{\phi,0}$, the probability $\text{Pr}(t_{i_0+1}-t_{i_0} = k)$ and stationary initial distribution $\pi^\infty$ can be derived the same as in Proposition \ref{proposition:Pr(ti0+1-ti0)} and \ref{proposition:initial distribution}.

\begin{corollary}\label{proposition:Pr(ti0+1-ti0)_M}
    Assume that the probability distribution of the hidden states starting from $I_d(t)=0$, i.e.,  $\pi^\infty$ is given. 
    In MRN with $L$ hops, the event that time interval between any two adjacent decoding time equals to $k$, i.e., $\text{Pr}(t_{i_0+1}-t_{i_0} = k)$ can derived as follows.
    For $k=1$, we have 
    \begin{equation}\label{equation:Pr(T=1)_M}
        \text{Pr}(t_{i_0+1}-t_{i_0} = k) = \pi^\infty \times \mathbf{T}_{0,0} \times \vec{\mathbf{1}}.
    \end{equation}
    For any $k\ge 2$, we have
    \begin{equation}\label{equation:Pr(T=k)_M}
        \text{Pr}(t_{i_0+1}-t_{i_0} = k) = \pi^\infty \times \mathbf{T}_{0,\phi} \times \mathbf{T}_{\phi,\phi}^{k-2} \times \mathbf{T}_{\phi,0}\times \vec{\mathbf{1}}.
    \end{equation}
\end{corollary}

\begin{corollary}\label{proposition:initial distribution_M}
    In MRN with $L$ hops, let $A = \left[\mathbf{I}_{\prod_{l = 0}^{L-1} m_l \cdot (m_{L}-1)}-\mathbf{T}_{\phi,\phi}\right]^{-1}$.
    The stationary distribution of the hidden states when information debt initials from zero, i.e., $\pi^\infty$, is the solution of the following equations:
\begin{equation}\label{equation:solution of initial_M}
    \begin{bmatrix}
    (T_{0\rightarrow 0} - \mathbf{I}_{\prod_{l = 0}^{L-1} m_l})^\top \\
     \begin{array}{ccc}
        1 & \cdots & 1 
     \end{array}
    \end{bmatrix}_{\big(\prod_{l = 0}^{L-1} m_l+1\big)\times\big(\prod_{l = 0}^{L-1} m_l\big)} \cdot 
\begin{bmatrix}
    {\pi^\infty}^\top
\end{bmatrix}_{\big(\prod_{l = 0}^{L-1} m_l\big)\times1}
=\begin{bmatrix}
    0 \\
    \vdots \\
    0 \\
    1
\end{bmatrix}_{\big(\prod_{l = 0}^{L-1} m_l+1\big)\times 1},
\end{equation}
where 
\begin{align}
    T_{0\rightarrow 0} = \mathbf{T}_{0,0} + \mathbf{T}_{0,\phi}\times A \times \mathbf{T}_{\phi,0}
\end{align}
is the joint transition matrix of initial distribution of the hidden states $\widehat{D}_0(t),\cdots,\widehat{D}_{L-1}(t)$ between any two adjacent decoding times.
\end{corollary}

Similar to Lemma \ref{lemma:denominator} and \ref{lemma:numerator}, the denominator and numerator of (\ref{equation:P_e_M}) can be also derived as follows.
\begin{corollary}\label{lemma:denominator_M}
    In MRN with $L$ hops, the denominator of (\ref{equation:P_e_M}) can be given by
    \begin{align}
        \mathbb{E}\bigg\{\frac{D_L(t_{i_0+1})}{K}\bigg\} =\pi^\infty  \Big[\mathbf{T}_{0,0} + \mathbf{T}_{0,\phi}  A \big(\mathbf{I}_{\prod_{l = 0}^{L-1} m_l \cdot (m_{L}-1)}+A\big) \mathbf{T}_{\phi,0}\Big]\vec{\mathbf{1}}.
    \end{align}
\end{corollary}

\begin{corollary}\label{lemma:numerator_M}
    Denote a $\big(1+\sum_{l=0}^{L-1}(m_l-1)\big)\times 1$ column vector $\vec{\gamma} = [\overbrace{1,\cdots,1}^{\Delta + 2},0,-1,-2,\cdots]^\top$, such that its first $\Delta + 2$ elements are all ones, and each of the rest element equals to its previous element minus one.
    Denote $B = diag\big(0,1,2,3,\cdots\big)$ as diagonal matrix with proper size.
    In MRN with $L$ hops, the numerator of (\ref{equation:P_e_M}) can be given as follows.
    \begin{align}
        &\quad\mathbb{E}\left\{
        \min\bigg(
        \Big(t_{i_0+1} - t_{i_0} + \frac{\sum_{l=0}^{L-1} D_{l}(t_{i_0})}{K} - \Delta - 1\Big)^+ , \frac{D_L(t_{i_0+1})}{K}
        \bigg)
        \right\} \nonumber\\
        &= \pi^\infty \mathbf{Q} \mathbf{P} (\mathbf{T}_{0,0} + \mathbf{T}_{0,\phi} A \mathbf{T}_{\phi,0}) \mathbf{Q} \vec{\gamma} 
        + \pi^\infty \mathbf{Q} \mathbf{P} \mathbf{T}_{0,\phi} \mathbf{T}_{\phi,\phi}^{\Delta + 1} A^2 \mathbf{T}_{\phi,0} \mathbf{Q} \vec{\mathbf{1}}
        + \pi^\infty \mathbf{Q} B \mathbf{P} \mathbf{T}_{0,\phi}\mathbf{T}_{\phi,\phi}^{\Delta + 1} A\mathbf{T}_{\phi,0} \mathbf{Q} \vec{\mathbf{1}}
        +  \nonumber\\
        & \qquad\qquad \pi^\infty \mathbf{Q} \sum_{k=1}^{\Delta+2}\begin{bmatrix}
        \mathbf{0}_{\Delta+2-k} & \\
         & B \end{bmatrix} \mathbf{P} \mathbf{T}_{0,\phi} \mathbf{T}_{\phi,\phi}^{k-2} \mathbf{T}_{\phi,0} \mathbf{Q} \vec{\mathbf{1}} - \sum_{k=1}^{\Delta+1} \pi^{\infty}(0:\Delta + 1 - k) \mathbf{Q} \mathbf{P}  T_{0\rightarrow 0}^{(k)}(0:\Delta+1-k,:) \mathbf{Q}\vec{\gamma},\label{equation:numerator_M}
    \end{align}
\end{corollary}

\begin{theorem}\label{theorem:MRN}
    In $L$-hop-MRN with delivery probability $q_0,\cdots,q_{L-1}$, for any decoding delay $\Delta$, the asymptotic slot error probability $p_e$ of large-field-size NRLSCs can be computed by assembling Corollary \ref{Proposition:characterization of the error event_M} to \ref{lemma:numerator_M}.
\end{theorem}

\section{Numerical Results}\label{section:numerical results}

In this section, we conduct numerical simulations on the theoretical results derived in this paper. 
In the first subsection, we examine the accuracy of our derived $p_e$ by comparing to some Monte-Carlo simulations.
In the second subsection, the performance of RLSCs with different values of parameters (decoding latency, erasure probability and coding rate) is compared.
In the third subsection, the performance of RLSCs is compared to the existing streaming code proposed in \cite{SLF 3hop first paper} for adversarial channel.

\subsection{Relative error of $p_e$ in Theorem \ref{theorem:TRN}}

\begin{figure}[thbp!]
    \centering
    \begin{minipage}[t]{0.49\linewidth}
        \centering
        \includegraphics[width=0.8\linewidth]{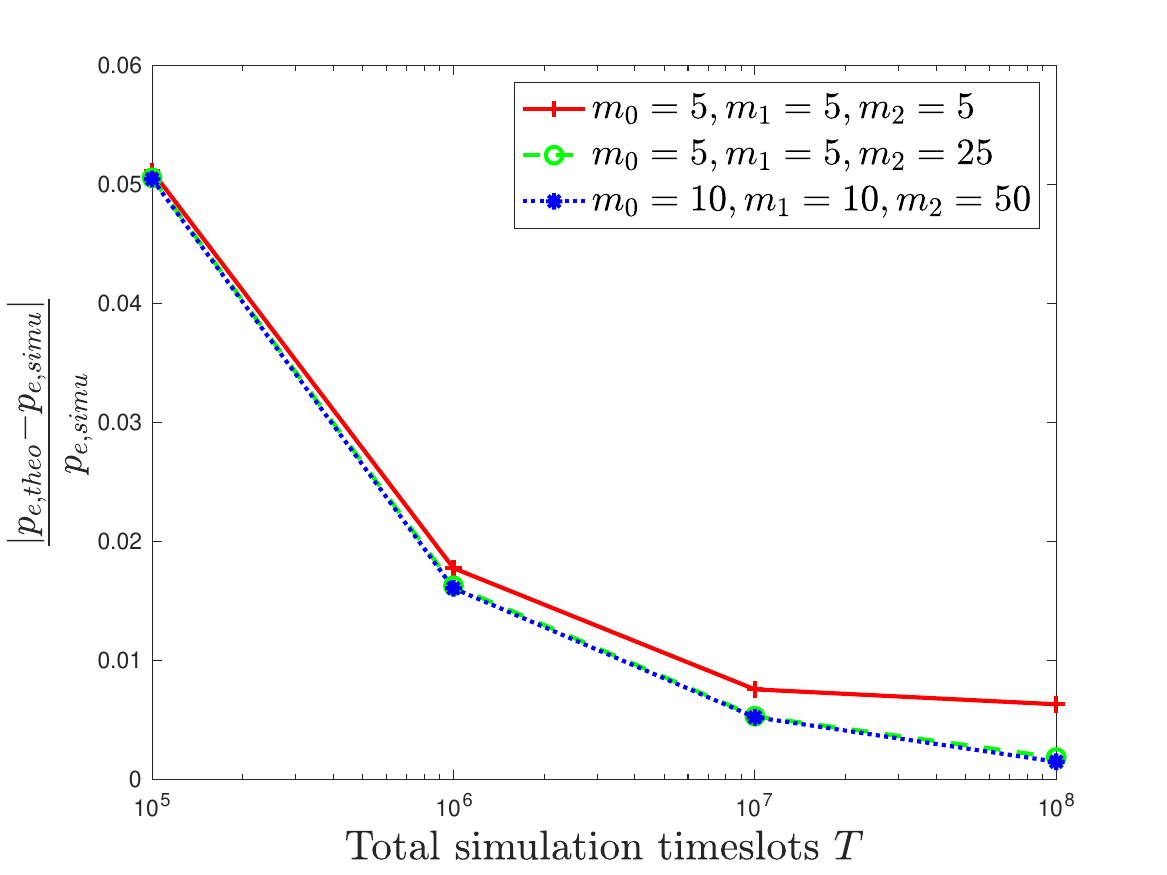}
        \caption{Relative error of $p_e$ in a two-hop relay network.}
        \label{fig:RelativeErrorOfPe_2hops_T}
    \end{minipage}
    \begin{minipage}[t]{0.49\linewidth}
        \centering
        \includegraphics[width=0.8\linewidth]{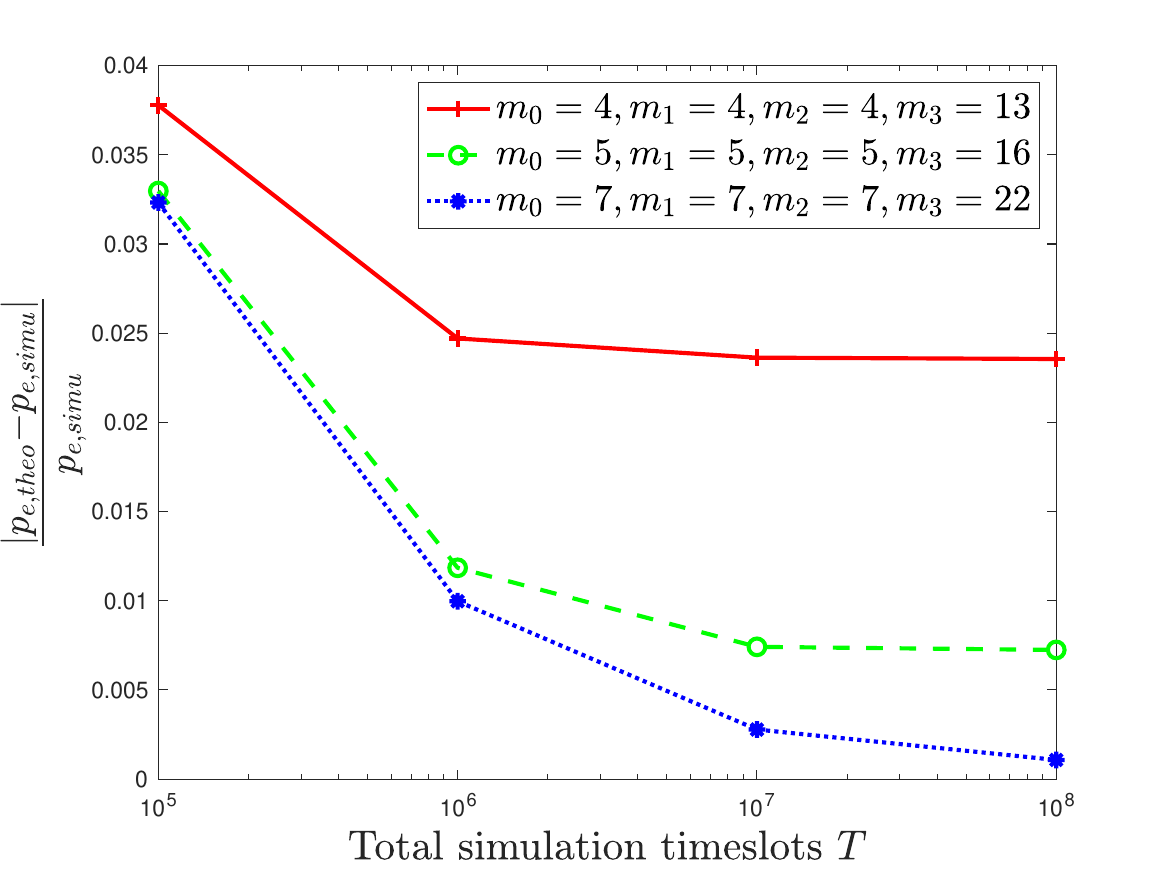}
        \caption{Relative error of $p_e$ in a three-hop relay network.}
        \label{fig:RelativeErrorOfPe_3hops_T}
    \end{minipage}
 \end{figure}

In this subsection, we verify the correctness of theoretical $p_e$ of RLSCs, denoted as $p_{e,theo}$, in multi-hop relay networks.
First we numerically compare $p_{e,theo}$ in two-hop relay network derived by Theorem \ref{theorem:TRN} to the corresponding $p_{e,simu}$ generated by Monte-Carlo simulation. 
The system parameters are set to $K=1,N_2=3,\Delta=2$. 
We consider two-hop relay network with i.i.d. packet erasure channel and symmetric successful delivery probability $q_0 = q_1 = 0.9$ in each hop.
For the Monte-Carlo simulation, in each experiment we sample $T$ timeslots of channel realizations and then determine the error events accordingly, and finally calculate the error probability $p_{e,simu}$. 
The number of timeslots $T$ varies logarithmically from $10^5$ to $10^8$. 
At each value of $T$, $p_{e,simu}$ is averaged over 100 experiments. 
In Fig. \ref{fig:RelativeErrorOfPe_2hops_T}, we plot the relative error of $p_e$, i.e., $\frac{|p_{e,theo}-p_{e,simu}|}{p_{e,simu}}$ versus, the sampling timeslots $T$. 
Recall that in the theoretical analysis, we use parameters $m_0,m_1,m_2$ to approximate the infinite number of states.
In the simulation, we also compare between different values of parameters $m_0,m_1,m_2$ to see their impact on the accuracy of approximation.
It is shown in Fig. \ref{fig:RelativeErrorOfPe_2hops_T} that the relative error decreases gradually as $T$ increases. 
It also shows that as the values of parameters $m_0,m_1,m_2$ increase, the corresponding relative error will decrease. 
More importantly, although larger values of $m_0,m_1,m_2$ are more accurate approximations on the infinity number of states, one can also notice that relatively small values of $m_0,m_1,m_2$ are capable of deriving a $p_e$ with sufficiently low error.
For example, at $T=10^8$, the relative error with small values $m_0=5,m_1=5,m_2=5$ is only 0.63\%.
Thus, our approximation of using parameters $m_0,m_1,m_2$ is accurate and effective.
And for $m_0=5,m_1=5,m_2=25$, the relative error at $T=10^8$ is 0.18\%, while for $m_0=10,m_1=10,m_2=50$, the relative error at $T=10^8$ is only 0.15\%.
It implies that the increase of $m_0,m_1,m_2$ has a marginal diminishing effect, such that excessively large value of $m_0,m_1,m_2$ only have limit reduction on the value of relative error. 
Then we also verify the extended theoretical results of multi-hop relay networks in Theorem \ref{theorem:MRN}.
In simulation, we assume a three-hop relay network with $K=1,N_3=3,\Delta=2$ and  successful delivery probability $q_0 = q_1 = q_2=0.9$ in each hop.
The relative error of $p_e$ in three-hop relay network is presented in Fig. \ref{fig:RelativeErrorOfPe_3hops_T}.
For $m_0=7,m_1=7,m_2=7,m_3=22$, the relative error at $T=10^8$ is only 0.11\%.
These results verify the correctness of our theoretical derivations.

\subsection{Performance of RLSCs with different decoding latency, erasure probability and coding rate}

\begin{figure}[thbp!]
    \centering
    \begin{minipage}[t]{0.49\linewidth}
        \centering
        \includegraphics[width=0.9\linewidth]{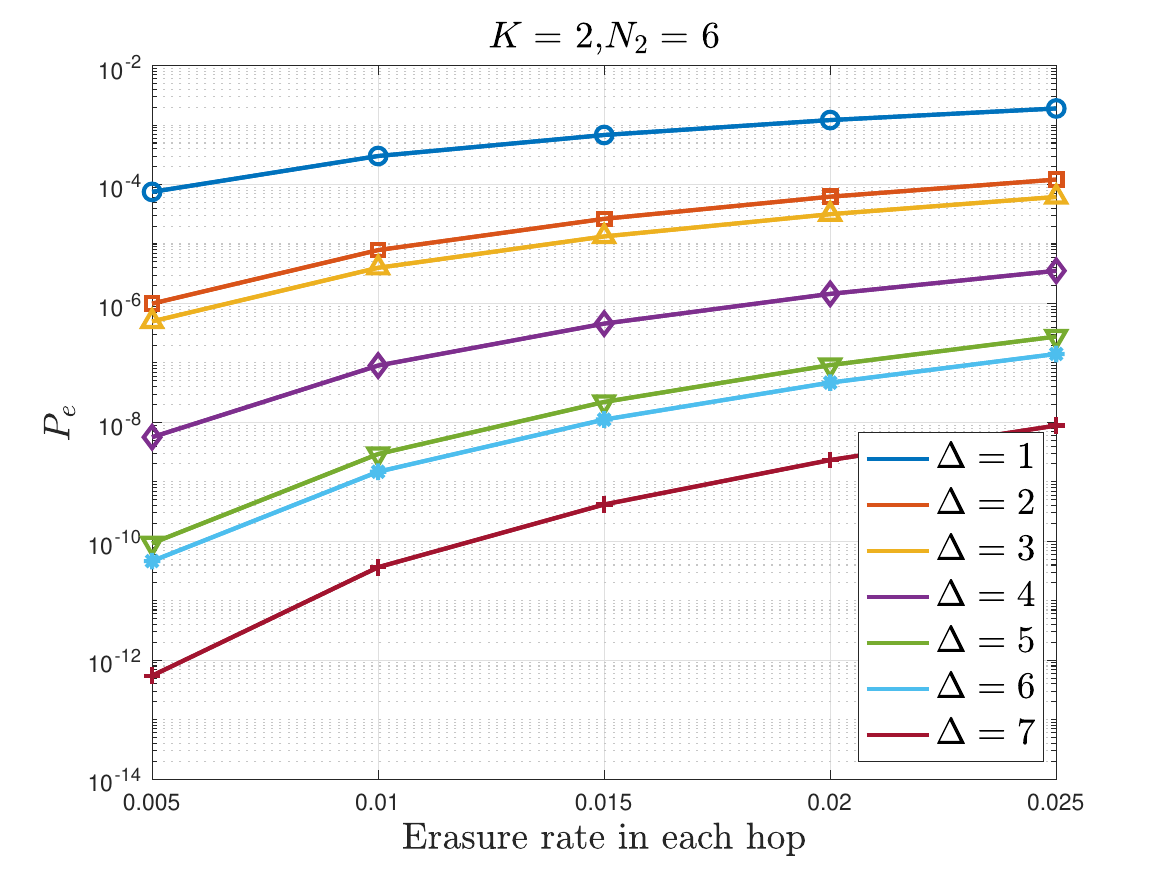}
        \caption{$p_e$ versus erasure probability $\epsilon$ in each hop when coding rate $\frac{K}{N_2} = 1/3$.}
        \label{figure:pe vs delta_K2N6}
    \end{minipage}
    \begin{minipage}[t]{0.49\linewidth}
        \centering
        \includegraphics[width=0.9\linewidth]{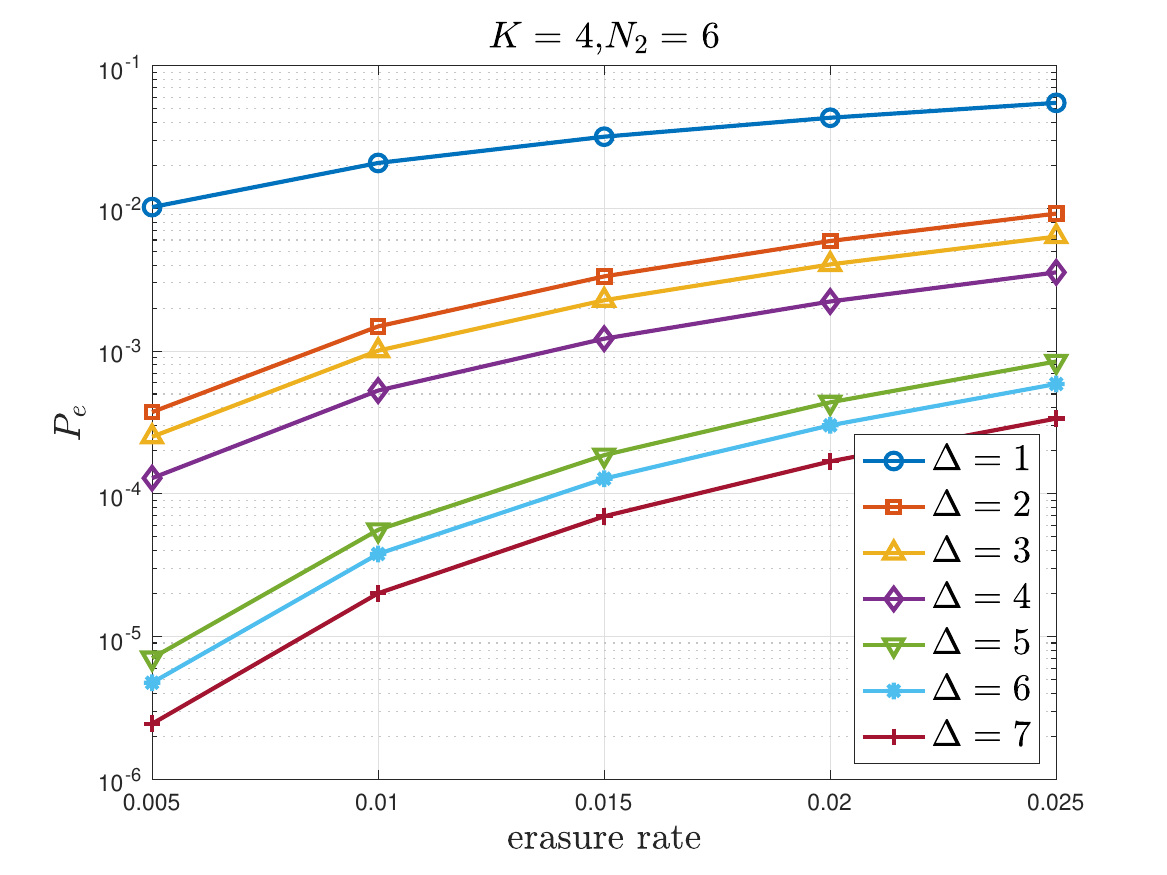}
        \caption{$p_e$ versus erasure probability $\epsilon$ in each hop when coding rate $\frac{K}{N_2} = 2/3$.}
        \label{figure:pe vs delta_K4N6}
    \end{minipage}
 \end{figure}

In this subsection, we conduct simulation on the performance of RLSCs with respect to decoding delay $\Delta$, erasure rate in each hop $1-q_l$, and coding rate $\frac{K}{N_2}$.   
The maximum values are set to $m_0=m_1=7,m_2=35$. 
Also assume symmetric delivery probability $q_0=q_1$ in this setting. 
And the erasure rate denoted by $\epsilon$ in each hop equals $1-q_0$. 
In Fig. \ref{figure:pe vs delta_K2N6} and Fig. \ref{figure:pe vs delta_K4N6}, the theoretical error probability $p_e$ is plotted against the erasure rate $\epsilon$ in each hop, with different values of $\Delta$ and coding rates $\frac{1}{3}$ and $\frac{2}{3}$, respectively. 
One can notice that as $\Delta$ increases by one, $p_e$ can decrease exponentially.
For different coding rates $\frac{1}{3}$ and $\frac{2}{3}$, the decreasing of $p_e$ also present different but regular patterns. 
Specifically, there appears ``periodic patterns" in Fig. \ref{figure:pe vs delta_K2N6} and Fig. \ref{figure:pe vs delta_K4N6} with a period of 3 with respect to the increase of $\Delta$.
These regular patterns are mainly due to the expression of the numerator in Lemma \ref{lemma:numerator}.

\subsection{Performance comparison to the streaming code proposed in \cite{SLF 3hop first paper}}

In this subsection, we compare the error probability of RLSCs with existing work \cite{SLF 3hop first paper} in two-hop relay networks and aim to explain the possible underlying causes.
Recall that in \cite{SLF 3hop first paper}, a novel symbol-wise Decode-and-Forward (DF) scheme is proposed. 
In their simulations, performance of the symbol-wise DF scheme, message-wise DF scheme, instantaneous forwarding (IF) strategy and also the derived upper bound on stochastic performance of symbol-wise DF scheme are compared.
The symbol-wise DF and message-wise DF schemes are developed by first constructing point-to-point streaming codes from block codes and then constructing DF schemes by concatenating two point-to-point streaming codes.
The IF strategy uses a point-to-point streaming code over the three-node relay network as if the network is a point-to-point channel, where the relay directly forwards every symbol received from the source in each timeslot. 
Thus, the overall point-to-point channel induced by the IF strategy experiences an erasure if either one of the channels experiences an erasure. 

\begin{figure}
    \centering
    \includegraphics[width=0.8\linewidth]{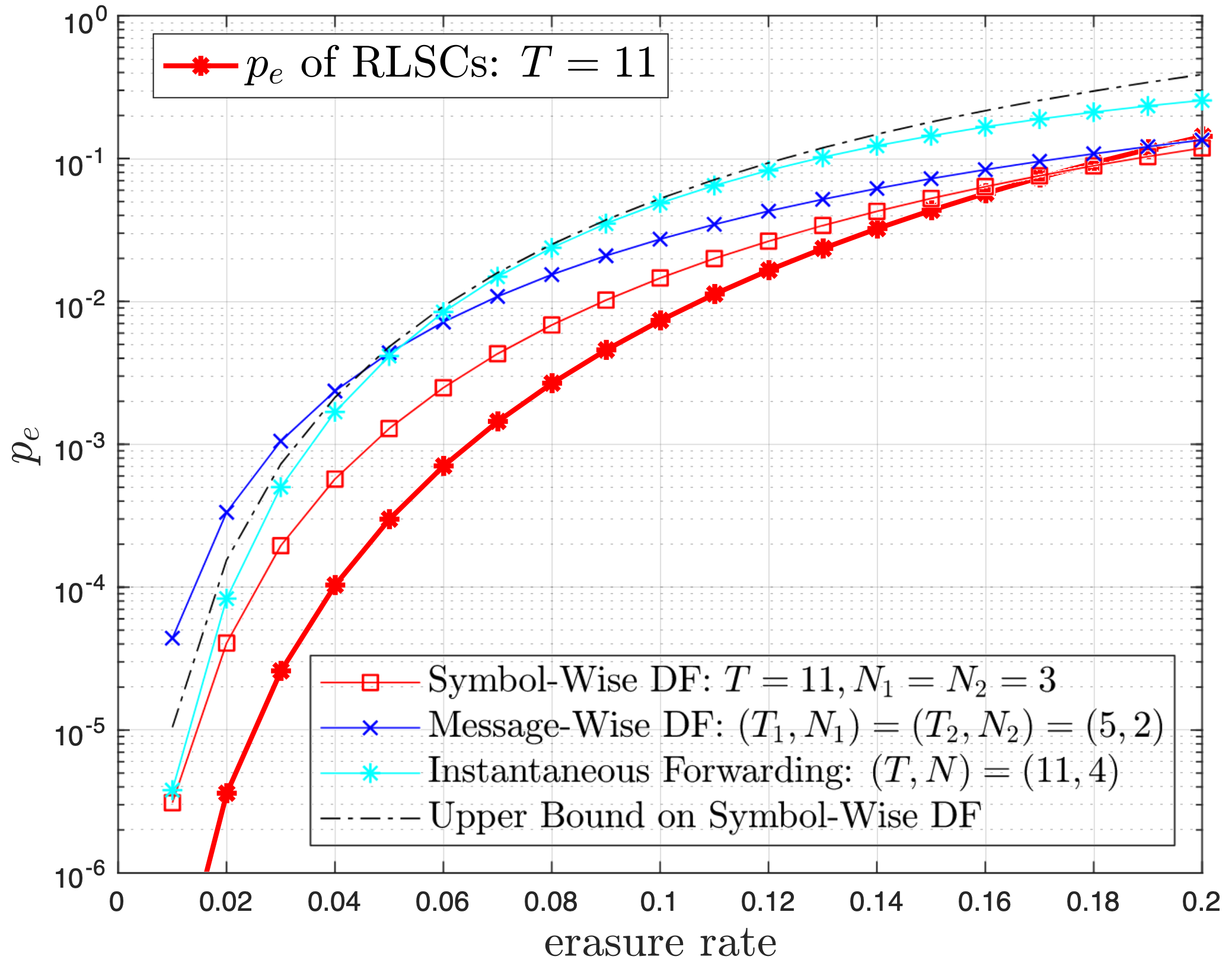}
    \caption{Comparison between large-field-size RLSCs and the results in \cite{SLF 3hop first paper}.}
    \label{fig:ComparisonToExistingCodes}
\end{figure}

In Fig. \ref{fig:ComparisonToExistingCodes}, we borrow the simulation results in \cite{SLF 3hop first paper} and compare them with the $p_e$ of RLSCs. 
The parameters\footnote{The notations in this figure are different from that in our paper. To be in line with \cite{SLF 3hop first paper}, in this figure, $\Delta$ is denoted as $T$. $N_1,N_2$ in this figure are the maximum number of arbitrary erasures that can occur in the first and the second hop. $T_1,T_2$ in this figure are the maximum decoding decoding that can be tolerated in the first and the second hop.} of RLSCs are set to $K=8,N_2=12$ and $\Delta=11$.
The erasure rate $\epsilon$ is also symmetric and plotted against error probability $p_e$. 
One can notice from Fig. \ref{fig:ComparisonToExistingCodes} that, in most regime of the erasure rate, i.e., $0.01\le\epsilon\le 0.18$, the large-field-size RLSCs presents a lower $p_e$ than the symbol-wise DF scheme. 
The reduction on $p_e$ could be attributed to the difference on the coding structures of streaming codes. 
In our model, the source symbols can be transmitted through the relay network in a ``pipeline fashion". In other words, the information of each symbol will be continuously transmitted from source to destination (if not being erased), without waiting for specific symbol to be decoded at any intermediate relay nodes. 
Therefore, the end-to-end decoding latency of large-field-size RLSCs depends on the decodability of source symbols upon receiving sufficient information at the destination.
In this case, each node can greedily forward all the information it has ever received in the form of linear equations, regardless of 
the decodability of source symbols at intermediate nodes. 
In contrast, for the Decode-and-Forward strategy, a symbol should be explicitly decoded at a node for further delivery to the next node, such that the end-to-end decoding latency equals to the summation of the decoding delay at each hop. 
In this case, the decoding delays between adjacent nodes can be guaranteed by the proposed coding structure\cite{SLF 3hop first paper} if the erasure patterns only occur in a predefined set. 
However, for the erasure patterns beyond the predefined set, the decoding failure of source symbols at some relay nodes could impede the successive transmission of information to the downstream nodes. 
It was also reported in \cite{DOO multihop DAF} that in multi-hop networks, the linearly growing delay is an artifact of the existing Decode-and-Forward strategy.
Last but not least, we have to mention that the comparison is not completely fair
, since the finite field chosen in \cite{SLF 3hop first paper} could be relatively small compared to the large-finite-field assumption in this paper. 

\section{Conclusion}\label{section:conclusion}

In this paper, we explore the stochastic performance limit of large-field-size RLSCs in multi-hop relay networks. 
Different from the previous work focusing on the adversarial channel, we concentrate on stochastic channel that injects i.i.d. packet erasures.
When the probabilistic erasure is considered, the analysis should be averaged over all possible erasure patterns along multiple hops of linear network. 
As the number of hops increases, the stochastic analysis becomes more analytically challenging. 
We first consider the simplest two-hop relay network.
The error event of large-field-size RLSCs is characterized by proposing a novel framework to quantify the number of source symbols detained at each node. 
Then the error probability is derived by constructing transition matrices with nested structure and carefully analyzing the expectation terms.
The results in two-hop network are finally extended to linear network with arbitrary number of hops by further exploiting the nested structure.  
Future work can be stochastic analysis of RLSCs on network with different topologies.

\begin{appendices}
\section{Proof of Proposition \ref{Proposition:characterization of the error event}}\label{Appendix:characterization of the error event}

\begin{figure}
    \centering
    \includegraphics[width=0.9\linewidth]{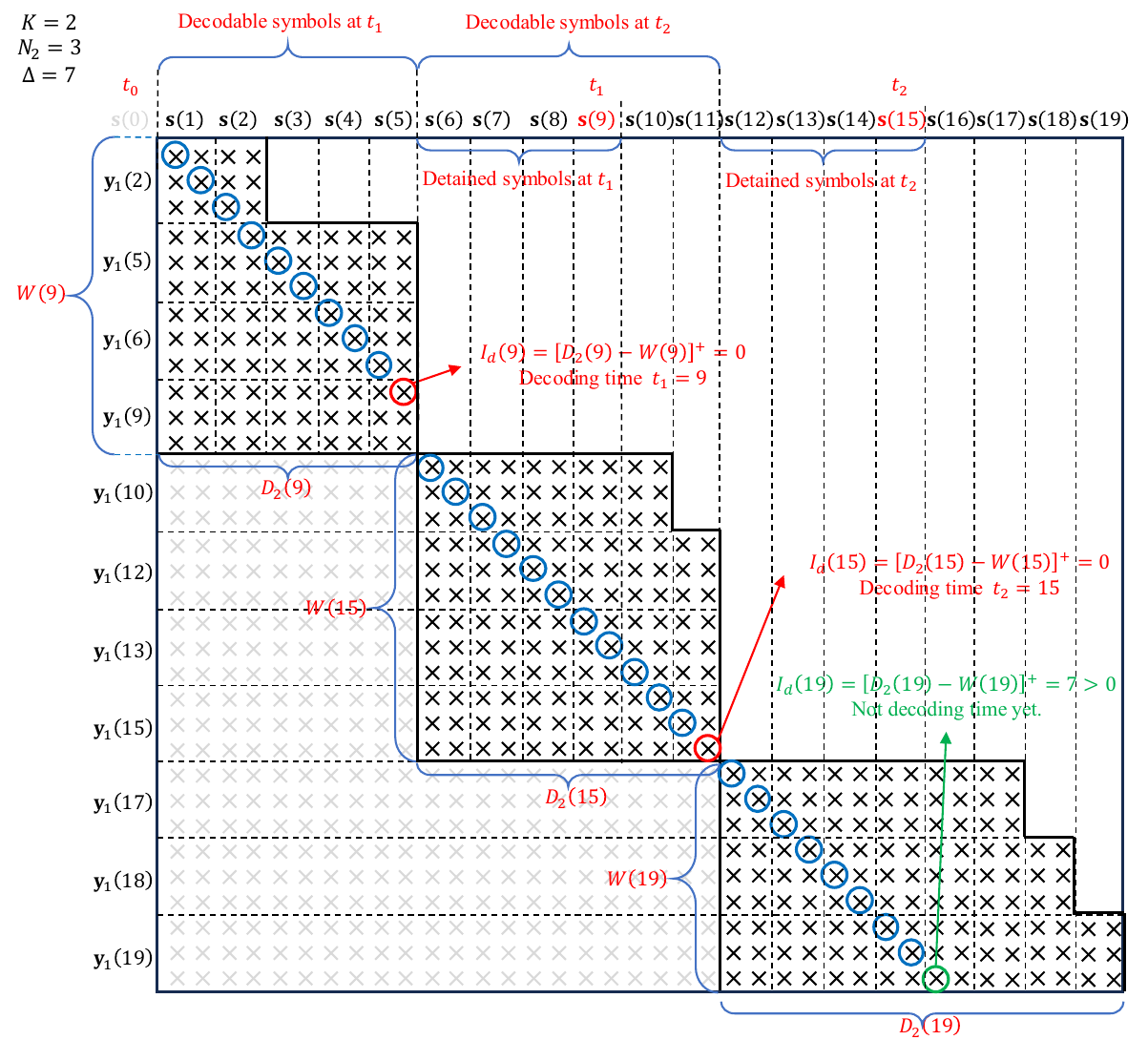}
    \caption{An example of $\mathcal{H}(19)$ with parameters $K=2,N_2=3,\delta = 7$. The cross mark ``×" shows the non-zero entries. The channel erasure pattern in the first hop is that $e_0(2) = e_0(3) = e_0(5) = e_0(10) = e_0(11) = e_0(16) = e_0(17) = e_0(18) = e_0(19) = 0$, and the channel erasure pattern in the second hop is that $e_1(2)=e_1(5)=e_1(6)=e_1(9)=e_1(10)=e_1(12)=e_1(13)=e_1(15)=e_1(17)=e_1(18)=e_1(19)=0$. At timeslot $t_1=9$, $r_2$ can decode source symbols $\mathbf{s}(1:5)$, while $\mathbf{s}(6:9)$ are still detained. At timeslot $t_2=15$, $r_2$ can decode source symbols $\mathbf{s}(6:11)$, while $\mathbf{s}(12:15)$ are still detained. Consider the decoding delay $\Delta = 7$, then the $\Delta$-decodable symbols in $[1,5]$ are $[1:5] \bigcap [9-7,9]=[2,5]$ and the $\Delta$-decodable symbols in $[6,11]$ are $[6:11] \bigcap [15-7,15]=[8,11]$.}
    \label{fig:Induction_1}
\end{figure}

First we demonstrate that Proposition \ref{Proposition:characterization of the error event} holds for indices $i_0 = 0,1$.
Without loss of generality, assume a large-enough-timeslot $t^*$, which is large enough to include the first and the second decoding times $t_1$ and $t_2$.
Then we analyze the decodability of source symbols $\mathbf{s}(1:t^*)$ for any given erasure patterns in the first and the second hops.
For any given erasure patterns in each hop $e_0(1:t^*),e_1(1:t^*)$, one can obtain the overall cumulative receiver matrix $\mathcal{H}(t^*)$ by equations (\ref{equation:overall receiver matrix}), the process of which was once illustrated in Fig. \ref{fig:Example_1_7} and Fig. \ref{fig:Example_1_8}.
Then we analyze $\mathcal{H}(t^*)$ for the decodability of source symbols $\mathbf{s}(1:t^*)$. An example of $\mathbf{s}(1:t^*)$ for parameters $K=2,N_2=3,\Delta=7,t^*=19$ is presented in Fig. \ref{fig:Induction_1}. The cross mark ``×" shows the non-zero entries, while the cross marks highlighted by blue circles at a 45-degree angle downward represents the equality of unknowns and equations. 
In other words, if the 45-degree-angle line of blue circles reaches the right boundary of the non-zero elements (which is marked in red circles in Fig. \ref{fig:Induction_1}), the number of equations is large than (or at least equal to) the number of unknowns in this round, thus the source symbols contained in these equations can be decoded. 
For example, in Fig. \ref{fig:Induction_1}, source symbols $\mathbf{s}(1:5)$ can be decoded by the received encoded packets $\mathbf{y}_1(2),\mathbf{y}_1(5),\mathbf{y}_1(6),\mathbf{y}_1(9)$ at timeslot 9.
Thus we have the first decoding time $t_1 = 9$.
However one can notice that at timeslot 9, not all existing source symbols $\mathbf{s}(1:9)$ can be decoded due to the channel erasures. Specifically, the received encoded packets $\mathbf{y}_1(2),\mathbf{y}_1(5),\mathbf{y}_1(6),\mathbf{y}_1(9)$ contain only the information of $\mathbf{s}(1:5)$, while $\mathbf{s}(6:9)$ are still detained at the source $r_0$ due to erasures in the first hop (note that $e_0(6)=e_0(7)=e_0(8)=e_0(9)=1$ in this example). 
And the decodability of $\mathbf{s}(6:9)$ should wait for $\mathbf{y}_1(10:t^*)$ to be determined. 
Continue the decodability analysis of this example for the second round. Note that in the first round, $\mathbf{s}(1:5)$ have been decoded at $r_2$, thus the impact of $\mathbf{s}(1:5)$ can be eliminated from the equations $\mathbf{y}_1(10:t^*)$ (which are represented by the gray cross marks in Fig. \ref{fig:Induction_1}, in contrast to the black cross marks representing the actually effective elements).
Therefore, the 45-degree-angle line of blue circles will start from the first unused equation, i.e., $\mathbf{y}(10)$, at the position of the earliest undecoded source symbols, i.e., $\mathbf{s}(6)$ in this example.
In the second round, similarly, the 45-degree-angle line of blue circles indicates that source symbols $\mathbf{s}(6:11)$ can be decoded from equations $\mathbf{y}_1(10),\mathbf{y}_1(11),\mathbf{y}_1(13),\mathbf{y}_1(15)$ at timeslot $t_2=15$, while $\mathbf{s}(12:15)$ are still detained at the source due to the first-hop-erasures $e_0(12)=e_0(13)=e_0(14)=e_0(15)=1$. 
And the decodability of $\mathbf{s}(12:15)$ should wait for $\mathbf{y}_1(16:t^*)$ to be determined. 
In the third round, at timeslot $t^*=19$, $\mathbf{y}_1(17),\mathbf{y}_1(18),\mathbf{y}_1(19)$ are not containing enough equations to decode unknowns $\mathbf{s}(12:19)$.
This is illustrated by the green circle at the end of the blue circles line that reaches the lower boundary of non-zero elements in Fig. \ref{fig:Induction_1}.
Thus $t^*=19$ is not a decoding time yet.

Then we abstract the analysis from this example to the general case. 
The number of equations in each round is directly determined by the erasure pattern at the second hop.
Let $W(t)$ denote the number of equations at timeslot $t$. 
Every time the transmission at the second hop succeeds, $N_2$ equations will be added to $W(t)$, regardless of how many source symbols are contained in the $N_2$ equations.
When it comes to the decoding time (which is equivalent to $I_d(t) = 0$ and will be explained later), $W(t)$ will be reset to zero since that after decoding process, the impact of these $W(t)$ decoded source symbols will can be eliminated from the subsequent equations. See $W(9)$ and $W(15)$ in Fig. \ref{fig:Induction_1} for example.
Thus, $W(t)$ can be derived as (\ref{equation:W}) in Definition \ref{definition:informtion debt}.
On the other hand, denoted by $D_2(t)$, the number of source symbols contained in the $W(t)$ equations not only depends on the erasures in two hops, but also depends on the number of source symbols at $r_0,r_1$.
Specifically, $D_2(t)$ equals to the number of undecoded source symbols that are contained in the latest received encoded packet from $r_1$.
Note that at each timeslot $t$, the relay $r_1$ will encode all packets $\mathbf{y}_0(1:t)$ that it ever received into the transmission packet $\mathbf{x}_1(t)$.
When transmission in the second hop succeed, $D_2(t)$ will be equal to the number of source symbols that is contained in $r_1$'s memory, otherwise, $D_2(t)$ stays the same. 
Therefore, we introduce the concept ``number of detained symbols" to describe the difference of number of source symbols that are contained in the memories of adjacent nodes.
Let $D_1(t)$ be the difference of number of source symbols that are contained in the memories of $r_1$ and $r_2$ and note that $D_2(t)$ will be also reset to zero after each decoding time similar to $W(t)$. 
Therefore, $D_2(t)$ can be derived as (\ref{equation:D2}) in Definition \ref{definition:informtion debt}.
Similarly, let $D_0(t)$ be the difference of number of source symbols that are contained in the memories of $r_0$ and $r_1$.
Note that the change of $D_1(t)$ not only depends on $D_0(t)$, but also depends on the erasures $e_0(t)$ and $e_1(t)$ in two hops.
Specifically, when the second hop succeeds, all source symbols contained in $r_1$ will be delivered to $r_2$ thus $D_1(t)$ will be reset to zero.
If the second hop fails, the change of $D_1(t)$ will depend on the first hop. If the first hop also fails, $D_1(t)$ will stay the same. Otherwise if first hop succeeds, the number of source symbols contained in $r_1$'s memory will be equal to the number of source symbols contained in $r_0$'s memory.
Thus $D_1(t)$ will be increased by $D_0(t)$ and then $D_0(t)$ will be reset to zero. 
Recall that there are $K$ symbols arrive at the source at each timeslot.
Therefore, $D_0(t)$ and $D_1(t)$ can be derived as (\ref{equation:D0}) and (\ref{equation:D1}) in Definition \ref{definition:informtion debt}, respectively.

With the above definitions, the decoding event can be easily characterized. Every time when $W(t) \ge D_2(t)$, due to the GMDS condition assumed in our model, $r_2$ have received enough information to decode $D_2(t)$ source symbols. See the two red circles in Fig. \ref{fig:Induction_1} for example. On the other hand, when $W(t) < D_2(t)$, also due to the GMDS condition, $D_2(t)$ source symbols are completely mixed up with each other and no source symbols can be decoded. See the green circle in Fig. \ref{fig:Induction_1} for example.
Therefore, the information debt $I_d(t) = \left[D_2(t) - W(t)\right]^+$ can be used as a indicator of decoding timeslot, and thus we have the equation (\ref{equation:Id}) in Definition \ref{definition:informtion debt} and the zero-hitting times $t_i,i\in[1,\infty]$ in Definition \ref{definition:hitting time old}.

Now we are ready to prove that Proposition \ref{Proposition:characterization of the error event} holds for indices $i_0 = 0,1$. 
At the beginning timeslot $t_0 = 0$, no source symbol exists in each node. Thus we have $D_0(0)=D_1(0)=D_2(0)=0$.
At timeslot $t_1$, the first $D_2(t_1)$ symbols $\mathbf{s}(1:D_2(t_1))$ can be decoded from the first $W(t_1)$ equations. 
Further consider the decoding latency, only symbols within the regime $[t_1-\Delta,t_1]$ are $\Delta$-decodable. 
Therefore, the $\Delta$-decodable regime in the first round is $t \in \left[1,\frac{D_2(t_1)}{K}\right] \bigcap \Big[t_1-\Delta,t_1\Big]$, satisfying equation (\ref{equation:error event}).
At the beginning of the second round $t=t_1+1$, source symbols $\mathbf{s}(t_1 - \frac{D_0(t_1)+D_1(t_1)}{K}:t_1)$ are still detained outside the destination $r_2$. At the end of the second round $t=t_2$, $D_2(t_2)$ source symbols $\mathbf{s}(t_1 - \frac{D_0(t_1)+D_1(t_1)}{K}+1:t_1 - \frac{D_0(t_1)+D_1(t_1)}{K}+\frac{D_2(t_1)}{K})$ can be decoded from $W(t_2)$ received packets. 
Further consider the decoding latency, the $\Delta$-decodable regime in the second round is $t \in \left[t_1 - \frac{D_0(t_1)+D_1(t_1)}{K}+1:t_1 - \frac{D_0(t_1)+D_1(t_1)}{K}+\frac{D_2(t_1)}{K}\right] \bigcap \Big[t_2-\Delta,t_2\Big]$, satisfying equation (\ref{equation:error event}). Thus Proposition \ref{Proposition:characterization of the error event} holds for indices $i_0 = 0,1$.

For any index $i_0\ge 2$, Proposition \ref{Proposition:characterization of the error event} can be proved by induction. Simply put, at each decoding time $t_{i_0}$, the impact of decoded symbols can be eliminated from later equations. It is as if the timeline has been reset, with $t_{i_0}$ being shifted back to $t_1$.

\begin{remark}\label{remark:infinite memory}
    \textit{(Discussion on the assumption of asymptotically large memory)} In the model statement, we assume that the memory size of each node is asymptotically large. 
    This assumption is to maintain the information relationship between the former and the latter source symbols. 
    With this assumption, any source symbol can be recovered at the destination as long as enough encoded packets have been received.
    Thus we can omit the corner case that due to some specific erasure patterns (especially the bursty erasures with a long period of time), some former source symbols may lose connection with latter symbols in the encoded packets, such that they are unable to be recovered no matter how many packets are received at the destination.
    
    However, in practical application, the model can be slightly modified, so that the assumption can be removed. 
    In the above, we demonstrate that the impact of symbols that have been decoded can be eliminated from the unused equations. 
    A non-time-sensitive uplink (from the destination to the source node) feedback, which is similar to the ACK messages carrying indices of decoded symbols can effectively reduce the requirement of the memory size in our model. 
    In the example of Fig. \ref{fig:Induction_1}, suppose a feedback carrying information that $\mathbf{s}(1:5)$ have been decoded at destination at timeslot 9 is transmitted to the source, and is received by the source at timeslot 15. Then the source can evict $\mathbf{s}(1:5)$ from its memory and the encoded packets $\mathbf{x}_0(t),t\ge 16$ can be encoded only from $\mathbf{s}(6:t)$. Similar operation is also applicable to the intermediate relay nodes.
\end{remark}

\section{Proof of Proposition \ref{proposition:initial distribution}}\label{Appendix:initial distribution}

We first derive $T_{0\rightarrow 0}$, the joint transition matrix of initial distribution of the hidden states $\widehat{D}_0(t)$ and $\widehat{D}_1(t)$ between any two adjacent decoding times $t_{i_0}$ and $t_{i_0+1}$.
Recall the definition of its entry 
\begin{equation}
    t(i,j;v,w) \triangleq \text{Pr}\big(\widehat{D}_0(t_{i_0+1})=v,\widehat{D}_1(t_{i_0+1})=w \big| \widehat{D}_0(t_{i_0})=i,\widehat{D}_1(t_{i_0})=j\big).
\end{equation}
Further denote $T_{0\rightarrow 0}^{(k)}$ the joint transition matrix of initial distribution of the hidden states $\widehat{D}_0(t)$ and $\widehat{D}_1(t)$ between any two adjacent decoding times $t_{i_0}$ and $t_{i_0+1}$, while the event $t_{i_0+1} - t_{i_0} = k$ occurs. Formally, its entries are defined as 
\begin{equation}
    t^{(k)}(i,j;v,w) \triangleq \text{Pr}\big(\widehat{D}_0(t_{i_0+1})=v,\widehat{D}_1(t_{i_0+1})=w, t_{i_0+1} - t_{i_0} = k \big| \widehat{D}_0(t_{i_0})=i,\widehat{D}_1(t_{i_0})=j\big).
\end{equation}
Also recall that we denoted the probability distribution of hidden states at timeslot $t$ as 
\begin{align}
        \pi^t =  \Big[\overbrace{\pi_{0,0}^t,\cdots,\pi_{0,m_1-1}^t}^{m_1},\cdots,\overbrace{\pi_{m_0,0}^t,\cdots,\pi_{m_0,m_1-1}^t}^{m_1}\Big],
\end{align}
where 
\begin{equation}
    \pi_{i,j}^t = \text{Pr}\big(\widehat{D}_0(t) = i, \widehat{D}_1(t) = j\big).
\end{equation}
For any index $l\ge 1$, $\pi^{t_l}$ represents the probability distribution of the hidden states at timeslot $t_l$, where $t_l$ is the $l$-th time $I_d(t)$ hits zero.
Formally, 
\begin{equation}
    \pi_{i,j}^{t_l} \triangleq \text{Pr}\big(\widehat{D}_0(t_l) = i, \widehat{D}_1(t_l) = j\big).
\end{equation}
Further denote $\pi^{t_l,k} = \left[\pi_{i,j}^{t_l,k}\right]$ as the joint probability distribution of the hidden states at timeslot $t_l$ while the event $t_l - t_{l-1} = k$ occurs.
Formally, 
\begin{equation}
    \pi_{i,j}^{t_l,k} \triangleq \text{Pr}\big(\widehat{D}_0(t_l) = i, \widehat{D}_1(t_l) = j,t_l - t_{l-1} = k\big).
\end{equation}
By the above definitions, we have 
\begin{align}
    \pi^{t_{l+1}} &= \pi^{t_l} \cdot T_{0\rightarrow 0} \label{equation:55}\\
    \pi^{t_{l+1},1} &= \pi^{t_l} \cdot T_{0\rightarrow 0}^{(1)} \\
    \pi^{t_{l+1},k} &= \pi^{t_l} \cdot T_{0\rightarrow 0}^{(k)}, k\ge 2.
\end{align}
By the law of total probability, $\pi^{t_l}$ can be derived as 
\begin{equation}
    \pi^{t_l} = \sum_{k=1}^\infty \pi^{t_l,k}.
\end{equation}

Then we focus on the transition of hidden states from timeslot $t_{l}$ to $t_{l+1} -1$, and derive $\pi^{t_{l+1},k}$ from $\pi^{t_l}$.
Recall the definition of joint transition matrices $\mathbf{T}_{0,0},\mathbf{T}_{0,\phi},\mathbf{T}_{\phi,\phi},\mathbf{T}_{\phi,0}$ defined in Definition \ref{definition:joint transition matrix}.
One can notice that 
\begin{equation}
    T_{0\rightarrow 0}^{(1)} = \mathbf{T}_{0,0}
\end{equation}
and 
\begin{equation}
    T_{0\rightarrow 0}^{(k)} = \mathbf{T}_{0,\phi}(\mathbf{T}_{\phi,\phi})^{k-2}\mathbf{T}_{\phi,0}.
\end{equation}
Therefore, for $k=1$, we have
\begin{equation}
\pi^{t_{l+1},1} = \pi^{t_l} \cdot\mathbf{T}_{0,0}.
\end{equation}
For $k\ge 2$, we have 
\begin{equation}
    \pi^{t_{l+1},k} = \pi^{t_l} \cdot \mathbf{T}_{0,\phi}(\mathbf{T}_{\phi,\phi})^{k-2}\mathbf{T}_{\phi,0}.
\end{equation}
Sum over all possible lengths $k\in[1,\infty]$ of a round, we obtain 
\begin{align}
        \pi^{t_{l+1}} &= \sum_{k=1}^\infty \pi^{t_{l+1},k} \\
        &= \pi^{t_l} \cdot\left[\mathbf{T}_{0,0} + \mathbf{T}_{0,\phi}\sum_{k=2}^\infty(\mathbf{T}_{\phi,\phi})^{k-2}\mathbf{T}_{\phi,0}\right]\label{equation:6?}\\
    &= \pi^{t_l}\left[\mathbf{T}_{0,0} + \mathbf{T}_{0,\phi}\times A \times \mathbf{T}_{\phi,0}\right],\label{equation:pil+1topil}
\end{align}
where $A = \left[\mathbf{I}_{m_0 \cdot m_1 \cdot (m_2-1)}-\mathbf{T}_{\phi,\phi}\right]^{-1}$.
Note that $\mathbf{T}_{\phi,\phi}$ is a matrix with each of its row summing no larger than 1.
Therefore, the spectral radius of 
$\mathbf{T}_{\phi,\phi}$ satisfies $\rho(\mathbf{T}_{\phi,\phi})<1$, and thus we have $\lim_{k\rightarrow\infty}\left(\mathbf{T}_{\phi,\phi}\right)^k = \mathbf{0}$.
Therefore, the summation in (\ref{equation:6?}) converges and thus (\ref{equation:pil+1topil}) holds. 
From (\ref{equation:55}) and (\ref{equation:pil+1topil}) one can notice that 
\begin{align}
    T_{0\rightarrow 0} = \mathbf{T}_{0,0} + \mathbf{T}_{0,\phi}\times A \times \mathbf{T}_{\phi,0}.
\end{align}

When the Markov chain becomes stationary, or equivalently, when $l\rightarrow \infty$, we have $\pi^\infty = \pi^\infty \cdot T_{0\rightarrow 0}$. 
Notice that $\pi^\infty \cdot \vec{\mathbf{1}} = 1$.
Thus, one can derive $\pi^\infty$ by solving linear equations 
\begin{equation}\label{equation:solution of initial_rewrite}
    \begin{bmatrix}
    (T_{0\rightarrow 0} - \mathbf{I}_{m_0 m_1})^\top \\
     \begin{array}{ccc}
        1 & \cdots & 1 
     \end{array}
    \end{bmatrix}_{(m_0 m_1+1)\times(m_0 m_1)} \cdot 
\begin{bmatrix}
    {\pi^\infty}^\top
\end{bmatrix}_{(m_0 m_1)\times1}
=\begin{bmatrix}
    0 \\
    \vdots \\
    0 \\
    1
\end{bmatrix}_{(m_0 m_1+1)\times1},
\end{equation}
Therefore, Proposition \ref{proposition:initial distribution} is proved.

\section{Proof of Lemma \ref{lemma:denominator}}\label{Appendix:denominator}

To derive $\mathbb{E}\left\{\frac{D_2(t_{i_0+1})}{K}\right\}$, we first introduce an important identity equation in  Proposition \ref{proposition:identity}.

\begin{prop}\label{proposition:identity}
For any index $i_0$, the number of detained source symbols $D_i(t_{i_0}),i\in[0,2]$ satisfies that 
\begin{equation}\label{equation:identity}
    \frac{D_2(t_{i_0+1})}{K} = t_{i_0+1}-t_{i_0} - \left[\frac{D_0(t_{i_0+1}) + D_1(t_{i_0+1})}{K} - \frac{D_0(t_{i_0}) + D_1(t_{i_0})}{K}\right].
\end{equation}
\end{prop}

\textit{Proof:} One can easily observe this identical equation from Fig. \ref{fig:Characterization_of_error_event}. The physical meaning of this identical equation is explained as follows.
Note that $t_{i_0+1}-t_{i_0}$ is the total timeslots in this round.
$D_2(t_{i_0+1})$ is the number of source symbols that are decodable at this round. 
$D_0(t_{i_0}) + D_1(t_{i_0})$ is the total number of source symbols that are detained at source and relay at the end of the previous round, which still remains potential decodability in this round.
$D_0(t_{i_0+1}) + D_1(t_{i_0+1})$ is the total number of source symbols that are detained at source and relay at the end of this round, which are un-decodable since the destination has never received any packets containing any information of them.
And these symbols should wait for the next round to determine their decodability. 
Therefore, the number of all decodable source symbols in this round equals 
\begin{equation}\label{equation:identity_1}
    \big[D_0(t_{i_0}) + D_1(t_{i_0})\big] + K\cdot (t_{i_0+1}-t_{i_0}) - \big[D_0(t_{i_0+1}) + D_1(t_{i_0+1})\big],
\end{equation}
which is also equal to $D_2(t_{i_0+1})$.
Normalized (\ref{equation:identity_1}) by $K$, then Proposition \ref{proposition:identity} is proved.
\QEDA

Proposition \ref{proposition:identity} transition the problem of deriving $\mathbb{E}\Big\{\frac{D_2(t_{i_0+1})}{K}\Big\}$ to problems of deriving two terms $\mathbb{E}\big\{t_{i_0+1}-t_{i_0}\big\}$ and $\mathbb{E}\left\{\frac{D_0(t_{i_0+1}) + D_1(t_{i_0+1})}{K} - \frac{D_0(t_{i_0}) + D_1(t_{i_0})}{K}\right\}$. 
Next, we will show that $\mathbb{E}\left\{\frac{D_0(t_{i_0+1}) + D_1(t_{i_0+1})}{K} - \frac{D_0(t_{i_0}) + D_1(t_{i_0})}{K}\right\} = 0$.
Intuitively, $\mathbb{E}\left\{\frac{D_0(t_{i_0+1}) + D_1(t_{i_0+1})}{K}\right\}$ and $\mathbb{E}\left\{\frac{D_0(t_{i_0}) + D_1(t_{i_0})}{K}\right\}$ are directly determined by the stationary initial probability distribution of the hidden states $\pi^\infty$.
When the initial distribution has become stationary, it will no more change with respect to the index $i_0$. 
Therefore, we have $\mathbb{E}\left\{\frac{D_0(t_{i_0+1}) + D_1(t_{i_0+1})}{K}\right\} = \mathbb{E}\left\{\frac{D_0(t_{i_0}) + D_1(t_{i_0})}{K}\right\}$.
For a rigorous proof, reader can refer to Proposition \ref{proposition:identity=zero} below.

\begin{prop}\label{proposition:identity=zero}
    For any index $i_0$, the number of detained source symbols $D_i(t_{i_0}),i\in[0,1]$ satisfies that
    \begin{align}\label{equation:identity=zero}
        \mathbb{E}\left\{\frac{D_0(t_{i_0+1}) + D_1(t_{i_0+1})}{K} - \frac{D_0(t_{i_0}) + D_1(t_{i_0})}{K}\right\} = 0.
    \end{align}
\end{prop}

\textit{Proof:} 
Let $\frac{D_0(t_{i_0}) + D_1(t_{i_0})}{K} = \alpha$ and $\frac{D_0(t_{i_0+1}) + D_1(t_{i_0+1})}{K} = \beta$ for briefness, satisfying $\alpha,\beta \ge 0$. 
Recall that the joint transition matrix of initial distribution of hidden states $\widehat{D}_0(t)$ and $\widehat{D}_1(t)$ between any two adjacent times that $I_d(t)$ hits zero, i.e., $T_{0\rightarrow 0}$ is derived in Proposition \ref{proposition:initial distribution}.
Based on $T_{0\rightarrow 0}$, one can further derive a transition matrix $\widetilde{T}_{0\rightarrow 0}$ on initial distribution of \textit{the summation of two hidden states $\widehat{D}_0(t)$ and $\widehat{D}_1(t)$} between any two adjacent times that $I_d(t)$ hits zero. 
Specifically, denote the entries of $\widetilde{T}_{0\rightarrow 0}$ as $\left[\widetilde{t}(j,l)\right],j,l\in[0,m_0+m_1-2]$, where $\widetilde{t}(j,l) \triangleq \text{Pr}\big(\widehat{D}_0(t_{i_0+1}) + \widehat{D}_1(t_{i_0+1})=l \big| \widehat{D}_0(t_{i_0}) + \widehat{D}_1(t_{i_0})=j\big)$.
Since the summations on rows and columns are linear operations, $\widetilde{T}_{0\rightarrow 0}$ can be generated through left-multiplying $T_{0\rightarrow 0}$ by a $(m_0+m_1-1)\times (m_0 m_1)$-matrix $\mathbf{P}$ and right-multiplying $T_{0\rightarrow 0}$ by a $(m_0 m_1)\times (m_0+m_1-1)$-matrix $\mathbf{Q}$, i.e., $\widetilde{T}_{0\rightarrow 0} = \mathbf{P} T_{0\rightarrow 0} \mathbf{Q}$.
For simplicity, the explicit form of $\mathbf{P},\mathbf{Q}$ are omitted. 
Recall the definition $\widetilde{t}(j,l) = \text{Pr}(\beta = l|\alpha = j)$, we have 
\begin{align}
        \mathbb{E}\left\{\beta - \alpha\right\} &= \sum_{j=0}^\infty \text{Pr}(\alpha = j) \cdot \sum_{l = -j}^{+\infty} \text{Pr}(\beta -\alpha = l|\alpha = j)\\
        &= \sum_{j=0}^\infty \text{Pr}(\alpha = j) \cdot \sum_{l = -j}^{+\infty} \text{Pr}(\beta = l + j|\alpha = j)\\
        &= \sum_{j=0}^\infty \text{Pr}(\alpha = j) \cdot \sum_{l = -j}^{+\infty} l\cdot \widetilde{t}(j,l+j).\label{equation:50}
\end{align}

Denote the second term of (\ref{equation:50}) as $a_j \triangleq \sum_{l = -j}^{+\infty} l\cdot \widetilde{t}(j,l+j)$ and $\vec{J} \triangleq \begin{bmatrix}
    a_0\\
    a_1\\
    a_2\\
    \vdots
\end{bmatrix}$. We have 
\begin{align}
    a_j &= \Big[\widetilde{t}(j,0),\widetilde{t}(j,1),\widetilde{t}(j,2),\cdots\Big] \cdot \begin{bmatrix}
        -j\\
        -j+1\\
        -j+2\\
        \vdots
    \end{bmatrix} \\
    &= \Big[\widetilde{t}(j,0),\widetilde{t}(j,1),\widetilde{t}(j,2),\cdots\Big] \cdot \left[\begin{bmatrix}
        0\\
        1\\
        2\\
        \vdots
    \end{bmatrix} - j\cdot \begin{bmatrix}
        1\\
        1\\
        1\\
        \vdots
    \end{bmatrix}\right] \\
    &\overset{(a)}{=} \Big[\widetilde{t}(j,0),\widetilde{t}(j,1),\widetilde{t}(j,2),\cdots\Big] \cdot \begin{bmatrix}
        0\\
        1\\
        2\\
        \vdots
    \end{bmatrix} - j,
\end{align}
where (a) is due to the fact that $\widetilde{T}_{0\rightarrow 0}$ is still a stochastic matrix and each of its row sums up to 1.
Then $\vec{J}$ can be written as 
\begin{align}
    \vec{J} &= \begin{bmatrix}
        \Big[\widetilde{t}(0,0),\widetilde{t}(0,1),\widetilde{t}(0,2),\cdots\Big] \cdot \begin{bmatrix}
        0\\
        1\\
        2\\
        \vdots
    \end{bmatrix} - 0 \\
    \Big[\widetilde{t}(1,0),\widetilde{t}(1,1),\widetilde{t}(1,2),\cdots\Big] \cdot \begin{bmatrix}
        0\\
        1\\
        2\\
        \vdots
    \end{bmatrix} - 1 \\
    \Big[\widetilde{t}(2,0),\widetilde{t}(2,1),\widetilde{t}(2,2),\cdots\Big] \cdot \begin{bmatrix}
        0\\
        1\\
        2\\
        \vdots
    \end{bmatrix} - 2 \\
    \vdots
    \end{bmatrix} \\
    &= \begin{bmatrix}
        \widetilde{t}(0,0)&\widetilde{t}(0,1)&\widetilde{t}(0,2)&\cdots\\
        \widetilde{t}(1,0)&\widetilde{t}(1,1)&\widetilde{t}(1,2)&\cdots\\
        \widetilde{t}(2,0)&\widetilde{t}(2,1)&\widetilde{t}(2,2)&\cdots\\
        \vdots &\vdots &\vdots &\ddots 
    \end{bmatrix}\cdot \begin{bmatrix}
        0\\
        1\\
        2\\
        \vdots
    \end{bmatrix} - \begin{bmatrix}
        0\\
        1\\
        2\\
        \vdots
    \end{bmatrix} \\
    &= \big(\widetilde{T}_{0\rightarrow 0} - \mathbf{I}_{m_0+m_1-1}\big) \cdot \begin{bmatrix}
        0\\
        1\\
        2\\
        \vdots
    \end{bmatrix}.
\end{align}

Thus, continue with (\ref{equation:50}), we have 
\begin{align}
        \mathbb{E}\left\{\beta - \alpha\right\} &= \sum_{j=0}^\infty \text{Pr}(\alpha = j) \cdot a_j \\
        &= [\text{Pr}(\alpha = 0),\text{Pr}(\alpha = 1),\text{Pr}(\alpha = 2),\cdots] \cdot \vec{J} \\
        &= [\text{Pr}(\alpha = 0),\text{Pr}(\alpha = 1),\text{Pr}(\alpha = 2),\cdots] \cdot \big(\widetilde{T}_{0\rightarrow 0} - \mathbf{I}_{m_0+m_1-1}\big) \cdot \begin{bmatrix}
        0\\
        1\\
        2\\
        \vdots
    \end{bmatrix}.\label{equation:59}
\end{align}

According to Proposition \ref{proposition:initial distribution}, we have 
\begin{align}
    \pi^\infty = \pi^\infty \cdot T_{0\rightarrow 0}.\label{equation:60}
\end{align}
Reform (\ref{equation:60}) by summing up the two hidden states, we have  
\begin{align}
    \pi^\infty \mathbf{Q}  = \pi^\infty \mathbf{Q} \cdot \mathbf{P} \cdot T_{0\rightarrow 0} \cdot \mathbf{Q},
\end{align}
which can be further written as 
\begin{align}\label{equation:62}
    [\text{Pr}(\alpha = 0),\text{Pr}(\alpha = 1),\text{Pr}(\alpha = 2),\cdots]  = [\text{Pr}(\alpha = 0),\text{Pr}(\alpha = 1),\text{Pr}(\alpha = 2),\cdots] \cdot \widetilde{T}_{0\rightarrow 0}.
\end{align}
Substitute (\ref{equation:62}) into (\ref{equation:59}), we have 
\begin{align}
    \mathbb{E}\left\{\beta - \alpha\right\} &= \vec{\mathbf{0}}^\top \cdot \begin{bmatrix}
        0\\
        1\\
        2\\
        \vdots
    \end{bmatrix} = 0.
\end{align}
Thus, Proposition \ref{proposition:identity=zero} is proved. \QEDA

With Proposition \ref{proposition:identity=zero}, we have $\mathbb{E}\Big\{\frac{D_2(t_{i_0+1})}{K}\Big\} = \mathbb{E}\big\{t_{i_0+1}-t_{i_0}\big\}$, which means the average number of source symbols decoded at each time equals to the expected time interval of a round. 
Recall that $A = \left[\mathbf{I}_{m_0 \cdot m_1 \cdot (m_2-1)}-\mathbf{T}_{\phi,\phi}\right]^{-1}$, we have 
\begin{align}
    \mathbb{E}\bigg\{\frac{D_2(t_{i_0+1})}{K}\bigg\}&= \mathbb{E}\big\{t_{i_0+1}-t_{i_0}\big\} \\
    &= \sum_{k = 1}^\infty k \cdot \text{Pr}(t_{i_0+1}-t_{i_0} = k) \\
    &= \pi^\infty \left[\mathbf{T}_{0,0} + \sum_{k = 2}^\infty k \cdot \mathbf{T}_{0,\phi}  \mathbf{T}_{\phi,\phi}^{k-2} \mathbf{T}_{\phi,0}\right] \vec{\mathbf{1}} \\
    &= \pi^\infty  \Big[\mathbf{T}_{0,0} + \mathbf{T}_{0,\phi}  A \big(\mathbf{I}_{m_0 m_1 (m_2-1)}+A\big) \mathbf{T}_{\phi,0}\Big]\vec{\mathbf{1}}.
\end{align}

Therefore, Lemma \ref{lemma:denominator} is proved.

\section{Proof of Lemma \ref{lemma:numerator}}\label{Appendix:numerator}

With Proposition \ref{proposition:identity}, the numerator of (\ref{equation:P_e}) can be rewritten as follows.
    \begin{align}
    &\mathbb{E}\left\{
        \min\bigg(
        \Big(t_{i_0+1} - t_{i_0} + \frac{D_0(t_{i_0}) + D_1(t_{i_0})}{K} - \Delta - 1\Big)^+ , \frac{D_2(t_{i_0+1})}{K}
        \bigg)
        \right\} \nonumber\\
        =& \mathbb{E}\left\{
        \min\bigg(
        \Big(t_{i_0+1} - t_{i_0} + \frac{D_0(t_{i_0}) + D_1(t_{i_0})}{K} - \Delta - 1\Big)^+ , t_{i_0+1}-t_{i_0} + \frac{D_0(t_{i_0}) + D_1(t_{i_0})}{K} - \frac{D_0(t_{i_0+1}) + D_1(t_{i_0+1})}{K} 
        \bigg)
        \right\}.
    \end{align}
For ease of presentation, denote 
\begin{align}
    t_{i_0+1}-t_{i_0} &= \omega, \\
    \frac{D_0(t_{i_0}) + D_1(t_{i_0})}{K} &= \alpha, \\
    \frac{D_0(t_{i_0+1}) + D_1(t_{i_0+1})}{K} &= \beta\\
    \min\bigg(\Big(\omega + \alpha - \Delta - 1\Big)^+ , \omega + \alpha - \beta\bigg) &= \lambda(\omega,\alpha,\beta),\label{equation:lamda definition}
\end{align}
where $\lambda$ is a function of random variables $\omega\ge 1,\alpha\ge 0,\beta\ge 0$. 
Since that $\frac{D_2(t_{i_0+1})}{K} = \omega + \alpha - \beta$ represents the number of source symbols that can be decoded at timeslot $t_{i_0}$, which must be a non-negative integer, we naturally have $\omega + \alpha - \beta\ge 0$.
Thus, (\ref{equation:lamda definition}) can be further written as 
\begin{equation}
    \lambda(\omega,\alpha,\beta) =  \left\{
    \begin{aligned}
        &0 \qquad\qquad\qquad\qquad\qquad\qquad\qquad \text{if} \  \omega+\alpha \le \Delta+1\\
        &\omega+\alpha-\max(\Delta-1,l) \qquad\qquad\quad   \text{if} \ \omega+\alpha>\Delta+1 
    \end{aligned}
    \right
    .
\end{equation}
Therefore, the numerator can be derived as follows. For simplicity, we use term $\text{Pr}(k,j,l)$ in place of $\text{Pr}(\omega = k,\alpha = j,\beta = l)$ as shorthand, such that each index corresponds to specific random variable.

\usetikzlibrary{calc}
    \begin{align}
    &\mathbb{E}\left\{
        \min\bigg(
        \Big(t_{i_0+1} - t_{i_0} + \frac{D_0(t_{i_0}) + D_1(t_{i_0})}{K} - \Delta - 1\Big)^+ , \frac{D_2(t_{i_0+1})}{K}
        \bigg)
        \right\} \nonumber\\
        =& \mathbb{E}\left\{
        \min\bigg(\Big(\omega + \alpha - \Delta - 1\Big)^+ , \omega + \alpha - \beta\bigg)
        \right\}\\
        =& \mathbb{E}\{\lambda(\omega,\alpha,\beta)\}\\
        =& \sum_{k=1}^\infty \sum_{j=0}^\infty  \text{Pr}(k,j,l) \cdot \sum_{l=0}^\infty \lambda(k,j,l) \\
        =&  \bigg(\sum_{k=1}^{\Delta+2} \sum_{j=\Delta+2-k}^\infty + \sum_{k=\Delta + 3}^{\infty} \sum_{j=0}^\infty\bigg) \sum_{l=0}^\infty \text{Pr}(k,j,l) \cdot \big[k + j - \max(\Delta+1 , l)\big]\\
        =&  \bigg(\sum_{k=1}^{\Delta+2} \sum_{j=\Delta+2-k}^\infty + \sum_{k=\Delta + 3}^{\infty} \sum_{j=0}^\infty\bigg) \Big[\text{Pr}(k,j,l=0), \text{Pr}(k,j,l=1), \cdots\Big] \begin{tikzpicture}[baseline = (M.west)]
    \tikzset{brace/.style = {decorate, decoration = {brace, amplitude = 5pt}, thick}}
    \matrix(M)
    [
        matrix of math nodes,
        left delimiter = (,
        right delimiter = )
    ]
    {
        k+j-\Delta -1 \\
        \vdots \\
        k+j-\Delta -1 \\
        k+j-\Delta -2 \\
        k+j-\Delta -3 \\
        \vdots \\
    };
    \draw[brace]
        ($(M-1-1.north east) + (0.45, 0)$)
        -- node[right = 5pt]{$\Delta +2$}
        ($(M-3-1.south east) + (0.45, 0)$);
\end{tikzpicture}
    \end{align}
Let the terms of the first two summations be $P_1$ and the terms of the second two summations be $P_2$, respectively.
$P_1$ can be further written as 
\begin{align}
    & P_1 =  \sum_{k=1}^{\Delta+2} \sum_{j=\Delta+2-k}^\infty  \text{Pr}(j) \cdot \Big [\text{Pr}(k,l=0|j), \text{Pr}(k,l=1|j), \cdots\Big] \begin{tikzpicture}[baseline = (M.west)]
    \tikzset{brace/.style = {decorate, decoration = {brace, amplitude = 5pt}, thick}}
    \matrix(M)
    [
        matrix of math nodes,
        left delimiter = (,
        right delimiter = )
    ]
    {
        k+j-\Delta -1 \\
        \vdots \\
        k+j-\Delta -1 \\
        k+j-\Delta -2 \\
        k+j-\Delta -3 \\
        \vdots \\
    };
    \draw[brace]
        ($(M-1-1.north east) + (0.45, 0)$)
        -- node[right = 5pt]{$\Delta +2$}
        ($(M-3-1.south east) + (0.45, 0)$);
    \end{tikzpicture}\\
    &= \sum_{k=1}^{\Delta+2} \begin{bmatrix}
        \text{Pr}(j=\Delta+2-k) \\
        \text{Pr}(j=\Delta+3-k) \\
        \text{Pr}(j=\Delta+4-k) \\
        \vdots
    \end{bmatrix}^\top
    \begin{bmatrix}
        \Big[\text{Pr}(k,l=0|j=\Delta+2-k), \text{Pr}(k,l=1|j=\Delta+2-k),  \cdots\Big]\begin{tikzpicture}[baseline = (M.west)]
    \tikzset{brace/.style = {decorate, decoration = {brace, amplitude = 5pt}, thick}}
    \matrix(M)
    [
        matrix of math nodes,
        left delimiter = (,
        right delimiter = )
    ]
    {
        1 \\
        \vdots \\
        1 \\
        0 \\
        -1 \\
        \vdots \\
    };
    \draw[brace]
        ($(M-1-1.north east) + (0.5, 0)$)
        -- node[right = 5pt]{$\Delta +2$}
        ($(M-3-1.south east) + (0.5, 0)$);
    \end{tikzpicture}   \\
    \Big[\text{Pr}(k,l=0|j=\Delta+3-k), \text{Pr}(k,l=1|j=\Delta+3-k),  \cdots\Big]\begin{tikzpicture}[baseline = (M.west)]
    \tikzset{brace/.style = {decorate, decoration = {brace, amplitude = 5pt}, thick}}
    \matrix(M)
    [
        matrix of math nodes,
        left delimiter = (,
        right delimiter = )
    ]
    {
        2 \\
        \vdots \\
        2 \\
        1 \\
        0 \\
        \vdots \\
    };
    \draw[brace]
        ($(M-1-1.north east) + (0.5, 0)$)
        -- node[right = 5pt]{$\Delta +2$}
        ($(M-3-1.south east) + (0.5, 0)$);
    \end{tikzpicture}   \\
    \Big[\text{Pr}(k,l=0|j=\Delta+4-k), \text{Pr}(k,l=1|j=\Delta+4-k),  \cdots\Big]\begin{tikzpicture}[baseline = (M.west)]
    \tikzset{brace/.style = {decorate, decoration = {brace, amplitude = 5pt}, thick}}
    \matrix(M)
    [
        matrix of math nodes,
        left delimiter = (,
        right delimiter = )
    ]
    {
        3 \\
        \vdots \\
        3 \\
        2 \\
        1 \\
        \vdots \\
    };
    \draw[brace]
        ($(M-1-1.north east) + (0.5, 0)$)
        -- node[right = 5pt]{$\Delta +2$}
        ($(M-3-1.south east) + (0.5, 0)$);
    \end{tikzpicture}\\
    \vdots
    \end{bmatrix}\\
    &= \sum_{k=1}^{\Delta+2} \begin{bmatrix}
        \text{Pr}(j=\Delta+2-k) \\
        \text{Pr}(j=\Delta+3-k) \\
        \text{Pr}(j=\Delta+4-k) \\
        \vdots
    \end{bmatrix}^\top
    \begin{bmatrix}
        \Big[\text{Pr}(k,l=0|j=\Delta+2-k), \text{Pr}(k,l=1|j=\Delta+2-k),  \cdots\Big]
        \left[
        \begin{tikzpicture}[baseline = (M.west)]
    \tikzset{brace/.style = {decorate, decoration = {brace, amplitude = 5pt}, thick}}
    \matrix(M)
    [
        matrix of math nodes,
        left delimiter = (,
        right delimiter = )
    ]
    {
        1 \\
        \vdots \\
        1 \\
        0 \\
        -1 \\
        \vdots \\
    };
    \draw[brace]
        ($(M-1-1.north east) + (0.5, 0)$)
        -- node[right = 5pt]{$\Delta +2$}
        ($(M-3-1.south east) + (0.5, 0)$);
    \end{tikzpicture}  \!\!\!\!\!+  
    \begin{pmatrix}
        0\\
        \vdots\\
        0\\
        \vdots
    \end{pmatrix}
    \right]   \\
    \Big[\text{Pr}(k,l=0|j=\Delta+3-k), \text{Pr}(k,l=1|j=\Delta+3-k),  \cdots\Big]\left[
        \begin{tikzpicture}[baseline = (M.west)]
    \tikzset{brace/.style = {decorate, decoration = {brace, amplitude = 5pt}, thick}}
    \matrix(M)
    [
        matrix of math nodes,
        left delimiter = (,
        right delimiter = )
    ]
    {
        1 \\
        \vdots \\
        1 \\
        0 \\
        -1 \\
        \vdots \\
    };
    \draw[brace]
        ($(M-1-1.north east) + (0.5, 0)$)
        -- node[right = 5pt]{$\Delta +2$}
        ($(M-3-1.south east) + (0.5, 0)$);
    \end{tikzpicture}  \!\!\!\!\! + 
    \begin{pmatrix}
        1\\
        \vdots\\
        1\\
        \vdots
    \end{pmatrix}
    \right]   \\
    \Big[\text{Pr}(k,l=0|j=\Delta+4-k), \text{Pr}(k,l=1|j=\Delta+4-k),  \cdots\Big]\left[
        \begin{tikzpicture}[baseline = (M.west)]
    \tikzset{brace/.style = {decorate, decoration = {brace, amplitude = 5pt}, thick}}
    \matrix(M)
    [
        matrix of math nodes,
        left delimiter = (,
        right delimiter = )
    ]
    {
        1 \\
        \vdots \\
        1 \\
        0 \\
        -1 \\
        \vdots \\
    };
    \draw[brace]
        ($(M-1-1.north east) + (0.5, 0)$)
        -- node[right = 5pt]{$\Delta +2$}
        ($(M-3-1.south east) + (0.5, 0)$);
    \end{tikzpicture}  \!\!\!\!\! +  2\cdot
    \begin{pmatrix}
        1\\
        \vdots\\
        1\\
        \vdots
    \end{pmatrix}
    \right]\\
    \vdots
    \end{bmatrix}\\
    &= \sum_{k=1}^{\Delta+2} \begin{bmatrix}
        \text{Pr}(j=\Delta+2-k) \\
        \text{Pr}(j=\Delta+3-k) \\
        \text{Pr}(j=\Delta+4-k) \\
        \vdots
    \end{bmatrix}^\top \!\! \left[
    \begin{bmatrix}
        \text{Pr}(k,l=0|j=\Delta+2-k) & \text{Pr}(k,l=1|j=\Delta+2-k) & \cdots \\
        \text{Pr}(k,l=0|j=\Delta+3-k) & \text{Pr}(k,l=1|j=\Delta+3-k) & \cdots \\
        \text{Pr}(k,l=0|j=\Delta+4-k) & \text{Pr}(k,l=1|j=\Delta+4-k) & \cdots \\
        \vdots & \vdots & \ddots
    \end{bmatrix}\!\!\!
    \begin{bmatrix}
        1 \\
        \vdots \\
        1 \\
        0 \\
        -1 \\
        \vdots \\
    \end{bmatrix} \!\!+\!\! 
    \begin{bmatrix}
        \text{Pr}(k|j=\Delta+2-k) \cdot 0 \\
        \text{Pr}(k|j=\Delta+3-k) \cdot 1 \\
        \text{Pr}(k|j=\Delta+4-k) \cdot 2 \\
        \text{Pr}(k|j=\Delta+5-k) \cdot 3 \\
        \vdots
    \end{bmatrix}
    \right] \\
    &\overset{(b)}{=} \sum_{k=1}^{\Delta+2} \begin{bmatrix}
        \text{Pr}(j=\Delta+2-k) \\
        \text{Pr}(j=\Delta+3-k) \\
        \text{Pr}(j=\Delta+4-k) \\
        \vdots
    \end{bmatrix}^\top \left[
    \widetilde{T}_{0\rightarrow 0}^{(k)}(\Delta+2-k:+\infty,:)\cdot 
    \begin{bmatrix}
        1 \\
        \vdots \\
        1 \\
        0 \\
        -1 \\
        \vdots \\
    \end{bmatrix}  + 
    \begin{bmatrix}
        0 &   &   &   &\\
          & 1 &   &   &\\
          &   & 2 &   &\\
          &   &   & 3 &\\
          &   &   &   &\ddots \\
    \end{bmatrix}\cdot
    \widetilde{T}_{0\rightarrow 0}^{(k)}(\Delta+2-k:+\infty,:)
    \begin{bmatrix}
        1 \\
        \vdots \\
        1 \\
        1 \\
        1 \\
        \vdots \\
    \end{bmatrix}
    \right] \\
    &\overset{(c)}{=}  \sum_{k=1}^{\Delta+2} \begin{bmatrix}
        \text{Pr}(j=0) \\
        \text{Pr}(j=1) \\
        \text{Pr}(j=2) \\
        \vdots
    \end{bmatrix}^\top 
    \widetilde{T}_{0\rightarrow 0}^{(k)}
    \begin{bmatrix}
        1 \\
        \vdots \\
        1 \\
        0 \\
        -1 \\
        \vdots \\
    \end{bmatrix} - 
    \sum_{k=1}^{\Delta+1} \begin{bmatrix}
        \text{Pr}(j=0) \\
        \text{Pr}(j=1) \\
        \text{Pr}(j=2) \\
        \vdots\\
        \text{Pr}(j=\Delta + 1 - k)
    \end{bmatrix}^\top 
    \widetilde{T}_{0\rightarrow 0}^{(k)}(0:\Delta+1-k,:)
    \begin{bmatrix}
        1 \\
        \vdots \\
        1 \\
        0 \\
        -1 \\
        \vdots \\
    \end{bmatrix} \nonumber\\
    &\qquad\qquad\qquad\qquad\qquad\qquad\qquad\qquad\qquad\qquad\qquad\qquad+ \sum_{k=1}^{\Delta+2} \begin{bmatrix}
        \text{Pr}(j=0) \\
        \text{Pr}(j=1) \\
        \text{Pr}(j=2) \\
        \vdots
    \end{bmatrix}^\top 
    \begin{bmatrix}
        \mathbf{0}_{\Delta+2-k} & \\
         & B
    \end{bmatrix}
    \widetilde{T}_{0\rightarrow 0}^{(k)}
    \begin{bmatrix}
        1 \\
        \vdots \\
        1 \\
        1 \\
        1 \\
        \vdots \\
    \end{bmatrix}
    . \label{equation:P_1}
\end{align}

In equality (b), matrix $\widetilde{T}_{0\rightarrow0}^{(k)}$ is the joint transition matrix of the summation of the hidden states, while event $t_{i_0+1}-t_{i_0} = k$ occurs.
Formally, its entries are defined as 
$\widetilde{t}^{(k)}(j,l) = \text{Pr}\big(\omega = k,\beta = l\big|\alpha = j\big)$.
And $T_{0\rightarrow 0}^{(k)}(0:\Delta+1-k,:)$ is a submatrix of $T_{0\rightarrow0}^{(k)}$ consisting of  its first $\Delta+2-k$ rows (starting with index 0).
Equality (c) is derived by complementing the terms from $j=0$ to $j=\Delta+1-k$.
Recall that in Proposition \ref{proposition:identity=zero}, to duel with the transition of summation of the hidden states, we introduce matrix $\widetilde{T}_{0\rightarrow 0}=\left[\widetilde{t}(j,l)\right]$, where $\widetilde{t}(j,l) \triangleq \text{Pr}\big(\widehat{D}_0(t_{i_0+1}) + \widehat{D}_1(t_{i_0+1})=l \big| \widehat{D}_0(t_{i_0}) + \widehat{D}_1(t_{i_0})=j\big)$.
$\widetilde{T}_{0\rightarrow 0}$ can be generated from $T_{0\rightarrow 0}$ by multiplication $\widetilde{T}_{0\rightarrow 0} = \mathbf{P} \cdot T_{0\rightarrow 0} \cdot \mathbf{Q}$, where $\mathbf{P},\mathbf{Q}$ are the left-summation and right-summation matrices, respectively.
Similarly, $\widetilde{T}_{0\rightarrow0}^{(k)}$ can be also given by multiplying the summation matrices $\mathbf{P},\mathbf{Q}$ with $T_{0\rightarrow0}^{(k)}$, i.e., $\widetilde{T}_{0\rightarrow0}^{(k)} = \mathbf{P} \cdot T_{0\rightarrow0}^{(k)} \cdot \mathbf{Q}$.
Besides, we denote $ \big[\text{Pr}(j=0),\text{Pr}(j=1),\cdots\big] = \widetilde{\pi}^{\infty} = \pi^{\infty} \cdot \mathbf{Q}$.

$P_2$ can be further written as 
\begin{align}
    & P_2 =  \sum_{k=\Delta + 3}^{\infty} \sum_{j=0}^\infty  \text{Pr}(j) \cdot \Big [\text{Pr}(k,l=0|j), \text{Pr}(k,l=1|j), \cdots\Big] \begin{tikzpicture}[baseline = (M.west)]
    \tikzset{brace/.style = {decorate, decoration = {brace, amplitude = 5pt}, thick}}
    \matrix(M)
    [
        matrix of math nodes,
        left delimiter = (,
        right delimiter = )
    ]
    {
        k+j-\Delta -1 \\
        \vdots \\
        k+j-\Delta -1 \\
        k+j-\Delta -2 \\
        k+j-\Delta -3 \\
        \vdots \\
    };
    \draw[brace]
        ($(M-1-1.north east) + (0.45, 0)$)
        -- node[right = 5pt]{$\Delta +2$}
        ($(M-3-1.south east) + (0.45, 0)$);
    \end{tikzpicture} \\
    &= \sum_{k=\Delta+3}^{\infty} \begin{bmatrix}
        \text{Pr}(j=0) \\
        \text{Pr}(j=1) \\
        \text{Pr}(j=2) \\
        \vdots
    \end{bmatrix}^\top
    \begin{bmatrix}
        \Big[\text{Pr}(k,l=0|j=0), \text{Pr}(k,l=1|j=0),  \cdots\Big]\begin{tikzpicture}[baseline = (M.west)]
    \tikzset{brace/.style = {decorate, decoration = {brace, amplitude = 5pt}, thick}}
    \matrix(M)
    [
        matrix of math nodes,
        left delimiter = (,
        right delimiter = )
    ]
    {
        k-\Delta-1 \\
        \vdots \\
        k-\Delta-1 \\
        k-\Delta-2 \\
        k-\Delta-3 \\
        \vdots \\
    };
    \draw[brace]
        ($(M-1-1.north east) + (0.5, 0)$)
        -- node[right = 5pt]{$\Delta +2$}
        ($(M-3-1.south east) + (0.5, 0)$);
    \end{tikzpicture}   \\
    \Big[\text{Pr}(k,l=0|j=1), \text{Pr}(k,l=1|j=1),  \cdots\Big]\begin{tikzpicture}[baseline = (M.west)]
    \tikzset{brace/.style = {decorate, decoration = {brace, amplitude = 5pt}, thick}}
    \matrix(M)
    [
        matrix of math nodes,
        left delimiter = (,
        right delimiter = )
    ]
    {
        k-\Delta \\
        \vdots \\
        k-\Delta \\
        k-\Delta-1 \\
        k-\Delta-2 \\
        \vdots \\
    };
    \draw[brace]
        ($(M-1-1.north east) + (0.7, 0)$)
        -- node[right = 5pt]{$\Delta +2$}
        ($(M-3-1.south east) + (0.7, 0)$);
    \end{tikzpicture}   \\
    \Big[\text{Pr}(k,l=0|j=2), \text{Pr}(k,l=1|j=2),  \cdots\Big]\begin{tikzpicture}[baseline = (M.west)]
    \tikzset{brace/.style = {decorate, decoration = {brace, amplitude = 5pt}, thick}}
    \matrix(M)
    [
        matrix of math nodes,
        left delimiter = (,
        right delimiter = )
    ]
    {
        k-\Delta+1 \\
        \vdots \\
        k-\Delta+1 \\
        k-\Delta \\
        k-\Delta-1 \\
        \vdots \\
    };
    \draw[brace]
        ($(M-1-1.north east) + (0.5, 0)$)
        -- node[right = 5pt]{$\Delta +2$}
        ($(M-3-1.south east) + (0.5, 0)$);
    \end{tikzpicture}\\
    \vdots
    \end{bmatrix}\\
    &= \sum_{k=\Delta+3}^{\infty} \begin{bmatrix}
        \text{Pr}(j=0) \\
        \text{Pr}(j=1) \\
        \text{Pr}(j=2) \\
        \vdots
    \end{bmatrix}^\top
    \begin{bmatrix}
        \Big[\text{Pr}(k,l=0|j=0), \text{Pr}(k,l=1|j=0),  \cdots\Big]
        \left[
        \begin{tikzpicture}[baseline = (M.west)]
    \tikzset{brace/.style = {decorate, decoration = {brace, amplitude = 5pt}, thick}}
    \matrix(M)
    [
        matrix of math nodes,
        left delimiter = (,
        right delimiter = )
    ]
    {
        k-\Delta-1 \\
        \vdots \\
        k-\Delta-1 \\
        k-\Delta-2 \\
        k-\Delta-3 \\
        \vdots \\
    };
    \draw[brace]
        ($(M-1-1.north east) + (0.5, 0)$)
        -- node[right = 5pt]{$\Delta +2$}
        ($(M-3-1.south east) + (0.5, 0)$);
    \end{tikzpicture}  \!\!\!\!\!+  
    \begin{pmatrix}
        0\\
        \vdots\\
        0\\
        \vdots
    \end{pmatrix}
    \right]   \\
    \Big[\text{Pr}(k,l=0|j=1), \text{Pr}(k,l=1|j=1),  \cdots\Big]\left[
        \begin{tikzpicture}[baseline = (M.west)]
    \tikzset{brace/.style = {decorate, decoration = {brace, amplitude = 5pt}, thick}}
    \matrix(M)
    [
        matrix of math nodes,
        left delimiter = (,
        right delimiter = )
    ]
    {
        k-\Delta-1 \\
        \vdots \\
        k-\Delta-1 \\
        k-\Delta-2 \\
        k-\Delta-3 \\
        \vdots \\
    };
    \draw[brace]
        ($(M-1-1.north east) + (0.5, 0)$)
        -- node[right = 5pt]{$\Delta +2$}
        ($(M-3-1.south east) + (0.5, 0)$);
    \end{tikzpicture}  \!\!\!\!\! + 
    \begin{pmatrix}
        1\\
        \vdots\\
        1\\
        \vdots
    \end{pmatrix}
    \right]   \\
    \Big[\text{Pr}(k,l=0|j=2), \text{Pr}(k,l=1|j=2),  \cdots\Big]\left[
        \begin{tikzpicture}[baseline = (M.west)]
    \tikzset{brace/.style = {decorate, decoration = {brace, amplitude = 5pt}, thick}}
    \matrix(M)
    [
        matrix of math nodes,
        left delimiter = (,
        right delimiter = )
    ]
    {
        k-\Delta-1 \\
        \vdots \\
        k-\Delta-1 \\
        k-\Delta-2 \\
        k-\Delta-3 \\
        \vdots \\
    };
    \draw[brace]
        ($(M-1-1.north east) + (0.5, 0)$)
        -- node[right = 5pt]{$\Delta +2$}
        ($(M-3-1.south east) + (0.5, 0)$);
    \end{tikzpicture}  \!\!\!\!\! +  2\cdot
    \begin{pmatrix}
        1\\
        \vdots\\
        1\\
        \vdots
    \end{pmatrix}
    \right]\\
    \vdots
    \end{bmatrix}\\
    &= \sum_{k=\Delta+3}^{\infty} \begin{bmatrix}
        \text{Pr}(j=0) \\
        \text{Pr}(j=1) \\
        \text{Pr}(j=2) \\
        \vdots
    \end{bmatrix}^\top \!\! \left[
    \begin{bmatrix}
        \text{Pr}(k,l=0|j=0) & \text{Pr}(k,l=1|j=0) & \cdots \\
        \text{Pr}(k,l=0|j=1) & \text{Pr}(k,l=1|j=1) & \cdots \\
        \text{Pr}(k,l=0|j=2) & \text{Pr}(k,l=1|j=2) & \cdots \\
        \vdots & \vdots & \ddots
    \end{bmatrix}\!\!\!
    \begin{bmatrix}
        k-\Delta-1 \\
        \vdots \\
        k-\Delta-1 \\
        k-\Delta-2 \\
        k-\Delta-3 \\
        \vdots 
    \end{bmatrix} \!\!+\!\! 
    \begin{bmatrix}
        \text{Pr}(k|j=0) \cdot 0 \\
        \text{Pr}(k|j=1) \cdot 1 \\
        \text{Pr}(k|j=2) \cdot 2 \\
        \text{Pr}(k|j=3) \cdot 3 \\
        \vdots
    \end{bmatrix}
    \right] \\
    &= \sum_{k=\Delta+3}^{\infty} \begin{bmatrix}
        \text{Pr}(j=0) \\
        \text{Pr}(j=1) \\
        \text{Pr}(j=2) \\
        \vdots
    \end{bmatrix}^\top \left[
    \widetilde{T}_{0\rightarrow 0}^{(k)}\cdot 
    \begin{bmatrix}
        k-\Delta-1 \\
        \vdots \\
        k-\Delta-1 \\
        k-\Delta-2 \\
        k-\Delta-3 \\
        \vdots
    \end{bmatrix}  + 
    \begin{bmatrix}
        0 &   &   &   &\\
          & 1 &   &   &\\
          &   & 2 &   &\\
          &   &   & 3 &\\
          &   &   &   &\ddots \\
    \end{bmatrix}
    \widetilde{T}_{0\rightarrow 0}^{(k)}\cdot \vec{\mathbf{1}}
    \right].\label{equation:P_2}
\end{align}

With (\ref{equation:P_1}) and (\ref{equation:P_2}), recall that $\vec{\gamma} = [\overbrace{1,\cdots,1}^{\Delta + 2},0,-1,-2,\cdots]^\top$, $\big[\text{Pr}(j=0),\text{Pr}(j=1),\cdots\big] = \pi^{\infty} \cdot \mathbf{Q}$, $T_{0\rightarrow 0}^{(1)}=\mathbf{T}_{0,0}$, and $T_{0\rightarrow 0}^{(k)} = \mathbf{T}_{0,\phi}\mathbf{T}_{\phi,\phi}^{k-2}\mathbf{T}_{\phi,0}, \forall k\ge 2$, the numerator can be derived by
\begin{align}
    &\mathbb{E}\{\lambda(k,j,l)\} = P_1 + P_2 \nonumber\\
    &= \sum_{k=1}^{\Delta+2} \widetilde{\pi}^{\infty}
    \widetilde{T}_{0\rightarrow 0}^{(k)} \vec{\gamma} - 
    \sum_{k=1}^{\Delta+1} \widetilde{\pi}^{\infty}(0:\Delta + 1 - k) \widetilde{T}_{0\rightarrow 0}^{(k)}(0:\Delta+1-k,:) \vec{\gamma} + 
    \sum_{k=1}^{\Delta+2} \widetilde{\pi}^{\infty} 
    \begin{bmatrix}
        \mathbf{0}_{\Delta+2-k} & \\
         & B
    \end{bmatrix}
    \widetilde{T}_{0\rightarrow 0}^{(k)} \vec{\mathbf{1}} \nonumber\\
    & \qquad\qquad\qquad\qquad\qquad\qquad\qquad+
    \sum_{k=\Delta+3}^{\infty} \widetilde{\pi}^{\infty}  \widetilde{T}_{0\rightarrow 0}^{(k)} \begin{bmatrix}
        k-\Delta-1 \\
        \vdots \\
        k-\Delta-1 \\
        k-\Delta-2 \\
        k-\Delta-3 \\
        \vdots
    \end{bmatrix} 
     + \sum_{k=\Delta+3}^{\infty} \widetilde{\pi}^{\infty} \cdot B \cdot \widetilde{T}_{0\rightarrow 0}^{(k)} \vec{\mathbf{1}}\\
     &= \pi^{\infty}\mathbf{Q} \mathbf{P}\left(\mathbf{T}_{0,0} + \mathbf{T}_{0,\phi}\sum_{k=2}^{\Delta+2}\mathbf{T}_{\phi,\phi}^{k-2}\mathbf{T}_{\phi,0}\right)\mathbf{Q} \vec{\gamma} + 
     \sum_{k=\Delta+3}^{\infty} \pi^{\infty}\mathbf{Q} \mathbf{P}T_{0\rightarrow 0}^{(k)} \mathbf{Q} \cdot k\cdot \begin{bmatrix}
        1 \\
        \vdots \\
        1 \\
        1 \\
        1 \\
        \vdots
    \end{bmatrix} -
     \sum_{k=\Delta+3}^{\infty} \pi^{\infty}\mathbf{Q}  \mathbf{P}T_{0\rightarrow 0}^{(k)}\mathbf{Q} \begin{bmatrix}
        \Delta+1 \\
        \vdots \\
        \Delta+1 \\
        \Delta+2 \\
        \Delta+3 \\
        \vdots
    \end{bmatrix} + \nonumber\\
    &\sum_{k=\Delta+3}^{\infty} \pi^{\infty}\mathbf{Q} B \mathbf{P} T_{0\rightarrow 0}^{(k)} \mathbf{Q} \vec{\mathbf{1}} + \sum_{k=1}^{\Delta+2} \pi^{\infty}\mathbf{Q} 
    \begin{bmatrix}
        \mathbf{0}_{\Delta+2-k} & \\
         & B
    \end{bmatrix} \mathbf{P}
    T_{0\rightarrow 0}^{(k)}\mathbf{Q} \vec{\mathbf{1}} - 
    \sum_{k=1}^{\Delta+1} \pi^{\infty}(0:\Delta + 1 - k)\mathbf{Q} \mathbf{P}T_{0\rightarrow 0}^{(k)}(0:\Delta+1-k,:)\mathbf{Q} \vec{\gamma} \\
    &= \pi^{\infty}\mathbf{Q}\mathbf{P} \left(\mathbf{T}_{0,0} + \mathbf{T}_{0,\phi} \big(\mathbf{I}-\mathbf{T}_{\phi,\phi}^{\Delta+1}\big)\big(\mathbf{I}-\mathbf{T}_{\phi,\phi}\big)^{-1}    
    \mathbf{T}_{\phi,0}\right) \mathbf{Q} \vec{\gamma} + 
    \pi^{\infty}\mathbf{Q} \mathbf{P}\mathbf{T}_{0,\phi} \mathbf{T}_{\phi,\phi}^{\Delta+1} \left[(\Delta+2)\mathbf{I}  \!+  \! \big(\mathbf{I}-\mathbf{T}_{\phi,\phi}\big)^{-1})\right] \!\!\big(\mathbf{I}-\mathbf{T}_{\phi,\phi}\big)^{-1} \mathbf{T}_{\phi,0}\mathbf{Q} \vec{\mathbf{1}} \nonumber\\
    &\quad + \pi^{\infty}\mathbf{Q}\mathbf{P} \mathbf{T}_{0,\phi} \mathbf{T}_{\phi,\phi}^{\Delta+1} \cdot \big(\mathbf{I}-\mathbf{T}_{\phi,\phi}\big)^{-1} \mathbf{T}_{\phi,0}\mathbf{Q} \left[-(\Delta+2)\begin{bmatrix}
        1 \\
        \vdots \\
        1 \\
        1 \\
        1 \\
        \vdots
    \end{bmatrix} + \begin{bmatrix}
        1 \\
        \vdots \\
        1 \\
        0 \\
        -1 \\
        \vdots
    \end{bmatrix}\right] +
    \pi^{\infty}\mathbf{Q} B \mathbf{P} \mathbf{T}_{0,\phi}  \mathbf{T}_{\phi,\phi}^{\Delta+1} \big(\mathbf{I}-\mathbf{T}_{\phi,\phi}\big)^{-1} \mathbf{T}_{\phi,0}\mathbf{Q} \vec{\mathbf{1}}  \nonumber\\
    & \qquad\qquad\qquad + \sum_{k=1}^{\Delta+2} \pi^{\infty} \mathbf{Q}
    \begin{bmatrix}
        \mathbf{0}_{\Delta+2-k} & \\
         & B
    \end{bmatrix} \mathbf{P}
    T_{0\rightarrow 0}^{(k)} \mathbf{Q} \vec{\mathbf{1}} - 
    \sum_{k=1}^{\Delta+1} \pi^{\infty}(0:\Delta + 1 - k) \mathbf{Q} \mathbf{P} T_{0\rightarrow 0}^{(k)}(0:\Delta+1-k,:) \mathbf{Q} \vec{\gamma} \\
    &= \pi^\infty \mathbf{Q} \mathbf{P} (\mathbf{T}_{0,0} + \mathbf{T}_{0,\phi} A \mathbf{T}_{\phi,0}) \mathbf{Q} \vec{\gamma} 
        + \pi^\infty \mathbf{Q} \mathbf{P} \mathbf{T}_{0,\phi} \mathbf{T}_{\phi,\phi}^{\Delta + 1} A^2 \mathbf{T}_{\phi,0} \mathbf{Q} \vec{\mathbf{1}}
        + \pi^\infty \mathbf{Q} B \mathbf{P} \mathbf{T}_{0,\phi}\mathbf{T}_{\phi,\phi}^{\Delta + 1} A\mathbf{T}_{\phi,0} \mathbf{Q} \vec{\mathbf{1}}
        +  \nonumber\\
        & \qquad\qquad \pi^\infty \mathbf{Q} \sum_{k=1}^{\Delta+2}\begin{bmatrix}
        \mathbf{0}_{\Delta+2-k} & \\
         & B \end{bmatrix} \mathbf{P} \mathbf{T}_{0,\phi} \mathbf{T}_{\phi,\phi}^{k-2} \mathbf{T}_{\phi,0} \mathbf{Q} \vec{\mathbf{1}} - \sum_{k=1}^{\Delta+1} \pi^{\infty}(0:\Delta + 1 - k) \mathbf{Q} \mathbf{P}  T_{0\rightarrow 0}^{(k)}(0:\Delta+1-k,:) \mathbf{Q}\vec{\gamma}.
\end{align}
The $\mathbf{I}$ in above equations are $\mathbf{I}_{m_0\cdot m_1 \cdot (m_2-1)}$. We omit the subscript for ease of presentation.
Thus, Lemma \ref{lemma:numerator} is proved.

\end{appendices}

\bibliographystyle{IEEEtran}
\begin{spacing}{1.0}
		
\end{spacing}

\end{document}